\documentclass[a4paper,11pt]{article}
\usepackage{epsfig}
\usepackage{cite}
\usepackage{url}
\usepackage{hyperref}
\usepackage{placeins}

\usepackage{epsfig}
\usepackage{amsmath}
\usepackage{amsfonts}
\usepackage{graphicx}
\usepackage{cite}
\usepackage{color}
\usepackage[dvipsnames]{xcolor}
\usepackage{multirow}

\usepackage{amssymb}

\date{\today}

\newcommand{\insertplot}[5]{\begin{figure}
 \hfill\hbox to 0.05in{\vbox to #5in{\vfill
 \inputplot{#1}{#4}{#5}}\hfill}
 \hfill\vspace{-.1in}
 \caption{#2}\label{#3}
 \end{figure}}
 \newcommand{\inputplot}[3]{
 \special{ps: plotfile #1}
}
\newcounter{fig}   

\numberwithin{equation}{section}

\newcommand{\ee}{\end{equation}}
\newcommand{\eea}{\end{eqnarray}}
\newcommand{\be}{\begin{equation}}
\newcommand{\bea}{\begin{eqnarray}}

\begin{document}

\title{{\bf  \LARGE Chaotic lensing around boson stars \\ and  Kerr black holes with scalar hair}}

 \author{
{\large P. V. P. Cunha}$^{1,2}$, \
{\large J. Grover}$^{3}$, \
{\large C. Herdeiro}$^{1}$, \\
{\large E. Radu}$^{1}$,  \
{\large H. R\'unarsson}$^{1}$
 and
{\large A. Wittig}$^{3}$
\\
\\
$^{1}${\small Departamento de F\'\i sica da Universidade de Aveiro and  } \\ {\small  Centre for Research and Development  in Mathematics and Applications (CIDMA),
 } \\ {\small    Campus de Santiago, 3810-183 Aveiro, Portugal}
 \\ \texttt{\small  cunhapcc@gmail.com; herdeiro@ua.pt; eugen.radu@ua.pt; helgi.runarsson@gmail.com}
 \\
 \\
$^{2}${\small CENTRA, Departamento de F\'\i sica, Instituto Superior T\'ecnico} \\ {\small Universidade de Lisboa, Avenida Rovisco Pais 1, 1049, Lisboa, Portugal}
 \\
 \\
$^{3}${\small ESA -- Advanced Concepts Team, European Space Research Technology Centre (ESTEC),
  } \\ {\small   Keplerlaan 1, Postbus 299, NL-2200 AG Noordwijk, The
Netherlands}
 \\ \texttt{\small  jai.grover@esa.int; alexander.wittig@esa.int}
}
\date{September 2016}
\maketitle

\begin{abstract}
In a recent letter~\cite{Cunha:2015yba}, it was shown that the lensing of light around rotating boson stars and Kerr black holes with scalar hair can exhibit chaotic patterns. Since no separation of variables is known (or expected) for geodesic motion on these backgrounds, we examine the 2D effective potentials for photon trajectories, to obtain a deeper understanding of this phenomenon. We find that the emergence of \textit{stable light rings} on the background spacetimes, allows the formation of ``pockets" in one of the effective potentials, for open sets of impact parameters, leading to an effective trapping of some trajectories, dubbed \textit{quasi-bound orbits}. We conclude that pocket formation induces chaotic scattering, although not \textit{all} chaotic orbits are associated to pockets.  These and other features are illustrated in a gallery of examples, obtained with a new ray-tracing code, \textsc{pyhole}, which includes tools for a simple, simultaneous visualization of the effective potential together with the  spacetime trajectory, for any given point in a lensing image. An analysis of photon orbits allows us to further establish a positive correlation between photon orbits in chaotic regions and those with more than one turning point in the radial direction; we recall that the latter is not possible around Kerr black holes. Moreover, we observe that the existence of several light rings around  a horizon (several \textit{fundamental orbits}, including a stable one),  is a central ingredient for the existence of multiple shadows of a single hairy black hole. We also exhibit the lensing and shadows by Kerr black holes with scalar hair, observed away from the equatorial plane, obtained with \textsc{pyhole}.
\end{abstract}

\newpage

\tableofcontents

\newpage

\section{Introduction}
The effect of gravitational lensing was considered by Einstein even prior to the completion of his General Theory of Relativity (GR)~(see~\cite{1997Sci...275..184R} for an historical account). In particular, in 1912, he derived the basic lensing equation and magnification factor for the intensity of the deflected light. These results, however, were only published in 1936, in a paper often considered the pioneering study on gravitational lensing~\cite{Einstein:1956zz}, where Einstein discusses that a gravitational lens can lead to both multiple images and ring shaped images (subsequently called \textit{Einstein rings}) of a star. Both these effects were actually discussed by other authors in the 1920s, the first one by Eddington~\cite{Eddington:1987tk} and the second by Chwolson~\cite{Chwolson}, thus before Einstein's 1936 paper~\cite{1997Sci...275..184R}.

At the time of Einstein, the prospects for observing this type of lensing were dim. By contrast, the phenomenon at the origin of the gravitational lensing, $i.e.$ the bending of light by a gravitational field -- and in particular that caused by the Sun --,  had been instrumental in establishing GR as a physical theory of the Universe. It was only with the discovery of quasars, in the 1960s~\cite{1963Natur.197.1040S}, that the subject of gravitational lensing was brought into the realm of observational astronomy. Being both very distant and very bright objects, quasars are ideal light sources for observing lensing effects, when a deflecting mass, typically a galaxy, is present along their line of sight. The first lensing effect of a distant quasar (a double image) was identified in 1979~\cite{1979Natur.279..381W} and, since then, many other systems with both multiple images and Einstein rings have been discovered (see $e.g.$~\cite{lensingcat}).

The largest lensing effects that have been observed, at present, in astrophysical objects (and cosmological contexts), are of the order of tens of arc seconds (see $e.g.$~\cite{2003Natur.426..810I}), corresponding to tiny local light bendings by (typically) lensing galaxies. Ultra-compact objects can, on the other hand, cause much more extreme local deflections of light.  Black holes (BHs), in particular, can possess \textit{light rings} and hence can bend light by an \textit{arbitrarily large angle}. For the paradigmatic Kerr BH spacetimes of GR, these light rings are \textit{unstable}. Their existence allows light to circle any number of times around the light ring before being scattered back to infinity (or fall into the BH). From the viewpoint of an observer which sees the BH lit by a distant celestial sphere, an infinite number of smaller and smaller copies of the whole celestial sphere accumulate near the edge of the absorption cross-section for light (at high frequencies) -- dubbed the~\textit{BH shadow}~\cite{Bardeen1973,Falcke:1999pj} -- in an organized self-similar structure -- see~\cite{Bohn:2014xxa} for striking visualizations of this effect in the Schwarzschild and Kerr BH spacetimes, and~\cite{Amarilla:2011fx,Yumoto:2012kz,Abdujabbarov:2012bn,Amarilla:2013sj,Nedkova:2013msa,Atamurotov:2013dpa,Atamurotov:2013sca,Li:2013jra,Tinchev:2013nba,Wei:2013kza,Tsukamoto:2014tja,Grenzebach:2014fha,Lu:2014zja,Papnoi:2014aaa,Sakai:2014pga,Psaltis:2014mca,Wei:2015dua,Abdolrahimi:2015rua,Moffat:2015kva,Grenzebach:2015uva,Vincent:2015xta,Grenzebach:2015oea,Abdujabbarov:2015xqa,Ortiz:2015rma,Ghasemi-Nodehi:2015raa,Ohgami:2015nra,Atamurotov:2015xfa, Perlick:2015vta,Bambi:2015rda,Atamurotov:2015nra,Yang:2015hwf,Tinchev:2015apf,Shipley:2016omi,Dolan:2016bxj,Amir:2016cen,Cunha:2015yba,Johannsen:2015hib,Abdujabbarov:2016hnw,Cunha:2016bpi,Huang:2016qnl,Dastan:2016vhb,Younsi:2016azx} for examples of recent investigations of BH shadows and lensing by compact objects in different models.

\bigskip

In a recent letter~\cite{Cunha:2015yba}, some of us have studied the lensing and shadows of a deformed type of Kerr BHs, known as Kerr BHs with scalar hair (KBHsSH)~\cite{Herdeiro:2014goa,Herdeiro:2015gia,Herdeiro:2014ima} (see also~\cite{Herdeiro:2014jaa,Brihaye:2014nba,Herdeiro:2015waa,Herdeiro:2015kha,Brito:2015pxa,Kleihaus:2015iea,Herdeiro:2015tia,Herdeiro:2016tmi,Herdeiro:2016gxs,Brihaye:2016vkv,Vincent:2016sjq,Ni:2016rhz,Delgado:2016zxv,Delgado:2016jxq,Dias:2011at} for generalizations and physical properties). These are solutions to Einstein's gravity minimally coupled to a simple and physically reasonable matter content: a complex, massive, free scalar field. KBHsSH interpolate between a (subset of) of vacuum Kerr BHs, when the scalar field vanishes, and horizonless, everywhere regular, gravitating scalar field configurations known as boson stars~\cite{Schunck:2003kk,Liebling:2012fv}, when the horizon vanishes.

\begin{figure}[ht]
\begin{center}
\includegraphics[width=6cm]{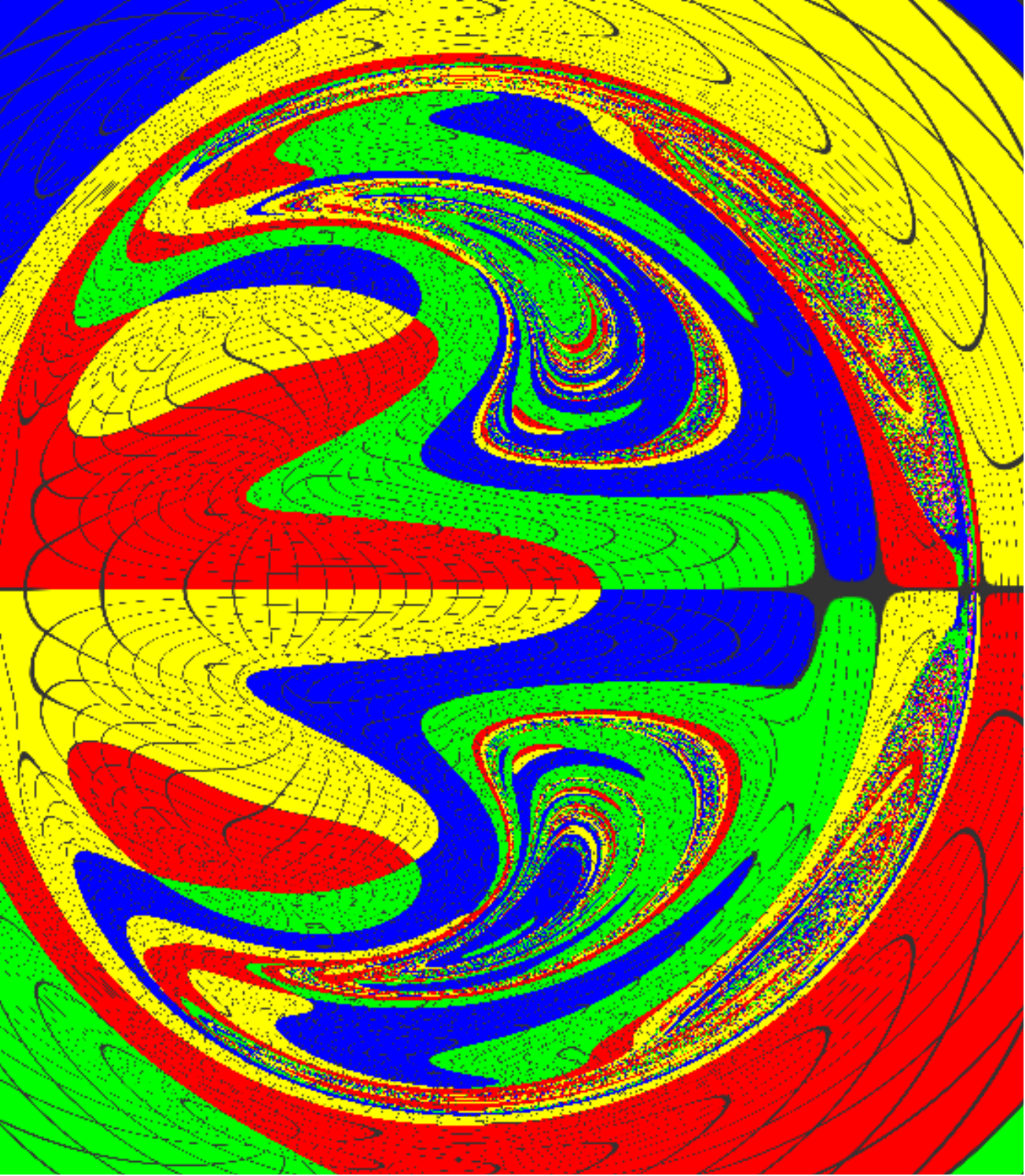}
\end{center}
\caption{\small Example of a RBS exhibiting chaotic scattering, which can be clearly seen in some fringes on the right hand side (wherein neighbouring pixels present different colours). The setup for this figure is explained in~\cite{Cunha:2015yba} ($cf.$ Section~\ref{sec3} below), and this image corresponds to configuration 11 therein (zoomed).}
\label{fig1}
\end{figure}

The lensing of both KBHsSH and their solitonic limit [rotating boson stars (RBSs)] was observed to exhibit chaotic patterns for solutions in some region of the parameter space, as illustrated by the example in Fig.~\ref{fig1}. Chaotic scattering in GR spacetimes has been observed and discussed in binary or multi-BH solutions -- see, $e.g.$,~\cite{Dettmann:1994dj,Yurtsever:1994yb,Dettmann:1995ex,Cornish:1996de,Sota:1995ms,deMoura:1999wf,deMoura:1999zd,Hanan:2006uf,Alonso:2007ts,Shipley:2016omi,Dolan:2016bxj} -- and is well known in the context of many body scattering in classical dynamics, for example the scattering of charged particles off magnetic dipoles~\cite{1992JPhA...25L.227A} and the 3-body problem (see $e.g.$~\cite{2011CeMDA.110...17M}). KBHsSH, or RBSs, provide an example of chaos in geodesic motion on the background of a single compact object, which moreover solves a simple and well defined matter model minimally coupled to GR.\footnote{Chaotic geodesic motion has also been reported around BHs surrounded by disks~\cite{Saa:1999je,Semerak:2012dw}. These models have some parallelism with KBHsSH, since the scalar field of the latter have a toroidal-type energy distribution, around the horizon.} Additionally, these objects possess a rich geometric structure, and may contain both multiple light rings~\cite{Cunha:2015yba}, including a stable one, as well as a structure of ergoregions~\cite{Herdeiro:2014jaa,Herdeiro:2016gxs}. The purpose of this paper is to investigate, in detail, chaotic scattering in this family of backgrounds and its interplay with the above geometric structure.

We start in Section~\ref{sec2} by performing an analysis of the effective potentials for (null) geodesic motion. We assume a stationary and axi-symmetric spacetime but no separation of variables; the latter is not known (or expected) in general for geodesic motion on RBSs or KBHsSH. Examining the 2D effective potentials for photon trajectories, we find that the emergence of \textit{stable light rings} on the background spacetimes, allows the formation of ``pockets" in one of the effective potentials, for open sets of impact parameters, leading to an effective trapping of the corresponding trajectories, dubbed \textit{quasi-bound orbits}. This analysis is analytical, with the exception of the explicit metric coefficients which are numerical for the examples exhibited.

Comparing this analysis with some of the lensing images obtained in~\cite{Cunha:2015yba} allows us to establish a correspondence between pocket formation and the emergence of chaotic patterns in the images. Then, searching for features shared by all trajectories in chaotic patches, we observe that they exhibit a positive correlation to the number of radial turning points of the geodesic motion, and their time delay. Note in particular that in the Kerr BH background a photon can only have one radial turning point~\cite{Wilkins1972}, while in our family of backgrounds more than one radial turning point can occur, corresponding, generically, to chaotic motion.

In Section~\ref{sec3} we exhibit a gallery of examples of lensing images obtained with a new ray-tracing code, \textsc{pyhole}, briefly described in Appendix~\ref{appendixb}. This code, based on \textsc{python}, includes tools for a simple, simultaneous visualization of the effective potential together with the spacetime trajectory, for any given point in a lensing image (and for an arbitrary numerical or analytical background metric). The results obtained with this code are in agreement with previous results~\cite{Cunha:2015yba,Vincent:2016sjq}, obtained with different ray-tracing codes, and it adds further tools that are useful for interpreting the results.

In Section~\ref{sec4} we present some conclusions. In Appendix~\ref{appendixao}, we illustrate a trajectory in phase space of a trapped photon. In Appendix~\ref{appendixa} the effective potentials for Kerr are discussed, without using separation of variables, which are useful for a comparison with those shown in Section~\ref{sec2} for RBSs and KBHsSH. In Appendix~\ref{appendixa2} we introduce an acceleration field and describe its connection to the number of radial turning points. In Appendix~\ref{appendixb}, some details on \textsc{pyhole} are discussed, and, as an application, we also exhibit the lensing and shadows by KBHsSH, observed away from the equatorial plane.

\section{ Effective potentials for photon motion}
\label{sec2}

\subsection{Preliminaries}
%
The geodesic motion of a photon on a background spacetime $(\mathcal{M},g_{\mu\nu})$, assuming minimal coupling between the (photon's) electromagnetic field and the geometry, is described by the Hamiltonian
\be
\mathcal{H}\equiv \frac{1}{2}p_\mu p_\nu\,g^{\mu\nu}=0 \ ,
\ee
where $p_\mu$ are the 4-momentum components of the photon orbit and $g^{\mu\nu}$ is the inverse metric. In this paper we shall be interested in RBSs and KBHsSH, described by the line element
\be
ds^2=-e^{2F_0}Ndt^2+e^{2F_1}\left(\frac{dr^2}{N}+r^2d\theta^2\right)+e^{2F_2}r^2\sin^2\theta(d\varphi-Wdt)^2 \ ,
\label{metric}
\ee
where $N=1-r_H/r$ and $r_H$ is the radial coordinate of the horizon (for RBSs $r_H=0$). The explicit form of the functions $F_0,F_1,F_2,W$, all of them functions of $(r,\theta)$, is only known numerically (examples can be found in~\cite{datakbhsh}). But the formalism we are describing is valid for any stationary and axi-symmetric spacetime, in coordinates adapted to these symmetries. For such a background, the Hamiltonian takes the form
\be
p_r^2g^{rr}+p_\theta^2g^{\theta\theta}+p_t^2g^{tt}+p_\varphi^2g^{\varphi\varphi}+2\,p_t\,p_\varphi\,g^{t\varphi}=0\ .\ee

Since the quantity
\be
T\equiv p_r^2g^{rr}+p_\theta^2g^{\theta\theta}\geqslant 0 \ ,
\ee
is positive definite, we can write the Hamiltonian condition in the form
\be
2\mathcal{H}=T+V=0\ ,
\ee
and identify the problem with a mechanical system with vanishing total energy, kinetic energy $T$ and  potential energy $V$
\begin{equation}V\equiv p_t^2g^{tt}+p_\varphi^2g^{\varphi\varphi}+2\,p_t\,p_\varphi\,g^{t\varphi}\leqslant 0 \ .\label{eq_V}\end{equation}
This inequality defines the \textit{allowed region} in the $(r,\theta)$--space.

The geodesic equations are obtained from Hamilton's equations:
\begin{equation}\dot{x}^\mu=\frac{\partial \mathcal{H}}{\partial p_\mu}\ ,\qquad \dot{p}_\mu=-\frac{\partial \mathcal{H}}{\partial x^\mu}\ ,\label{eq_Hamilton}\end{equation}
where the dot denotes differentiation with respect to an affine parameter. In coordinates adapted to the stationarity and axi-symmetry, $\mathcal{H}$ does not depend on $t$ and $\varphi$, and both $p_t$ and $p_\varphi$ are constants of the geodesic motion. We can then define the integrals of motion $E$ and $\Phi$ which are interpreted as the photon's energy and angular momentum, as measured by an asymptotic static observer (assuming asymptotic flatness):
\be
E\equiv -p_t\qquad \Phi\equiv p_\varphi\ .
\ee
Inserting these terms in equation (\ref{eq_V}) for $V$ we obtain
\be
V=-\frac{1}{D}\left(E^2 g_{\varphi\varphi} +2E\Phi g_{t\varphi} +\Phi^2g_{tt}\right)\leqslant 0\ ,
\ee
where
\be
D\equiv g_{t\varphi}^2-g_{tt}g_{\varphi\varphi}= Nr^2\sin^2\theta\,e^{2\left(F_2+F_0\right)} \ ,
\ee
which implies $D>0$ outside the horizon.
%
%

Since we are only interested in the geodesic motion outside the event horizon, and in order to introduce an explicit dependence on the  \textit{impact parameter}, $\eta$,
\be
\eta\equiv \frac{\Phi}{E} \ ,\ee
 we define the rescaled potential energy $\widetilde{V}$, such that:
\be
-\frac{D\,V}{E^2}\equiv \widetilde{V}=g_{\varphi\varphi} +2g_{t\varphi}\eta +g_{tt}\eta^2\geqslant 0\ ,\ee
which is a quadratic function of the impact parameter, with $(r,\theta)$-dependent coefficients. Factorizing this function leads to  \textit{two effective potentials}, that we now address.

\subsection{The two effective potentials}

The rescaled potential energy $\widetilde{V}$ can be written in the form:
\begin{equation}
\widetilde{V}=g_{tt}\left(\eta-h_+\right)\left(\eta-h_-\right)\geqslant 0\ ,\qquad (\textrm{with}\quad g_{tt}\neq 0)\ .\label{eq_Vtilde}
\end{equation}
This introduces the two functions $h_\pm(r,\theta)$ which we dub the two effective potentials. Their usefulness, is connected to the observation that $h_{\pm}=\eta \implies \widetilde{V}=0$.  Thus, the \textit{equipotential lines} of $h_\pm(r,\theta)$ give the boundary of the allowed region in the $(r,\theta)$--space, for each value of $\eta$. A similar effective potential was recently used in~\cite{Shipley:2016omi,Dolan:2016bxj}. Moreover, since the solutions of the quadratic equation are
\begin{equation}
h_{\pm}\equiv\frac{-g_{t\varphi}\pm\sqrt{D}}{g_{tt}}\ ,\label{eq_h}
\end{equation}
there is a regime transition when $g_{tt}$ changes sign, which is possible outside the event horizon when entering/exiting an ergoregion.

For the special case $g_{tt}=0$, we have
\be
\widetilde{V}=2g_{t\varphi}\left(\eta-\widetilde{h}\right)\geqslant 0\ ,\qquad (\textrm{with}\quad g_{tt}=0) \ ,\ee
where:
\be
\widetilde{h}\equiv-\frac{g_{\varphi\varphi}}{2g_{t\varphi}} \ .
\ee
In the limit $g_{tt}\to 0$, one of the functions $h_{\pm}$ diverges and the other converges to $\widetilde{h}$.

We remark that the asymptotic limit of the effective potentials (at spatial infinity) is:
\be
h_{\pm}\to \mp r\sin\theta \ .
\ee

In the following two subsections we analyse the effective potentials outside and inside an ergoregion respectively, and in the subsequent one we examine light rings and spherical orbits.

\subsubsection{Outside the ergoregion ($g_{tt}<0$)}
\label{out_ergo}

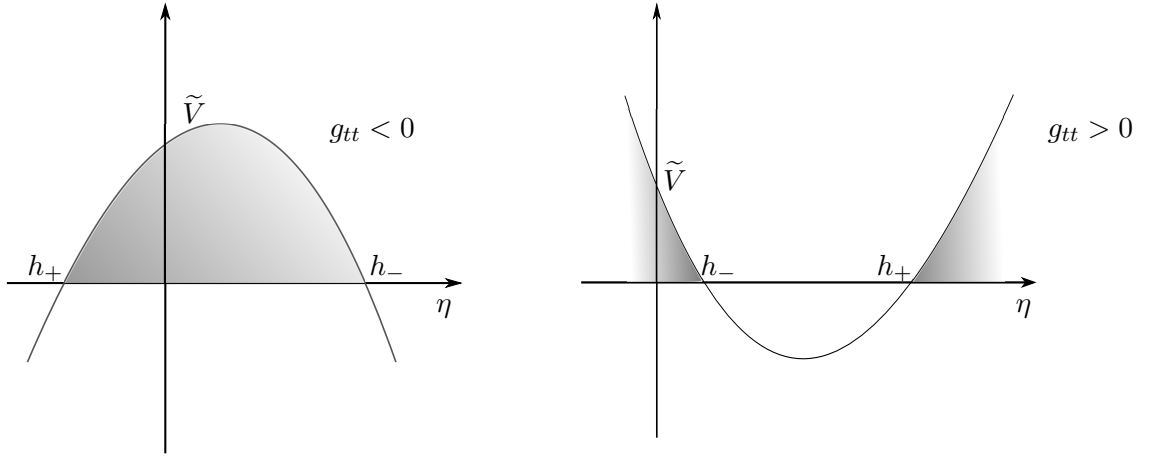
\begin{figure}[ht]
\begin{center}
\includegraphics[height=6cm,width=6cm]{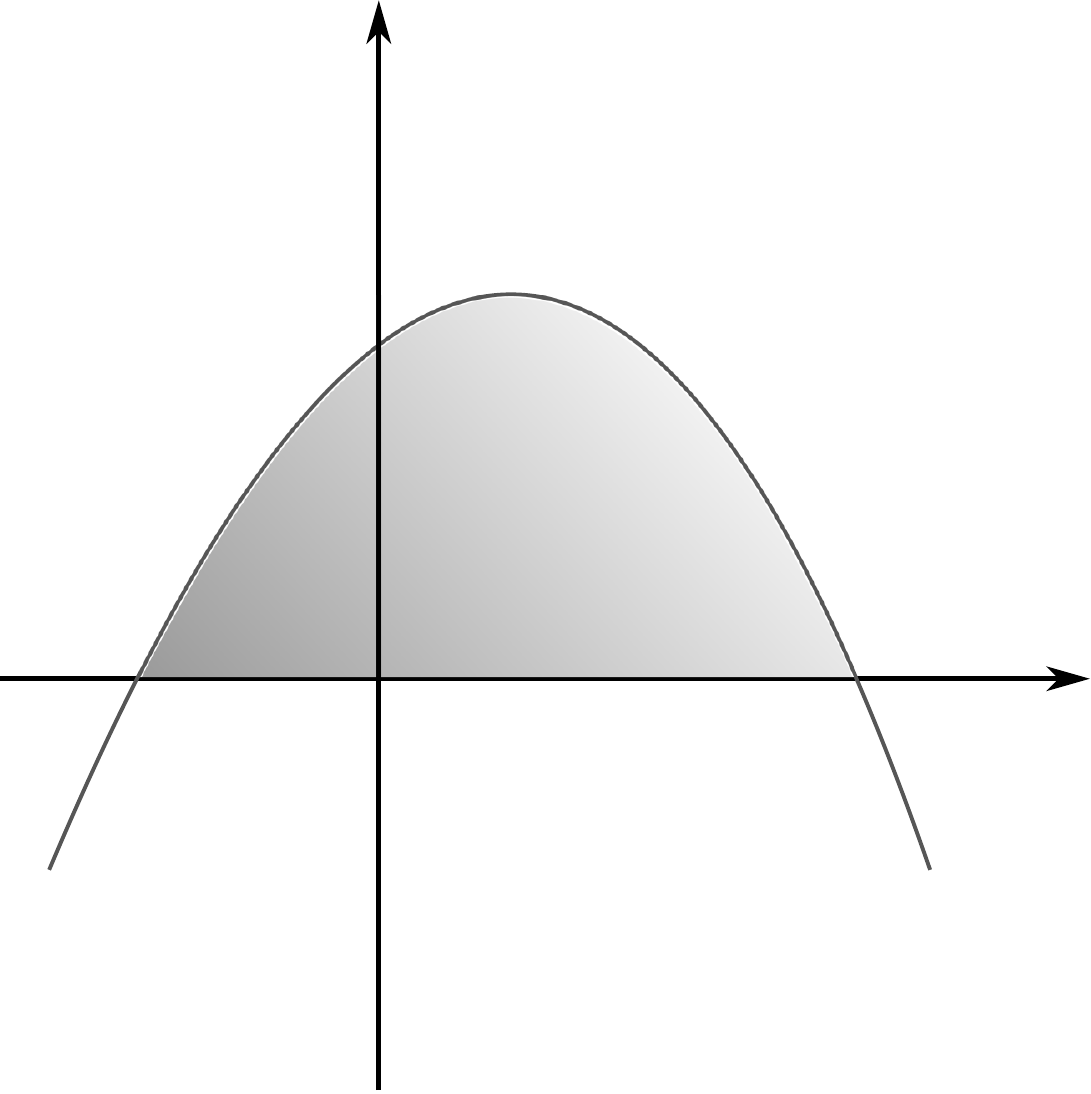} \hspace*{1.3cm} \includegraphics[trim=0cm -0.4cm -0cm -0cm, height=6cm,width=6cm]{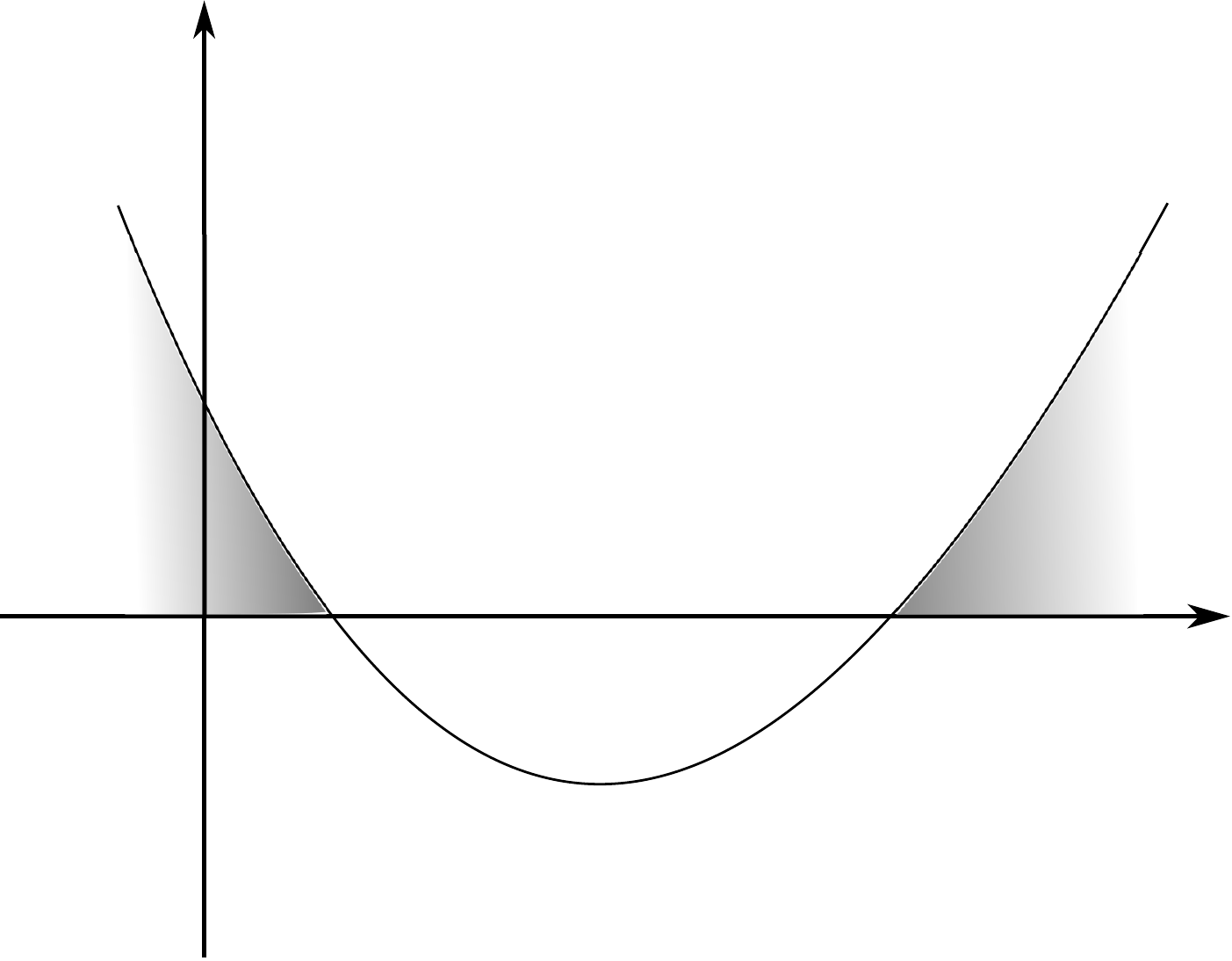}
\put(-320,125){$\widetilde{V}$}
\put(-225,54){$\eta$}
\put(-250,68){$h_-$}
\put(-378,68){$h_+$}
\put(-265,120){$g_{tt}<0$}
\put(-140,100){$\widetilde{V}$}
\put(-8,54){$\eta$}
\put(-60,68){$h_+$}
\put(-126,68){$h_-$}
\put(4,120){$g_{tt}>0$}
\end{center}
\caption{\small Dummy shape of the potential $\widetilde{V}$ (left panel) outside of the ergoregion, $g_{tt}<0$; (right panel) inside of the ergoregion, $g_{tt}>0$. The shaded region illustrates the allowed $\eta$ interval. In the first case, $h_+$ must always be negative and $h_-$ always positive. Due to the condition $\widetilde{V}\geqslant 0$, we have $h_+\leqslant\eta\leqslant h_-$. In the second case, if $W>0$ (spacetime with positive rotation), then $h_\pm$ is always positive, with $h_- < h_+$. Due to the condition $\widetilde{V}\geqslant 0$, we have $\eta\leqslant h_-$ or $h_+\leqslant\eta$. Since $h_+\to +\infty$ as $g_{tt}\to0^+$ the right region is not accessible from spatial infinity.}
\label{fig_VI}
\end{figure}

Since $g_{tt}<0$ holds outside the ergoregion, $-g_{tt}g_{\varphi\varphi}>0\implies g_{t\varphi}^2-g_{tt}g_{\varphi\varphi}>g_{t\varphi}^2\implies \sqrt{D}>|g_{t\varphi}|$,
where we assumed that $g_{\varphi\varphi}>0$ (absence of closed timelike curves). This condition is verified for all RBSs and KBHsSH that shall be studied in this work. As a consequence, the effective potentials read:
\be
h_+=\frac{-g_{t\varphi}+\sqrt{D}}{g_{tt}}<0\ ,\qquad h_-=\frac{-g_{t\varphi}-\sqrt{D}}{g_{tt}}>0\ .
\ee
A generic plot of $\widetilde{V}$ outside of the ergoregion can be found in Fig. \ref{fig_VI} (left panel). We conclude that the boundary of the forbidden region in the phase space $(r,\theta)$ is given by the equipotential lines defined as:
\be
h_+(r,\theta)=\eta,\quad\textrm{if }\eta<0\qquad\textrm{and}\qquad h_-(r,\theta)=\eta,\quad\textrm{if }\eta>0 \ .
\ee

%

\subsubsection{Inside the ergoregion ($g_{tt}>0$)}
\label{in_ergo}

Since $g_{tt}>0$ holds inside the ergoregion, $-g_{tt}g_{\varphi\varphi}<0\implies g_{t\varphi}^2-g_{tt}g_{\varphi\varphi}<g_{t\varphi}^2\implies \sqrt{D}<|g_{t\varphi}|$, where we again assumed that $g_{\varphi\varphi}>0$. In this case the sign of the $h$-functions will depend on the sign of the function $W$. For the sake of simplicity we will here assume\footnote{The case $W<0$ can be obtained by a transformation $W\to -W$, $h_\pm\to-h_\mp$ and $\eta\to -\eta$.} that $W>0\implies -g_{t\varphi}>0$. This will be the case for all configurations analysed afterwards. In such a situation:
\be
h_+=\frac{-g_{t\varphi}+\sqrt{D}}{g_{tt}}>0\ ,\qquad h_-=\frac{-g_{t\varphi}-\sqrt{D}}{g_{tt}}>0\ .
\ee
We remark that $h_+>h_-$ holds, regardless of the sign of $W$.

A generic plot of $\widetilde{V}$ inside of the ergoregion (with $W>0$) can be found in Fig. \ref{fig_VI} (right panel). Notice that as we go from the inside to the outside of the ergoregion, or in other words as we approach $g_{tt}\to 0^+$, we have that $h_+\to +\infty$. Since the impact parameter $\eta$ is a constant of motion for a given photon trajectory, the allowed region $h_+<\eta$ is not accessible from spatial infinity: as it turns out, it corresponds to bound states with negative energy ($cf.$ Section~\ref{seclr}). In fact, there are stable light rings around RBSs which can be populated by photons in such a state.

A boundary to a forbidden region only exists in this case for $\eta>0$ (if $W>0$):
\be
h_-(r,\theta)=\eta,\quad\textrm{(scattering state)}\qquad\textrm{and}\qquad h_+(r,\theta)=\eta,\quad\textrm{(bound state only)} \ .
\ee

\subsection{Effective potentials contour plots}

We will now exhibit contour plots of $h_+$ and $h_-$ for different spacetimes, namely three RBSs and one KBHSH. The solid lines (\textcolor{blue}{blue}) represent negative $\eta$ values, whereas dashed lines (\textcolor{red}{red}) represent positive values of $\eta$. Although the function $h_-$ is also relevant for defining the allowed region for some photon trajectories, the landscape of the function $h_+$ is richer, in particular as it leads to the appearance of a trapping region. We remark that in the following, the term \textit{light ring} will be used for photon orbits with $p_r= \dot{p}_r=0$ on the equatorial plane ($\theta=\pi/2\Rightarrow p_\theta=0$). We equally remark that light rings are related to extrema of $h_{\pm}$ ($cf.$ Section~\ref{seclr}). The following table summarizes the particular configurations (see Fig.~\ref{fig_overview}) for which the $h_\pm$  contour lines are shown below:

\begin{center}
\begin{tabular}{l*{4}{c}r}
\hline
\hline
Object  & Configuration in~\cite{Cunha:2015yba} & Light rings  & Ergoregions  & Chaos  & Fig.  \\
\hline
\hline
{\bf RBS}  & 9 ($w$=0.75 $\mu$) & No & No & No &   \ref{fig_75} \\
 \hline
{\bf RBS} & 10 ($w$=0.7 $\mu$) & 1 Stable + 1 Unstable & No & Yes & \ref{fig_7}, \ref{gallery10}  \\
 \hline
{\bf RBS} & 11 ($w$=0.65 $\mu$) & 1 Stable + 1 Unstable & Yes & Yes & \ref{fig1}, \ref{fig_65}, \ref{gallery11}   \\
\hline
{\bf KBHSH} & III  & 1 Stable + 3 Unstable & Yes & Yes &  \ref{fig_V}, \ref{galleryIIIa} -- \ref{galleryIIIc}    \\
 \hline
\end{tabular}
\end{center}

\bigskip

The value of $w$ (in units of the scalar field mass $\mu$), in the second column of the table, is the frequency in the scalar field ansatz, $cf.$ eqs. (4) in~\cite{Herdeiro:2014goa}, whereas the column ``Chaos" refers to the occurrence of chaotic patterns in the lensing images of that configuration. In the plots below a compactified radial coordinate $R\in[0;1]$ will be used,
\be
R=\frac{R^*}{1+R^*},\qquad\textrm{with}\quad R^*\equiv\sqrt{r^2-r_H^2} \ . \label{R_def}
\ee
In the remainder of this paper, configurations 9, 10 and 11 of RBSs, as well as configurations III (and also II) of KBHsSH, are the same as those considered in~\cite{Cunha:2015yba}. We keep this labelling here, to avoid confusion, even though we shall not discuss all configurations presented in~\cite{Cunha:2015yba}.

\begin{figure}[ht!]
\begin{center}
\includegraphics[width=10cm]{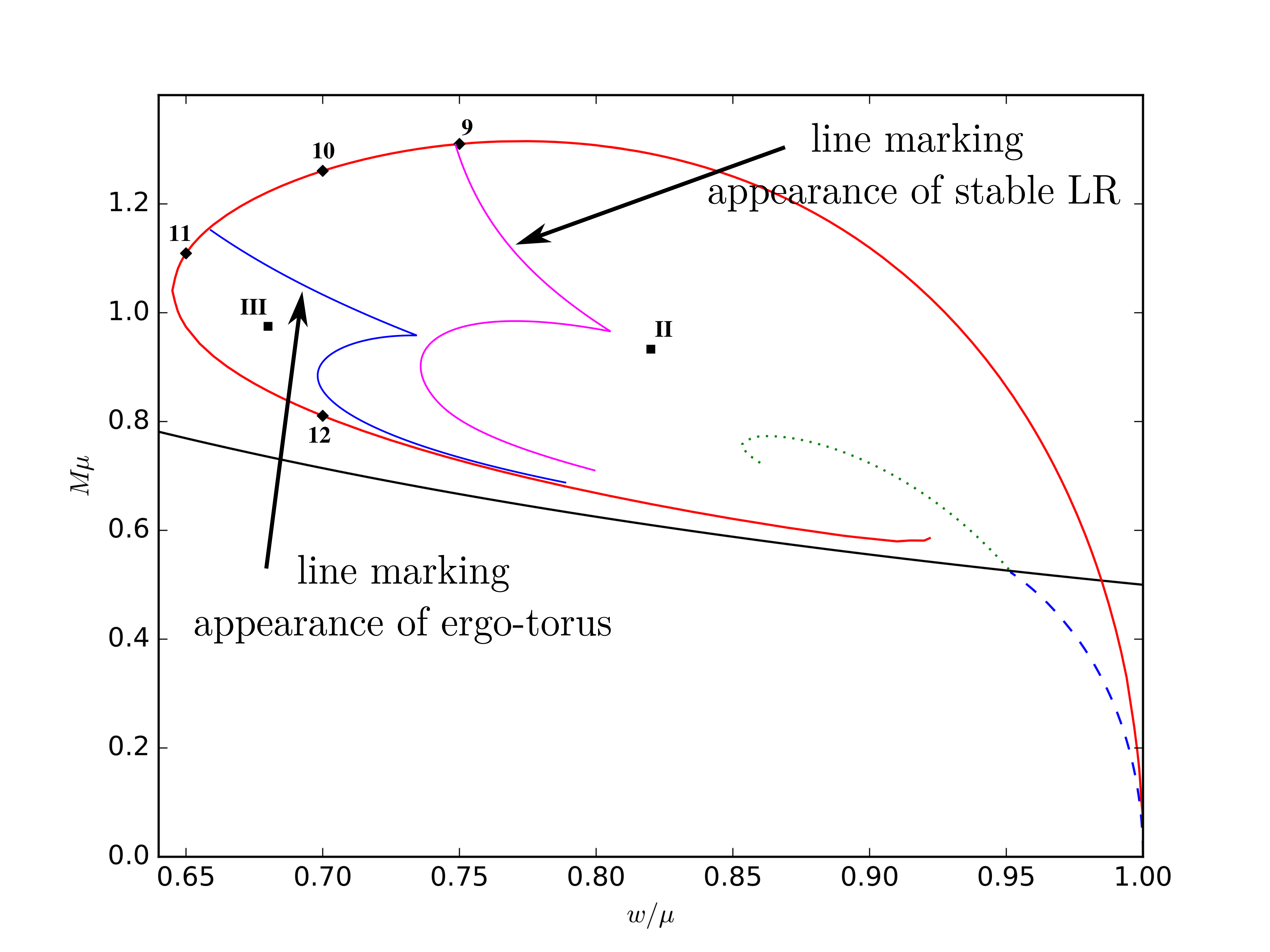}
\end{center}
\caption{\small RBS solutions (solid \textcolor{red}{red} spiral) in a ADM mass, $M_{\textrm{ADM}}$, $vs.$ scalar field frequency $w$ diagram. KBHsSH exist within the RBS spiral and are bounded by a subset of Kerr solutions (dashed \textcolor{blue}{blue} line), extremal KBHsSH (dotted \textcolor{OliveGreen}{green} line) and the RBS spiral itself. Points 9 - 12 (II - III) correspond to the BSs (KBHsSH) under discussion. Two extra lines mark the appearance of a stable light ring (LR) and an ergo-torus, always to the left of these lines. See~\cite{Cunha:2015yba,Herdeiro:2014goa,Herdeiro:2014jaa} for more details.}
\label{fig_overview}
\end{figure}

\begin{figure}[h!tb]
\begin{center}
\includegraphics[width=10cm]{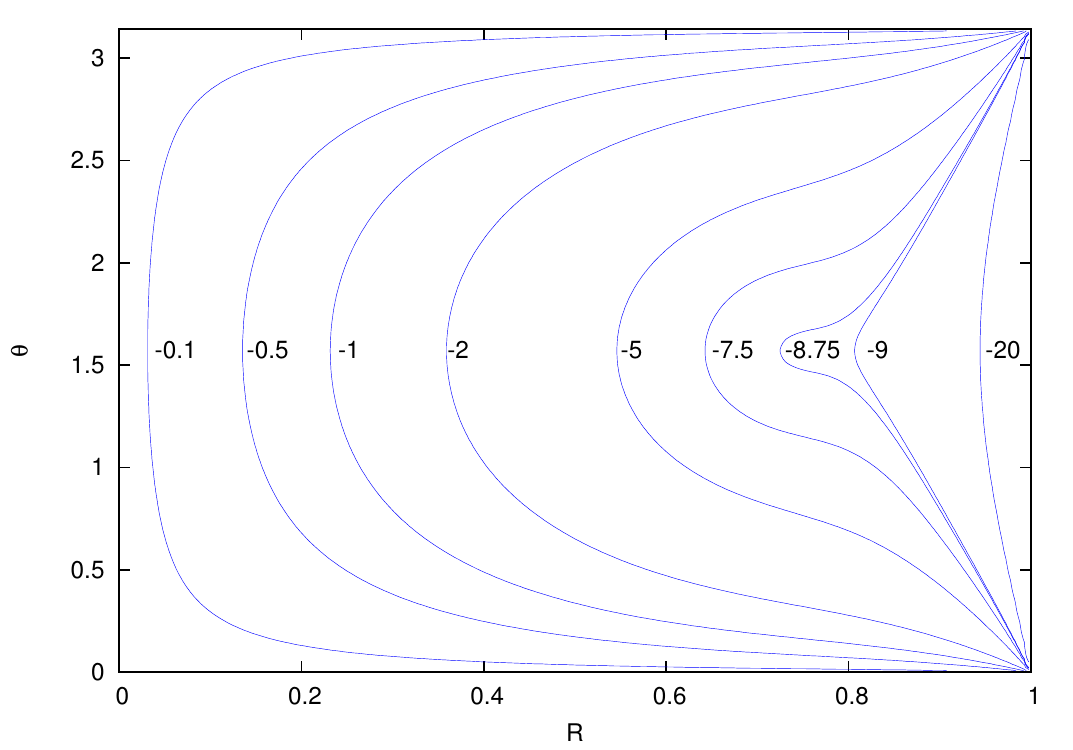} 
\put(0,110){$h_+$}
\end{center}
\caption{\small Contour plots of $h_+$ for the RBS configuration 9. In this and the next figures, the solid lines (\textcolor{blue}{blue}) represent negative $\eta$ values. It has no ergoregion or light rings. This configuration is very close in solution space to a RBS where light rings first appear. There is a  deformation of the $h_+$ lines, which will grow into a \textit{pocket} in the cases considered next.}
\label{fig_75}
\end{figure}

Fig. \ref{fig_75} exhibits the effective potentials contour plots for the RBS configuration 9. This background has no ergoregion or light rings, but it is very close, in solution space, to the RBS for which light rings first appear (see Fig.~\ref{fig_overview}). Each contour line of $h_+$ in Fig. \ref{fig_75} sets the boundary of the forbidden region in $(r,\theta)$ space for a given $\eta$. There is a distinct deformation of the $h_+$ contour lines, which will grow into a \textit{pocket} in the following cases to be analysed. Since $\partial_r h_\pm$ is never zero on the equatorial plane there are no light rings -- neither maxima, minima nor saddle points of $h_\pm$ exist. The contour plot of $h_-$ for this configuration is very similar to the one displayed in the bottom panel of Fig. \ref{fig_65} and hence it will not be shown.\\

\begin{figure}[h!tb]
\begin{center}
\includegraphics[width=10cm]{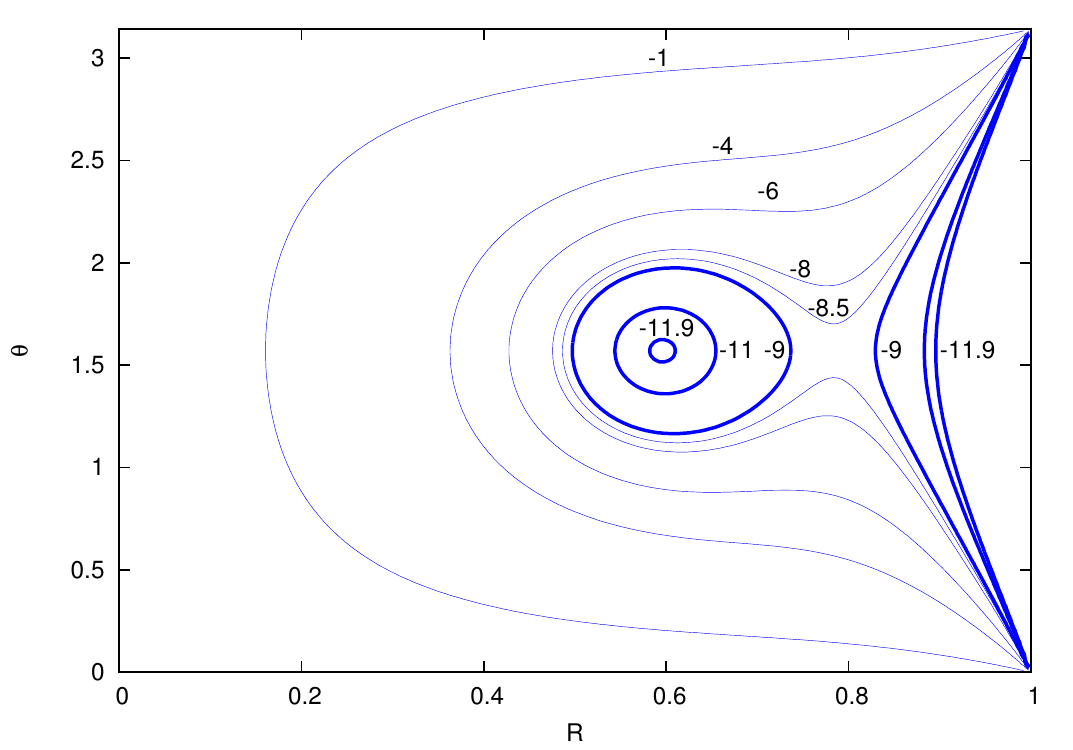} 
\put(0,110){$h_+$}
\end{center}
\caption{\small Contour plots of $h_+$ for the RBS configuration 10.  It has no ergoregion but it has two light rings, one stable and one unstable. There is a \textit{pocket} that can be closed below a certain impact parameter $\eta$ and form an allowed region which is disconnected from spatial infinity, leading to bound orbits.}
\label{fig_7}
\end{figure}

\begin{figure}[h!tb]
\begin{center}
\includegraphics[width=10cm]{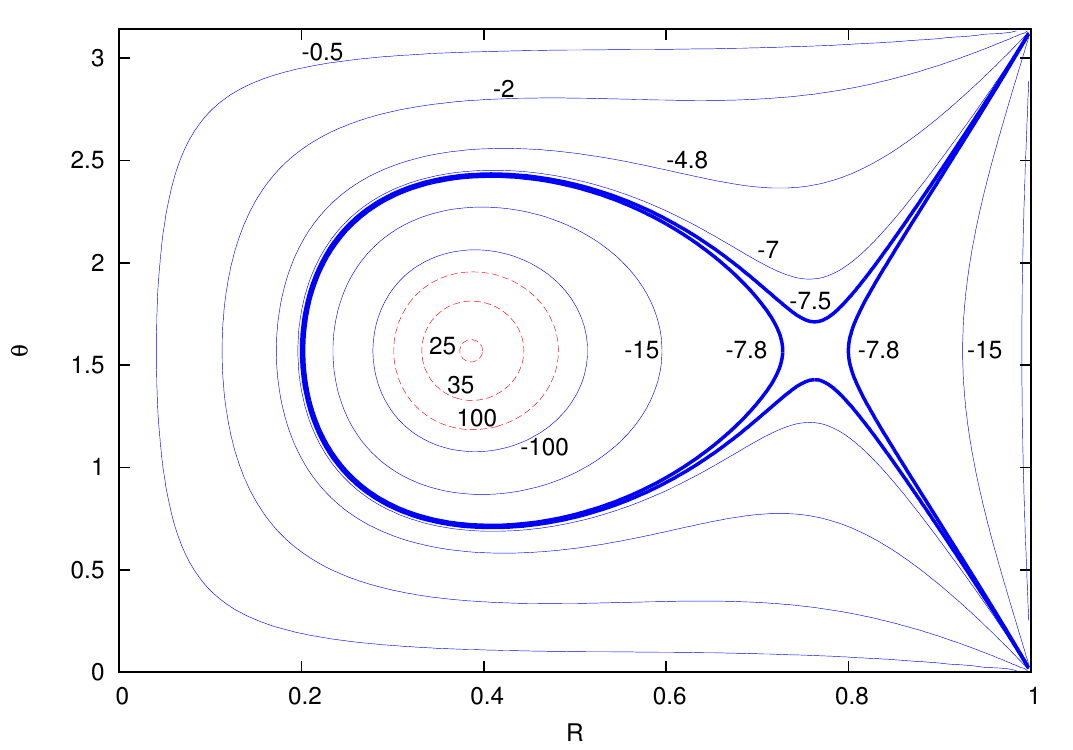} 
\put(0,110){$h_+$}\\
\includegraphics[width=10cm]{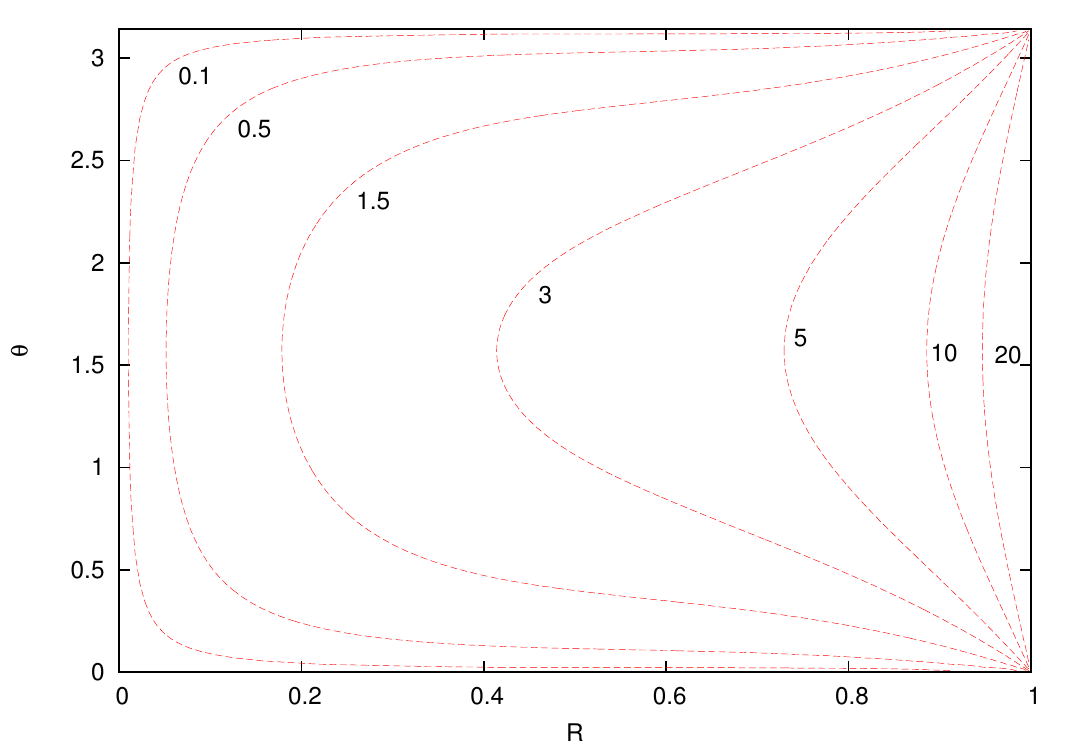}
\put(0,110){$h_-$}\\
\end{center}
\caption{\small Contour plots of $h_+=\eta$ (top panel) and $h_-$ (bottom panel) for the RBS configuration 11, which has two light rings and an ergoregion. The positive values of $h_+$ set the position of the ergoregion, with the minimum corresponding to the stable light ring. The saddle point corresponds to an unstable light ring (with negative $\eta$). The function $h_-$ has no light rings associated with it.}
\label{fig_65}
\end{figure}

The next case, shown  in Fig. \ref{fig_7}, corresponds to the RBS configuration 10. It has no ergoregion but it has two light rings, one stable and one unstable. The new feature in the $h_+$ contour lines is the existence of a \textit{pocket} that can be closed below a certain impact parameter $\eta$, and form an allowed region which is disconnected from spatial infinity (thus leading to \textit{bound orbits}). This can be seen in Fig. \ref{fig_7}. This disconnected region can in fact be made arbitrarily small until it becomes a single point on the equatorial plane for $\eta\simeq -11.97$, with $\partial_r h_+=0$ at that point. The latter actually corresponds to a \textit{stable light ring} since the motion is bounded. From Fig. \ref{fig_7}, we see clearly that a saddle point appears on the equatorial plane, which in that case corresponds to an \textit{unstable light ring}, since the photon can escape due to radial perturbations, for  $\eta\simeq -8.61$. The contour plot of $h_-$ for this configuration is very similar to the one displayed in the bottom panel of Fig. \ref{fig_65} and hence it will not be shown.\\

In Fig. \ref{fig_65} we consider the RBS configuration 11. This background has two light rings and one ergoregion (an ergo-torus~\cite{Herdeiro:2014jaa}). As discussed before, $h_+$ will diverge to $-\infty$ as the ergosurface is approached from the outside of the ergoregion. After entering the latter, $h_+$ will decrease from $+\infty$ to a minimum at positive $\eta$, which corresponds to a stable light ring. It turns out that such a light ring has negative energy ($cf.$ Section~\ref{seclr}). In Fig. \ref{fig_65} (top panel) are displayed the contour plots of $h_+$. Again, \textcolor{blue}{blue} solid lines represent negative values of $\eta$, whereas \textcolor{red}{red} dashed contour lines represent positive values. Notice the sharp transition of $h_+$ from -100 to +100, since the function diverges at the boundary of the ergoregion. Observe that the function $h_-$ (Fig. \ref{fig_65}, bottom panel) does not form a pocket; the corresponding $h_-$ functions of the previous configurations 9 and 10 were not displayed due to the strong similarity with the RBS 11 function $h_-$.

The existence of a pocket in the effective potential, whose opening can be made arbitrarily small, leads to \textit{trapped} or \textit{quasi-bound} orbits, from which a photon might only escape after a very long time. Such trapped orbits will be exemplified in the gallery of Section~\ref{secgallery}.

\begin{figure}[h!tb]
\begin{center}
\includegraphics[width=10cm]{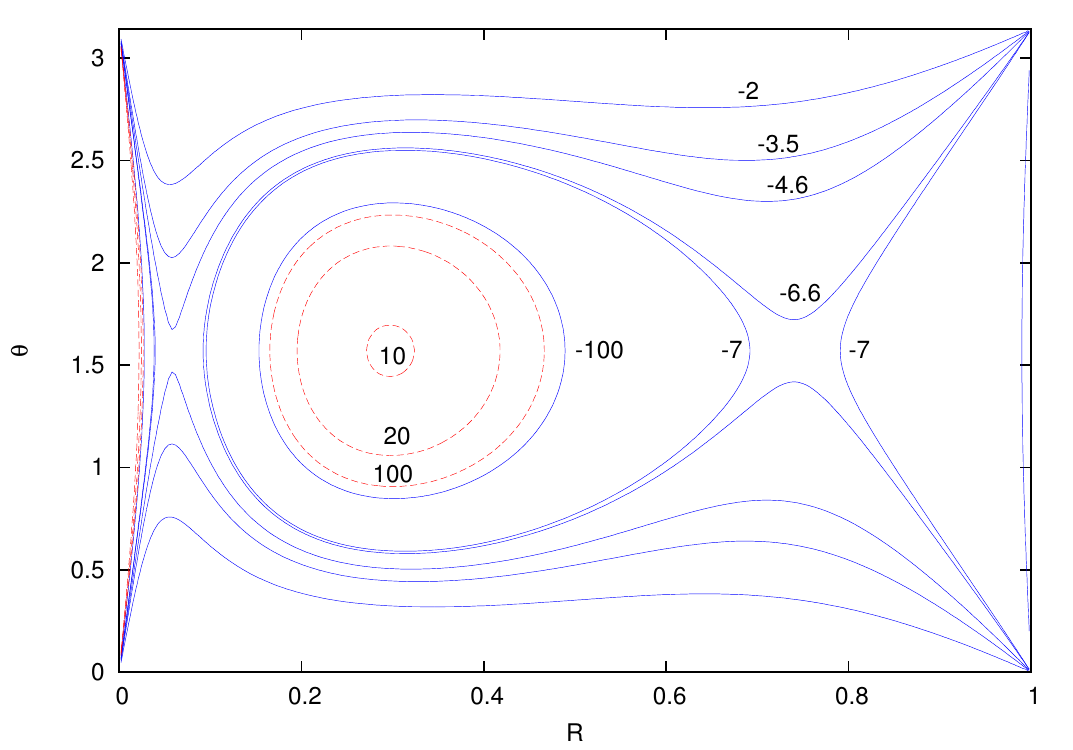} 
\put(0,110){$h_+$}\\
\includegraphics[width=10cm]{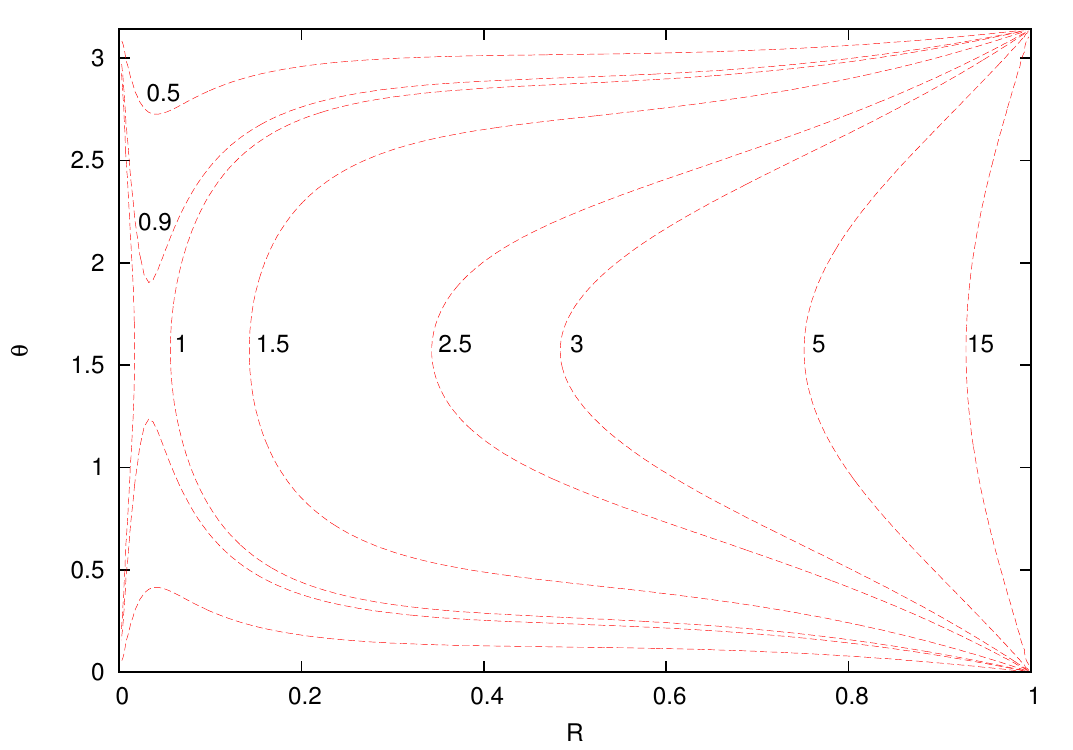}
\put(0,110){$h_-$}\\
\end{center}
\caption{\small Contour plots of $h_+$ (top panel) and $h_-$ (bottom panel), for the KBHSH with the \textit{hammer}-like shadow (configuration III). This configuration contains two ergoregions (Saturn-like topology) and four light rings, three unstable and one stable. The $h_+$ dashed (\textcolor{red}{red}) lines occur within the two disconnected ergoregions, one of which is near the horizon (at $R=0$). The function $h_-$ in this case has a saddle point on the equatorial plane, signaling the existence of an unstable light ring (bottom panel).}
\label{fig_V}
\end{figure}

Finally, in the top (bottom) panel of Fig. \ref{fig_V},  the $h_+$ ($h_-$) contour lines are shown for the KBHSH with the \textit{hammer}-like shadow -- configuration III in~\cite{Cunha:2015yba}. This spacetime contains two ergoregions (Saturn-like topology~\cite{Herdeiro:2014jaa}) and four light rings, three unstable and one stable. In Fig. \ref{fig_V} (top panel), as before, the sharp transition from negative to positive $\eta$ values marks the boundary of the ergoregion. As this boundary is approached from the outside (inside) of the ergoregion, $h_+$ diverges to negative (positive) values.  Inside the ergotorus there is a stable light ring for $R\simeq 0.3$. Clearly, there are also saddle points for $R\simeq 0.06$ and $R\simeq 0.74$ on the equatorial plane, corresponding to unstable light rings. Additionally, there is an ergoregion near the horizon (which is at $R=0$), amounting to a pileup of $h_+$ contour lines at $R\sim 0.02$ (on the equatorial plane), since $h_+$ diverges. Inside this ergoregion there are no light rings.

Fig. \ref{fig_V}, bottom panel, shows the $h_-$ contour plot, which reveals the existence of a saddle point at $R\simeq 0.032$ on the equatorial plane and hence an unstable light ring. Heuristically, this is the merging of the structure of both a Kerr-like BH and a RBS: a Kerr BH has an ergosphere and two unstable light rings ($cf.$ Appendix~\ref{appendixa}); a RBS such as configuration 11 has an ergotorus and two light rings, one stable and the other unstable.

\subsection{Pocket formation, chaos and turning points}
\label{poch}
The formation of pockets in the effective potential will lead to quasi-bound orbits, $i.e.$, orbits that stay in a confined spatial region for a long time. We shall now show that one can associate these orbits to the emergence of chaotic patterns in the lensing images, such as the one exhibited in Fig.~\ref{fig1}. In order to do so, we recall that for a given photon orbit the value of the impact parameter $\eta$ is a constant of motion. This value also fixes the photon's allowed spacetime region. Let us analyse the contour $\eta=$ constant in an image containing the lensing of a RBS or a KBHSH.

\begin{figure}[h!]
\begin{center}
\includegraphics[width=8cm]{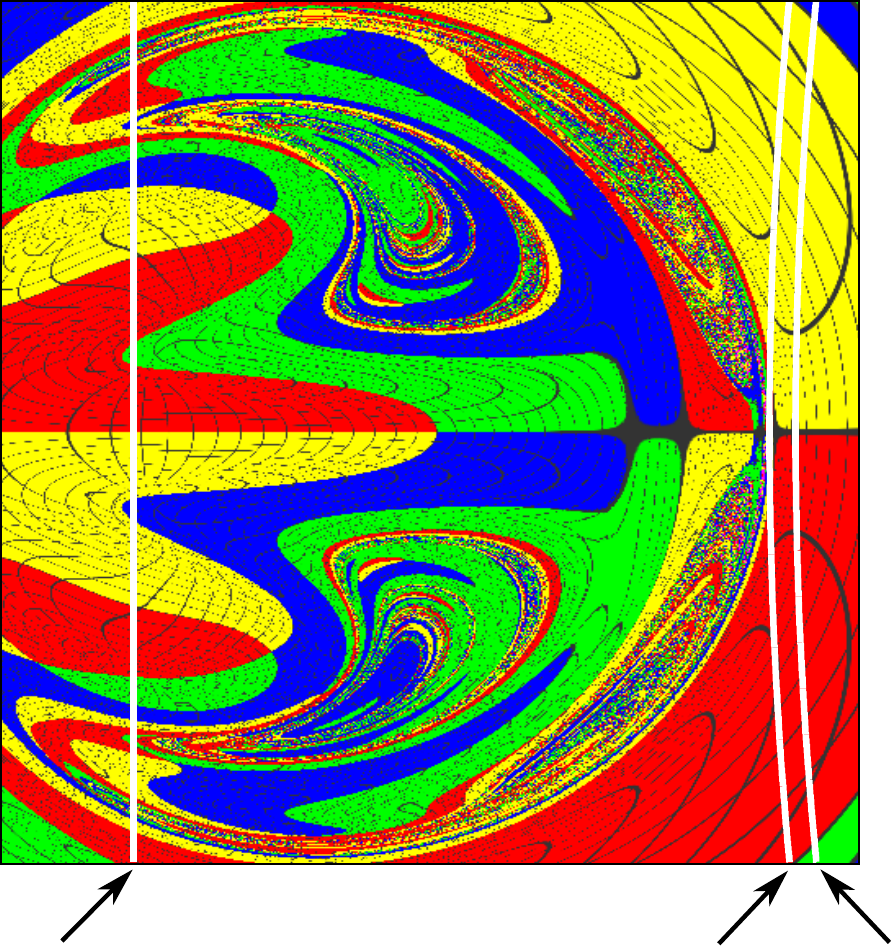}
\put(-219,-8){0.1}
\put(-56,-8){-7.5}
\put(-5,-8){-7.8}
\end{center}
\caption{\small Gravitational lensing of the RBS configuration 11 (zoomed). White contour lines of constant $\eta$ are shown for three values of $\eta$. Notice the transition from $\eta=-7.8$ to $\eta=-7.5$ leads in this image to an overlap with the chaotic region, whereas in the effective potential (Fig. \ref{fig_65}, top panel) it is connected to the appearance of a pocket. However, the line of constant $\eta=0.1$ crosses a chaotic region near the edges of the figure, but there is no pocket associated with it in the potential (Fig. \ref{fig_65}, bottom panel). }
\label{fig_IMG}
\end{figure}

In Fig.~\ref{fig_IMG} we exhibit three contour plots of $\eta=$ constant, with $\eta=-7.8$, $\eta=-7.5$ and $\eta=0.1$, in a detail of Fig.~\ref{fig1}, corresponding to the gravitational lensing of the RBS configuration 11, whose effective potentials are shown in~Fig. \ref{fig_65}. Observe that in the contour plot for $\eta=-7.8$ the pocket is not yet open (Fig. \ref{fig_65}, top panel); correspondingly, there is no chaos in the lensing image for this value of the impact parameter. For $\eta=-7.5$, on the other hand, the pocket is open and indeed the $\eta=$ constant contour line in the lensing image intersects a chaotic region. As the impact parameter becomes even larger, the pocket's opening becomes wider, explaining why the chaotic region expands to higher latitudes in the lensing image. This analysis suggests that pocket formation induces chaotic behaviour. One may wonder, however, if the existence of a pocket is \textit{necessary} for the occurrence of chaotic regions. It turns out that it is not. To establish this, observe that the line of constant $\eta=0.1$ crosses a chaotic region near the edges of the figure, but there is no pocket associated with it in the $h_-$ function (Fig. \ref{fig_65}, bottom panel). Thus, there are chaotic regions with no corresponding pocket in the effective potential. One way to understand these regions is via a different ``potential", \textit{the acceleration field} $\mathcal{F}_r$. However, in order to avoid distracting the reader from the main message, we leave the discussion of such concept to Appendix \ref{appendixa2}.

The relation between chaotic patterns on the image plane and the characteristics of the corresponding geodesic motion can be understood in a number of different ways. The manifestation of this chaos is the pixelated aspect of some image patches, which suggest that there is a sensitive dependence on initial conditions in the map between a camera pixel and a point on the celestial sphere; the map corresponding to the geodesic linking the two points. To quantify such chaoticity, one can then introduce a number of measures such as the Lyapunov exponent, entropy, the time delay function T associated to each pixel, or the number of radial turning points. In the following, we shall expand on the two last notions, as particularly well suited to measure chaoticity.\\

\begin{figure}[!htp]
\begin{center}
\includegraphics[width=9cm]{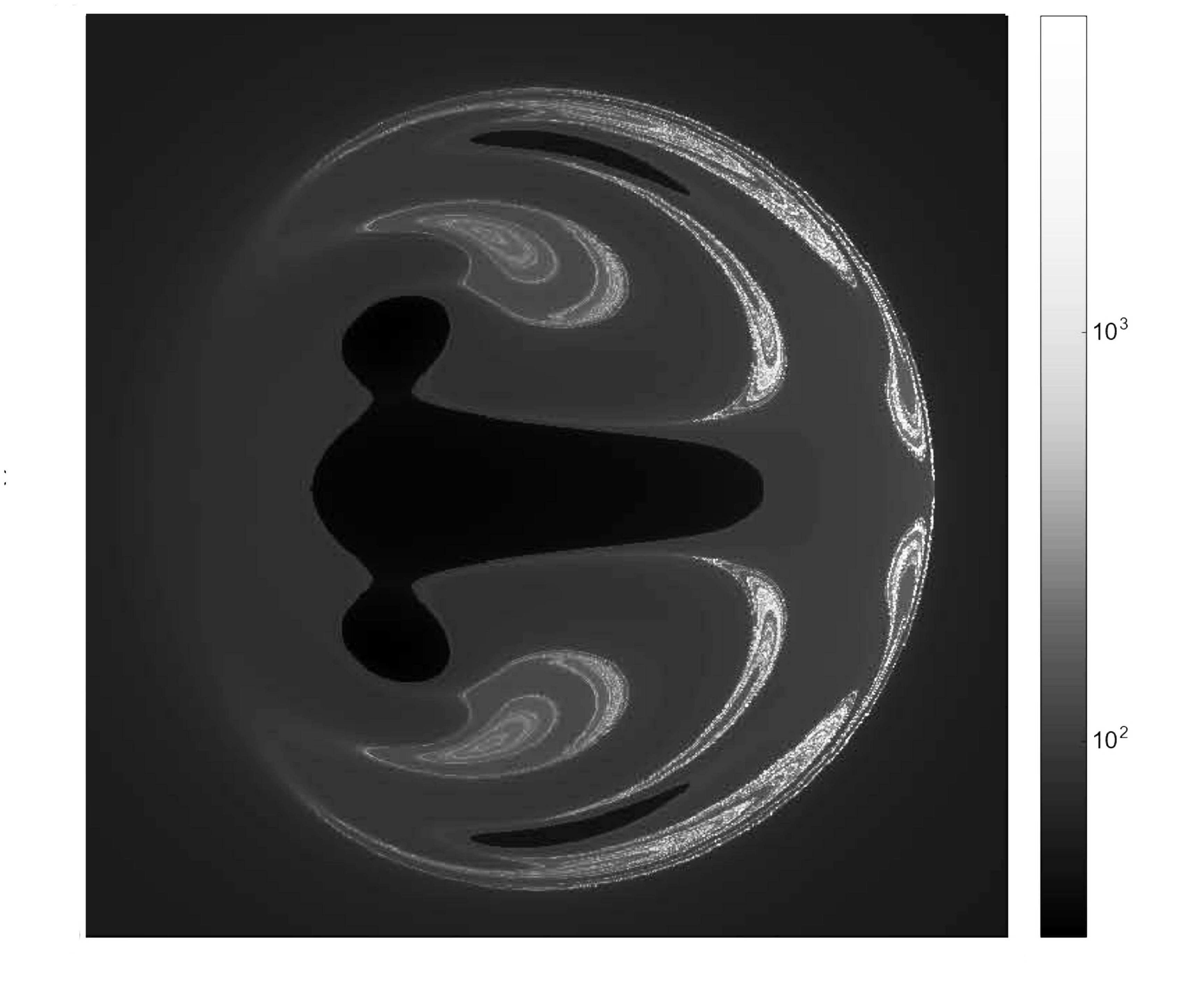}  
\put(-21,109){$t\mu$}   
\end{center}
\caption{\small Time delay heat map associated to scattering orbits for KBHSH configuration III (zoomed).}
\label{timedelay}
\end{figure}

\begin{figure}[tb]
    \begin{center}
        \includegraphics[trim=8cm 5.5cm 3cm 5.5cm,clip,height=6cm]{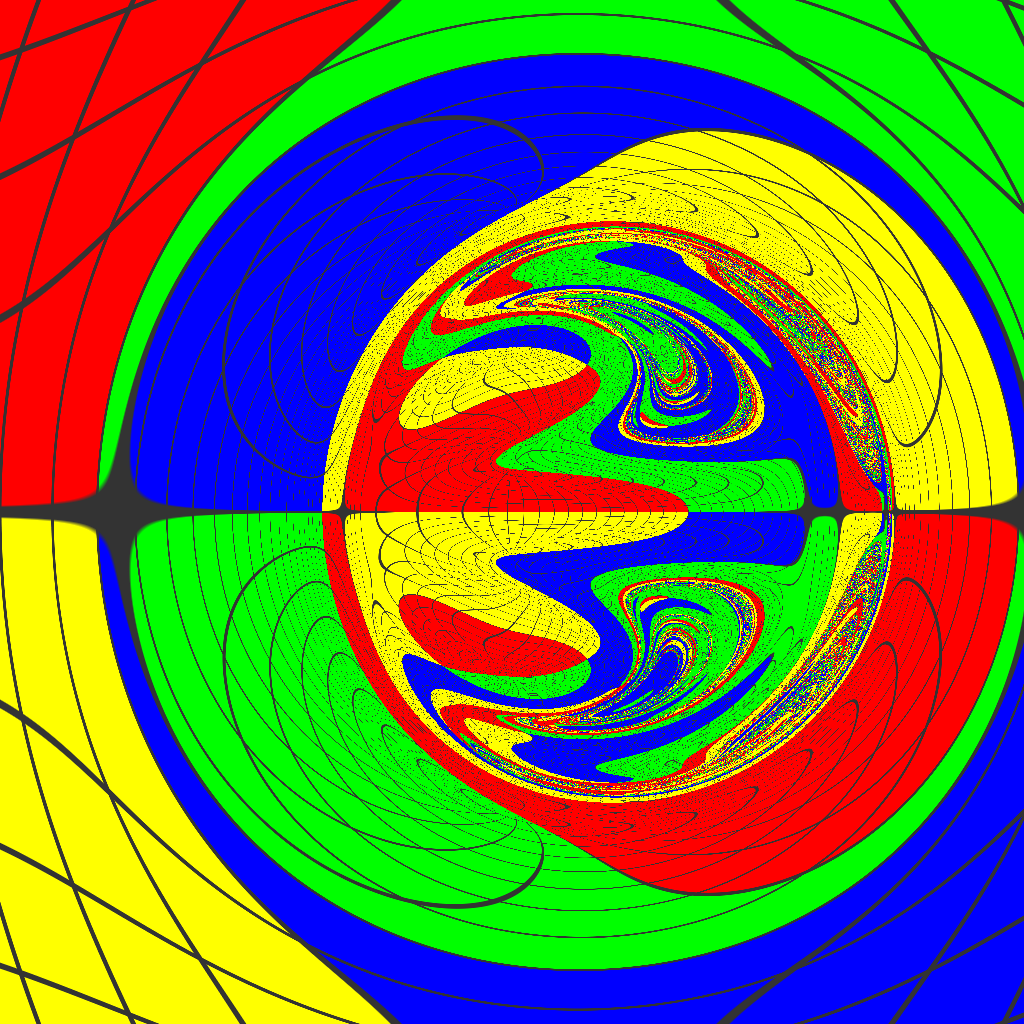}\hspace*{0.2cm}\includegraphics[trim=4cm 3.1cm 4cm 2.7cm,height=6cm,clip]{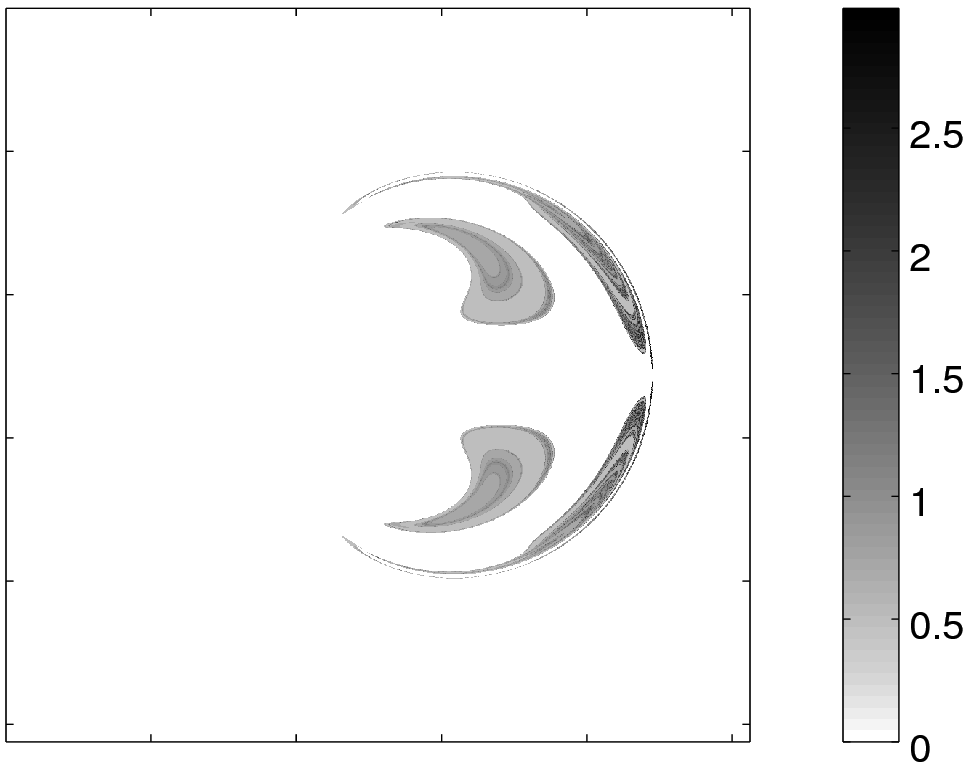}\includegraphics[trim=14cm 0.0cm 0cm 0cm,height=6.1cm,clip]{./turn-11.png}\put(22,84){(a)}\\
        \vspace*{0.2cm}
        \includegraphics[trim=8cm 5.5cm 3cm 5.5cm,clip,height=6cm]{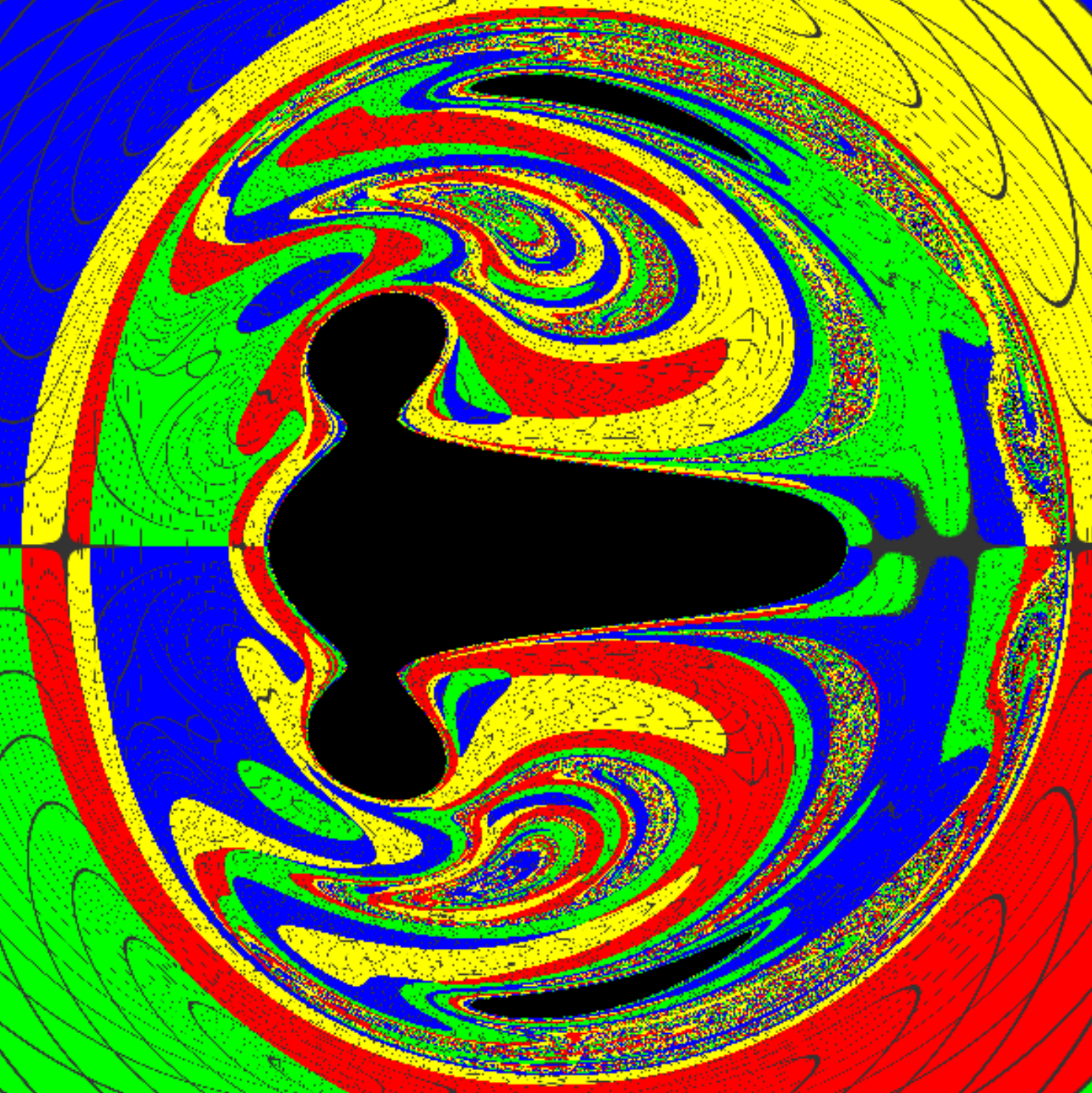}\hspace*{0.2cm}\includegraphics[trim=4cm 3.1cm 4cm 2.7cm,height=6cm,clip]{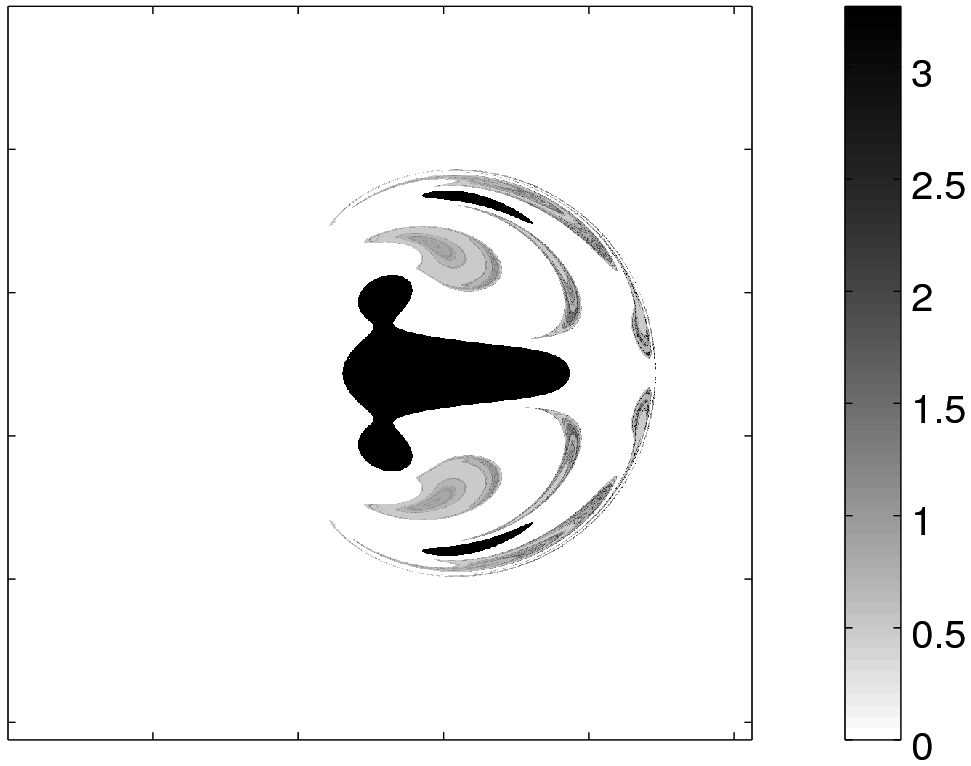}\includegraphics[trim=14.2cm 0.0cm 0cm 0cm,height=6.1cm,clip]{./turn-III.png}\put(22,84){(b)}\\
    \end{center}
    \caption{\small Zoomed turning point heat map (right panels) and lensed image (left panels) for the RBS 11 (row a) and the KBHSH III (row b). Clearly, there is a strong correlation between the chaotic patters (left) and the number of turning points $n$ (right). The logarithmic scale displayed is given by $\log_{10}(n)$, with $n\geqslant 1$; the exceptional case $n=0$ corresponds to the shadow points, shown in black.}
    \label{fig_turning}
\end{figure}

The \textit{time delay function} is defined as the variation of the coordinate time t, in units of $1/\mu$ (with $\mu$ the mass of the scalar field), required for the geodesic emanating from a particular pixel to reach a corresponding point on the celestial sphere or fall asymptotically into the black hole. The idea behind this function is that trajectories which are semi-permanently trapped in the pocket take much longer to escape, giving initially nearby orbits more time to diverge. In Fig. \ref{timedelay} the time delay for configuration III is portrayed as a heat map -- with the corresponding scale on the right of the image -- indicating the variation of the coordinate time for each trajectory to travel from the camera to the celestial sphere. The ``brighter'' regions on the time delay image can be seen to match the chaotic regions seen in the lensing image of this configuration -- see Fig.\ref{fig_turning}.\\

The \textit{number of radial turning points}, on the other hand, is defined as the number of times that $\dot{r}$ changes sign during the light ray's trajectory. Recall that null geodesics on a Kerr spacetime have at most one radial turning point \cite{Wilkins1972}; hence a violation of the latter can be interpreted as a deviation from Kerr.

On the panels to the right of Fig. \ref{fig_turning} we have a representation of the number of radial turning points as a grey level for the RBS configuration 11 (top row) and the KBHSH III (bottom row); a larger number of turning points corresponds to a darker shade in the image, with white connected to just one turning point. The shadow of KBHSH III is represented in black in order to ease visualization -- the number of turning points is actually zero in that case, both for the main shadow as well as the eyebrows. On the left panels of Fig. \ref{fig_turning} we have a representation of the gravitational lensing of the respective configurations; observe the correlation between the regions with a larger number of turning points (right panels) and the chaotic patterns (left  panels). This suggests that having more than one radial turning point is a necessary ingredient for chaos, a feature absent in geodesic motion around a Kerr BH. Note, however, this correlation is not an equivalence:  there are still some regular regions with more than one turning point.

Let us summarise the situation briefly. Chaotic patterns on the image plane correspond to trajectories that stay quasi-bound around the central object (RBS or KBHSH) and hence have a large time delay and numerous (more than one) radial turning points. Spacetimes that admit such trajectories/patterns are those that have a stable light ring (in addition to an unstable light ring as with Kerr). The existence of a stable light ring leads to the formation of pockets in the effective potential describing null geodesic motion; the opening of these pockets to infinity, through a narrow throat, signals the onset of a regime in which quasi-bound trajectories are possible (but not guaranteed). These pockets, while bottle-necked, have the effect of promoting quasi-bound motion. The widening of these pockets suppresses chaoticity but does not eliminate it, even when fully opened. This last feature can be understood intuitively by the effect of a stable light ring on an acceleration field, as defined in Appendix \ref{appendixa2}.

As a side note, we remark that the presence of an ergoregion is not necessary for quasi-bound motion. However the existence of an \textit{ergo-torus} is a sufficient condition for the existence of a stable light ring - this can be understood from the behaviour of the effective potential $h_+$ - and hence a sufficient condition for chaotic behaviour to manifest for a given spacetime.

%

\subsection{Light rings}
\label{seclr}

As we have seen, the existence of light rings is central to the properties of the effective potentials. In particular, the existence of stable light rings allows for pockets, which translate into spacetime quasi-trapping regions for photons. In this subsection we will investigate in detail the properties of the light rings for the above configurations that possess them.\\

Throughout this article, a light ring refers to a null geodesic on the equatorial plane ($\theta=\pi/2$) that satisfies $p_r=p_\theta=0$ and $\dot{p}_r=0$. These conditions are equivalent to $V=0$ and $\partial_r V=0$ ($cf.$ Appendix \ref{appendixa2}). Using equation \eqref{eq_Vtilde} and computing the derivative of $V$ with respect to $r$, enforcing $h_\pm=\eta\iff V=0$ at the end, we obtain:
\[\partial_r V=\pm\left(\frac{E^2}{D}g_{tt}\right)(h_+-h_-)\partial_r h_\pm.\]
Since $h_+\neq h_-$ outside the horizon we conclude\footnote{We would obtain the same result even if $g_{tt}=0$.} that $\partial_r V=0 \iff \partial_r h_\pm=0$. Moreover, since the radial condition for stable (unstable) light rings is $\partial_r^2 V>0$ $\quad(\partial_r^2 V<0)$, by a similar calculation one can then conclude that stable (unstable) light rings satisfy $\pm \partial_r^2 h_\pm >0$, $\quad(\pm \partial_r^2 h_\pm <0$). \\
Besides stability, light rings can also be categorized by their rotational direction. From the geodesic equations for $t$ and $\varphi$ we have that:
\be
\dot{\varphi}/E=-\frac{1}{D}\left(g_{t\varphi} +g_{tt}\eta\right) \ ,\qquad  \dot{t}/E=\frac{1}{D}\left(g_{\varphi\varphi} +g_{t\varphi}\eta\right)\ .
\ee
Their quotient yields
\[\Omega\equiv\frac{d\varphi}{dt}=-\frac{g_{t\varphi} +g_{tt}\eta}{g_{\varphi\varphi} +g_{t\varphi}\eta},\]
which describes the azimuthal rotation direction with respect to a static observer at spatial infinity. At a light ring, $V=0$ holds $g_{\varphi\varphi}+2g_{t\varphi}\eta +g_{tt}\eta^2=0$, which leads to:
\[g_{\varphi\varphi}+\eta g_{t\varphi}=-\eta\left(g_{t\varphi} +\eta g_{tt}\right) \implies \Omega=\frac{1}{\eta} \quad(\textrm{at a light ring}).\]
Hence, the rotational direction of the light ring is given by sign of the impact parameter $\eta$.
Additionally, at a light ring the expression for $\dot{\varphi}/E$ can be simplified using equation \eqref{eq_h}:
\[\dot{\varphi}/E=-\frac{1}{D}(g_{t\varphi}+ g_{tt}h_\pm)=\mp\frac{1}{D}\sqrt{D} \implies \mp\dot{\varphi}/E>0.\]
Since $\eta=1/\Omega=\dot{t}/\dot{\varphi}$ and recalling sections \ref{out_ergo} -- \ref{in_ergo}:
\begin{align*}
&g_{tt}<0 \implies \mp h_\pm>0\implies \mp\,\eta>0 \implies (\mp\,\eta)(\mp\,\dot{\varphi}/E)>0 \implies \dot{t}/E>0;\\
&g_{tt}>0 \implies h_\pm>0 \implies \eta>0 \implies \mp\dot{t}/E>0,
\end{align*}
where $g_{t\varphi}<0$ was assumed.
Hence we conclude that we can have $\dot{t}/E<0$ at a light ring only if it is inside an ergoregion with $h_+=\eta$. For physical photons, this actually implies that their energy is negative. Let us first detail this conclusion and then discuss its implications.\\

Consider a Zero Angular Momentum Observer (ZAMO) frame~\cite{Novikov1998} at the position of the light ring. The locally measured energy  of the photon, $p^{(t)}$, in this frame, is given by $p^{(t)}=E\Lambda -\gamma \Phi$,\cite{Cunha:2016bpi}  which must be positive for physical photons. Notice that in general $p^{(t)}$ is different from $E$, the latter being the photon's energy with respect to spatial infinity. The expressions for $\gamma$ and $\Lambda$ are given by:
\be
\gamma=-\frac{g_{t\varphi}}{g_{\varphi\varphi}}\sqrt{\frac{g_{\varphi\varphi}}{D}},\qquad\qquad\Lambda=\sqrt{\frac{g_{\varphi\varphi}}{D}}\ .
\label{gl}
\ee
Hence we have:
\be
p^{(t)}>0 \quad \Leftrightarrow \quad E\Lambda>\gamma\Phi \quad \Leftrightarrow \quad E>-\frac{g_{t\varphi}}{g_{\varphi\varphi}}\Phi\ .
\ee
The same relation is obtained if we require $\dot{t}>0$, $i.e.$, $\dot{t}=\frac{1}{D}\left(E g_{\varphi\varphi} +\Phi g_{t\varphi}\right)>0$. We used the fact that $D>0$ and $g_{\varphi\varphi}>0$. Hence $p^{(t)}>0  \Leftrightarrow  \dot{t}>0$.
Since it is always possible to construct a ZAMO frame at the position of the light ring, we conclude that $\dot{t}>0$ is a necessary condition for a physical photon. Then, the condition $\dot{t}/E<0$ implies $E<0$. Despite having a positive energy regarding a local observer, the photon has negative energy with respect to spatial infinity. Likely, the accumulation of negative energy states around this light ring is associated to an instability~\cite{Keir:2014oka}.\\

For the spacetime configurations already analysed, the signs of $\eta,\,\,\dot{t}/E$ and $\dot{\varphi}/E$ for different light rings (LR) are organized in the following table, together with other information.

\begin{center}
\begin{tabular}{l*{8}{c}cc}
\hline
\hline
Configuration in~\cite{Cunha:2015yba} & Fig. & LR & $R$ & stability & $\eta$ & $g_{tt}$ & $\dot{t}/E$ & $d\varphi/dt$\\
\hline
\hline
\multirow{2}{*}{RBS 10} & \multirow{2}{*}{\ref{fig_7}, \ref{gallery10}} & $h_+$ & 0.60 & stable & $-$ & $-$  & $+$ & $-$  \\
 & & $h_+$ & 0.79 & unstable & $-$ & $-$ & $+$ & $-$ \\
\hline
\multirow{2}{*}{RBS 11 } & \multirow{2}{*}{\ref{fig1}, \ref{fig_65}, \ref{gallery11}} & $h_+$ & 0.39 & stable & $+$ & $+$ & $-$ & $+$ \\
& & $h_+$ & 0.76  & unstable & $-$ & $-$ & $+$ & $-$ \\
\hline
\multirow{4}{*}{KBHSH III} & \multirow{4}{*}{\ref{fig_V}, \ref{galleryIIIa} -- \ref{galleryIIIc} } & $h_-$ & 0.03 & unstable & $+$ & $-$ & $+$ & $+$ \\
& & $h_+$ & 0.06  & unstable & $-$ & $-$ & $+$ & $-$ \\
& & $h_+$ & 0.30 & stable & $+$ & $+$ & $-$ & $+$ \\
& & $h_+$ & 0.74 & unstable & $-$ & $-$ & $+$ & $-$ \\
\hline
\end{tabular}
\end{center}
\vspace{4mm}

The value of $g_{tt}$  reveals whether a light ring is inside an ergoregion or not ($g_{tt}>0$ in the former case). This occurs for two of the cases displayed. In both cases, as expected, the light ring is co-rotating, from the viewpoint of the asymptotic observer, as appropriate for causal particles inside an ergoregion. For both these cases observe that $\dot{t}/E<0$, which implies from the previous discussion that $E<0$ for a physical state. We remark, however, that it is possible to have light rings inside an ergoregion with $E>0$. Indeed, the KBHSH dubbed configuration II in~\cite{Cunha:2015yba}, which is not discussed in detail in this paper, has two unstable light rings, one of which is inside an ergoregion with $\dot{t}/E>0 \implies E>0$.\\

Throughout this article, light rings are only defined on the equatorial plane ($\theta=\pi/2$). We remark that it seems plausible such concept can be continued outside the equatorial plane, by a set of  \textit{spherical orbits}. Such orbits would involve oscillatory motion in the $\theta$ coordinate, with light rings being a subset of the latter ($cf.$ Appendix \ref{appendixao}, \ref{appendixa} and \ref{appendixa2}). We shall not, however, make a detailed study of these spherical orbits in this work.

\section{Lensed images with PYHOLE}
\label{sec3}
In the previous section we provided various insights for the emergence of chaotic patterns in lensing images of RBSs and KBHsSH. In particular we have established that the presence of stable light rings allows for the existence of pockets in the effective potential leading to quasi-bound orbits, which are strongly correlated to the chaotic patterns. In this section we will exhibit a gallery of examples of spacetime orbits, represented together with the effective potential and the corresponding point in the lensing image, for a sample of solutions of RBSs and KBHsSH. To do so, we shall use a new code for ray tracing, \textsc{pyhole}, which provides a simple interactive graphical user interface with tools that facilitate the interpretation of the results (see Appendix\ref{appendixb}). Of particular importance is the ability within this code, to select any point in the lensing image and obtain, within seconds, the corresponding visualization of the spacetime trajectory, as well as the trajectory in the effective potential. This feature is used in this section to analyse the emergence of chaotic scattering.

\subsection{Ray-tracing setup for lensed images}

Our setup is the same as in~\cite{Cunha:2015yba}, inspired by the construction in~\cite{Bohn:2014xxa}. We consider a coordinate system centered around the region we wish to study -- a RBS or a KBHSH. An observer (or camera) is placed off-centre in the spacetime and it receives light from a collection of far away sources, emitting isotropically in all directions, which we call the \textit{celestial sphere}. We assume from the outset that this is a four dimensional stationary (axisymmetric) background and take our coordinates to be adapted to these symmetries: the coordinates are spherical-polar $\{ t,r,\theta, \phi \}$ as defined in~\cite{Herdeiro:2014goa}, where ${\bf k}=\partial/\partial t$ and ${\bf m}=\partial/\partial \phi$ are the Killing vector fields which generate respectively the stationarity and axi-symmetry of the spacetime. We fix $t= \phi = 0$ for the camera's position, at some $(r,\theta)$ coordinates $(r_{\rm obs},\theta_{\rm obs})$. The emitting celestial sphere surrounds both the central region and the observer, being  placed at a large radial coordinate $r_{\rm cs}$.

In particular for this work, unless otherwise stated, we have placed the camera on the equatorial plane, $\theta = \pi/2$, and at a fixed radial distance specified differently for the RBS and KBHSH solutions: For RBS solutions we keep the camera at a circumferential radius of $\rho_{\rm obs} = 22.5 / \mu$ with $\mu$ the mass of the scalar field (taken to be $1$). For KBHsSH we place the camera at a circumferential radius of $\rho_{\rm obs} = 15 M$ where M is the ADM mass of the BH. Here the circumferential radius $\rho$ is defined as
\be
\rho = \frac{1}{2\pi}\oint d\varphi \sqrt{g_{\varphi\varphi}} \ ,
\ee
where $g_{\varphi\varphi}$ is evaluated on the equatorial plane on a spacelike slice. The celestial sphere is then placed at $\rho_{\rm cs} = 2\rho_{\rm obs}$. These relations implicitly fix $r_{\rm obs}$ and $r_{\rm cs}$.

The camera is attached to a ZAMO with frame: $e_1 \propto \partial_{\phi}, \; e_2 \propto \partial_{\theta}, \; e_3 \propto -\partial_r$; $e_0$ is directed perpendicular to the constant time hypersurface. We then define standard spherical coordinates on the observer's sky in terms of a polar angle with respect to $e_3$ and an azimuthal angle in the $e_1 - e_2$ plane, measured with respect to $e_1$.

Given a light ray incident on the camera, it is straightforward to relate its momentum, at the camera position, to these spherical coordinates~\cite{Cunha:2016bpi}. To obtain an image, a scan over observation angles is performed, tracing the corresponding light rays backwards on the background, starting at the camera position and ending, heuristically, either at a point on the distant celestial sphere or at the horizon, in case there is one.

One further step is required, which is a projection from observation angles to the image plane, and this is described briefly in Appendix \ref{appendixb}. The net result is a map, from camera pixels to points on the celestial sphere (or vice versa), that allows us to construct the final lensed image.

\subsection{Rotating boson stars}
\label{secgallery}

\begin{figure}[h!]
\begin{center}
\includegraphics[trim={2.5cm 0cm 1.5cm 1.2cm}, clip, height=.28\textheight, angle =0]{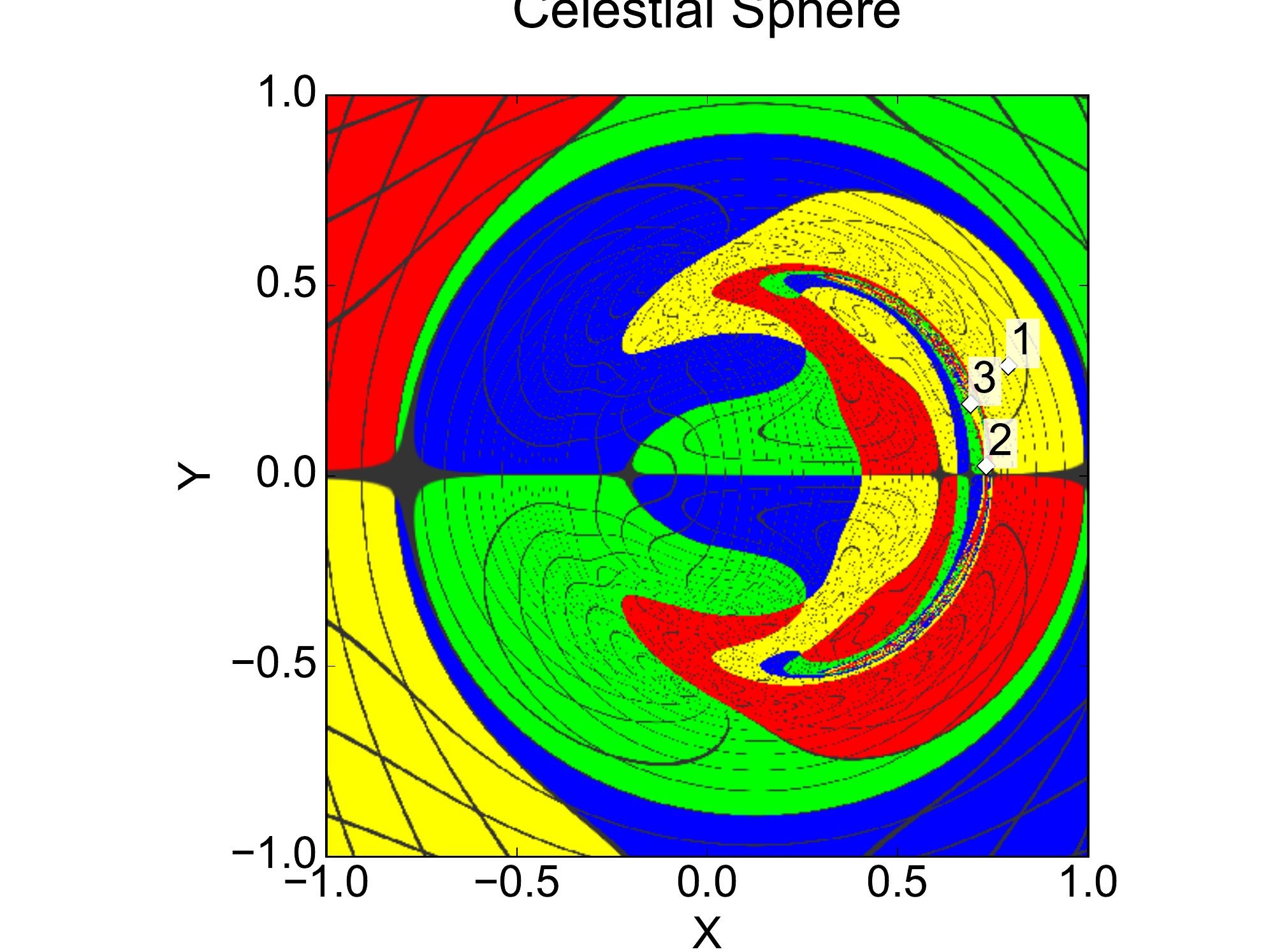}

\vspace{0.6cm}

\includegraphics[trim={0.2cm 0cm 1.5cm 1.6cm}, clip, height=.18\textheight, angle =0]{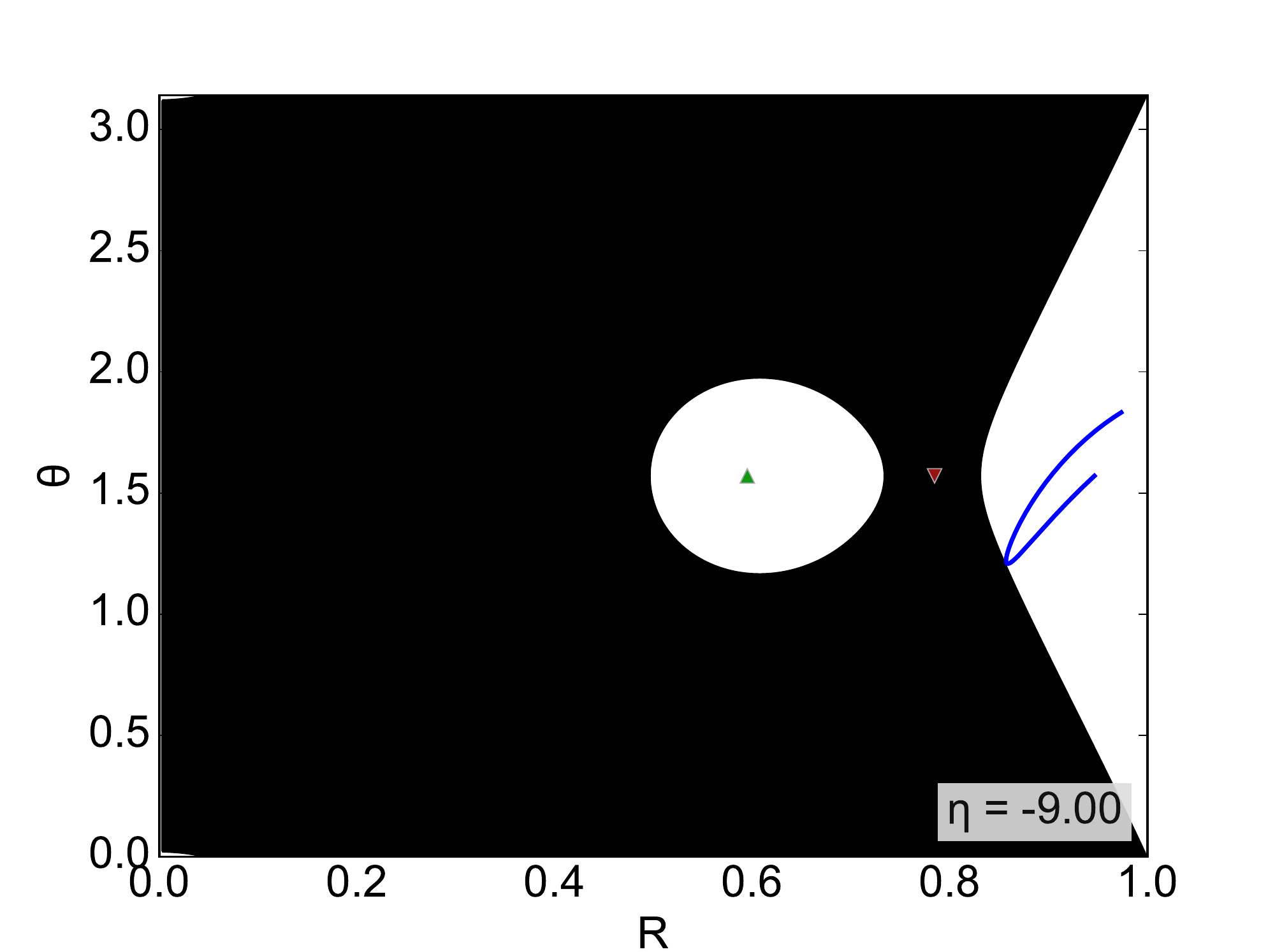}\
\includegraphics[trim={3.0cm 0.3cm 0.6cm 2.3cm}, clip, height=.16\textheight, angle =0]{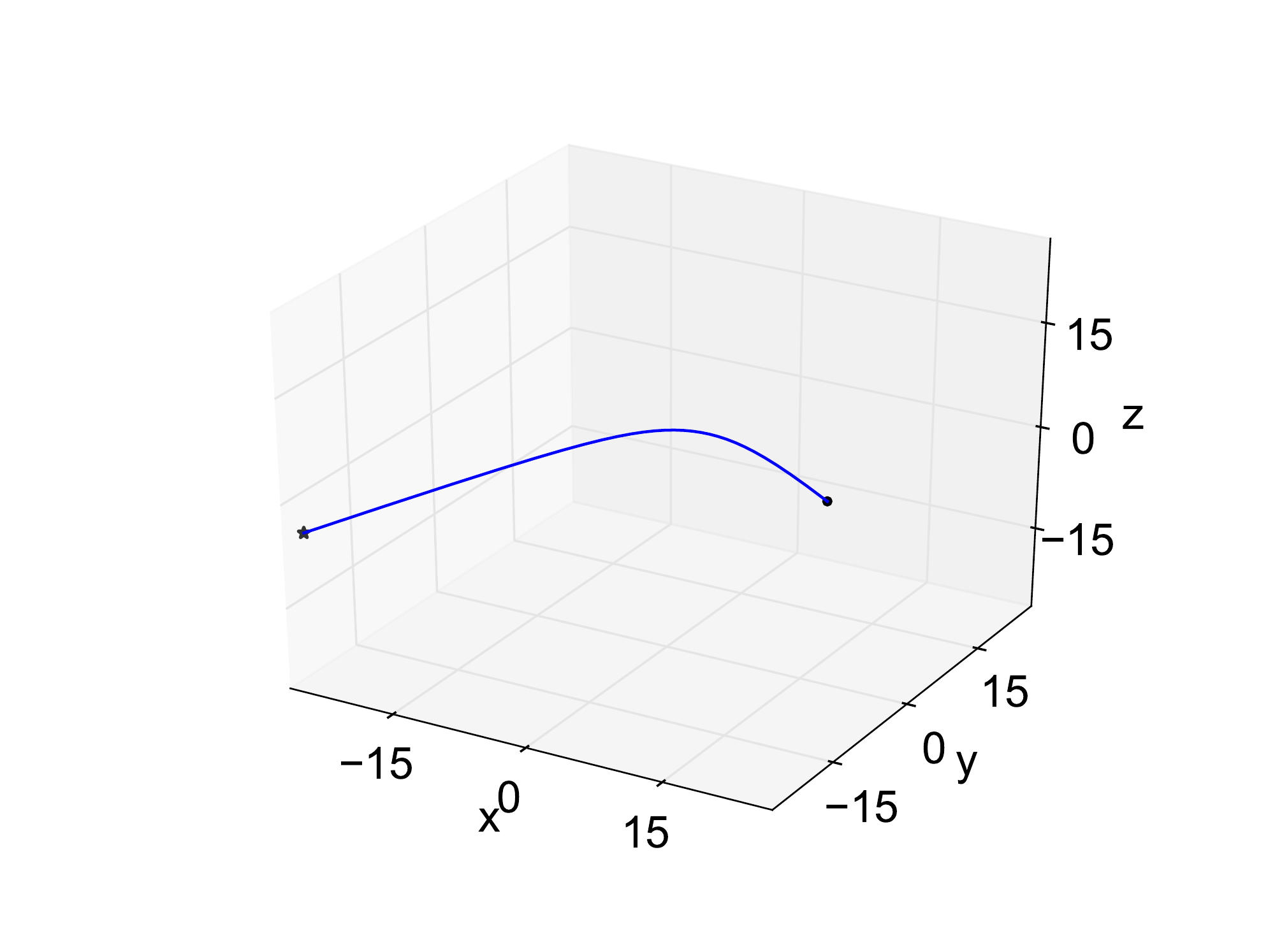}\
\put(0,50){${\bf 1}_{10}$}

\vspace{0.3cm}

\includegraphics[trim={0.2cm 0cm 1.5cm 1.6cm}, clip, height=.18\textheight, angle =0]{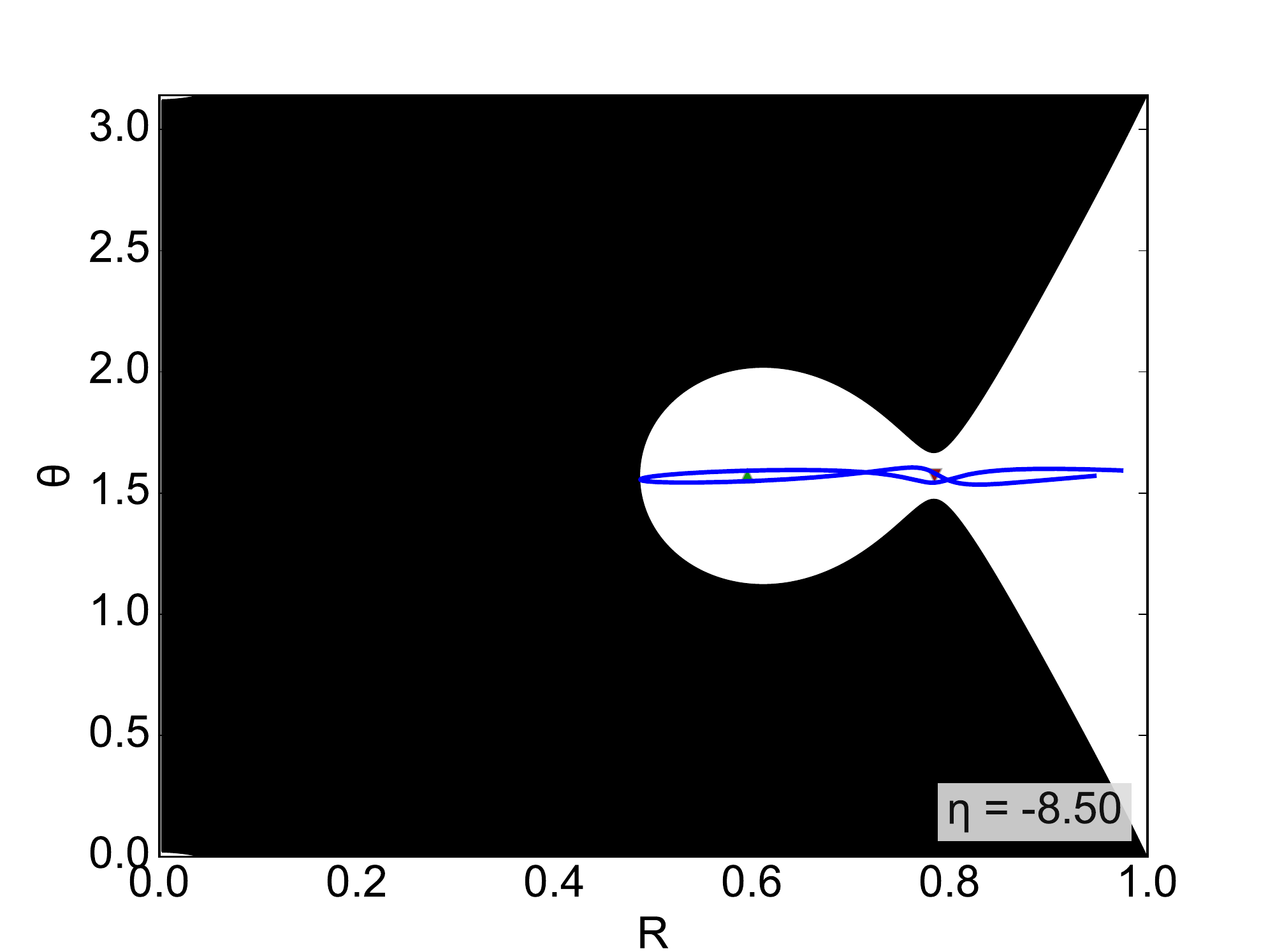}\
\includegraphics[trim={3.0cm 0.3cm 0.6cm 2.3cm}, clip, height=.16\textheight, angle =0]{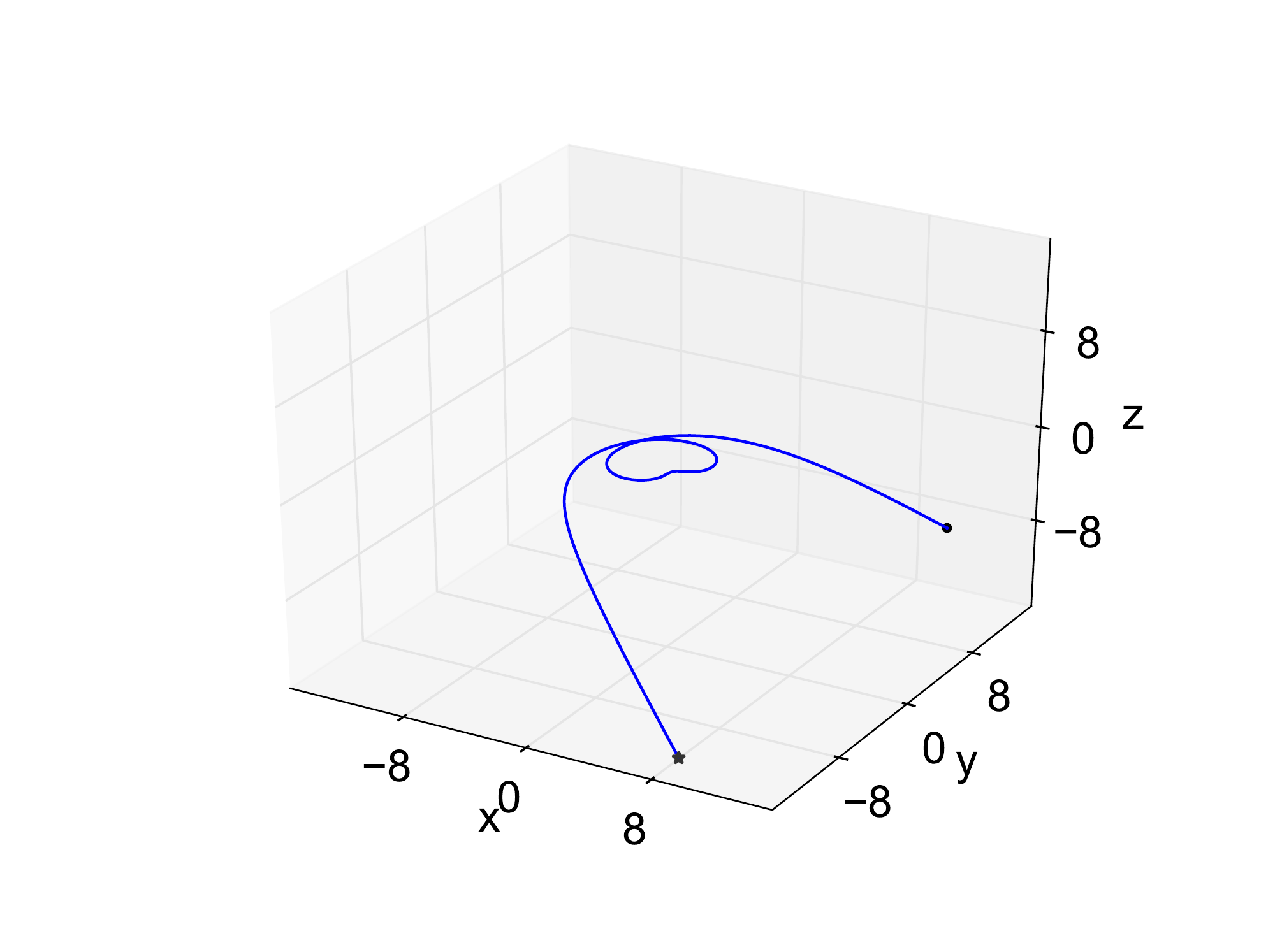}\
\put(0,50){${\bf 2}_{10}$}

\vspace{0.3cm}

\includegraphics[trim={0.2cm 0cm 1.5cm 1.6cm}, clip, height=.18\textheight, angle =0]{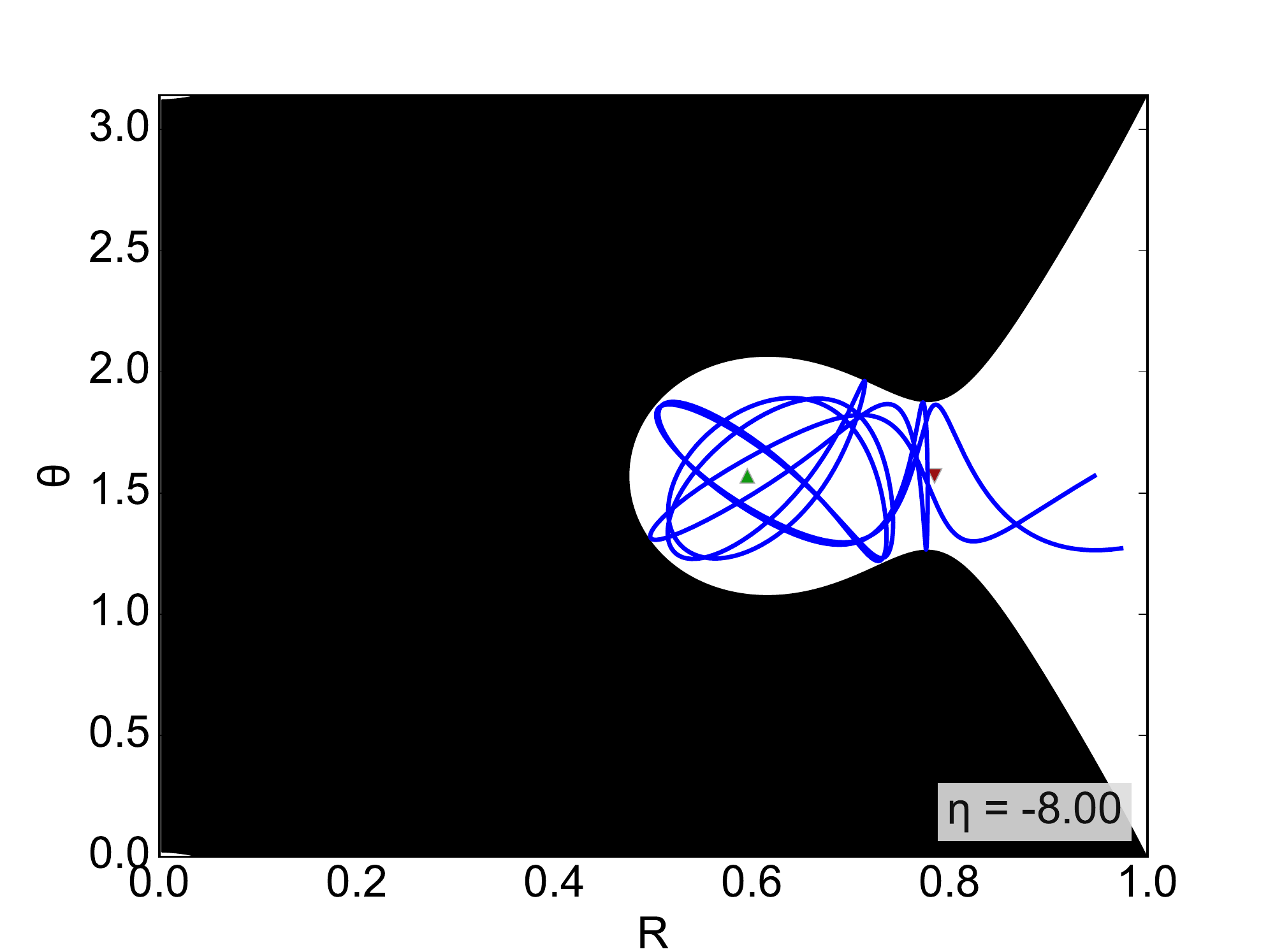}\
\includegraphics[trim={3.0cm 0.3cm 0.6cm 2.3cm}, clip, height=.16\textheight, angle =0]{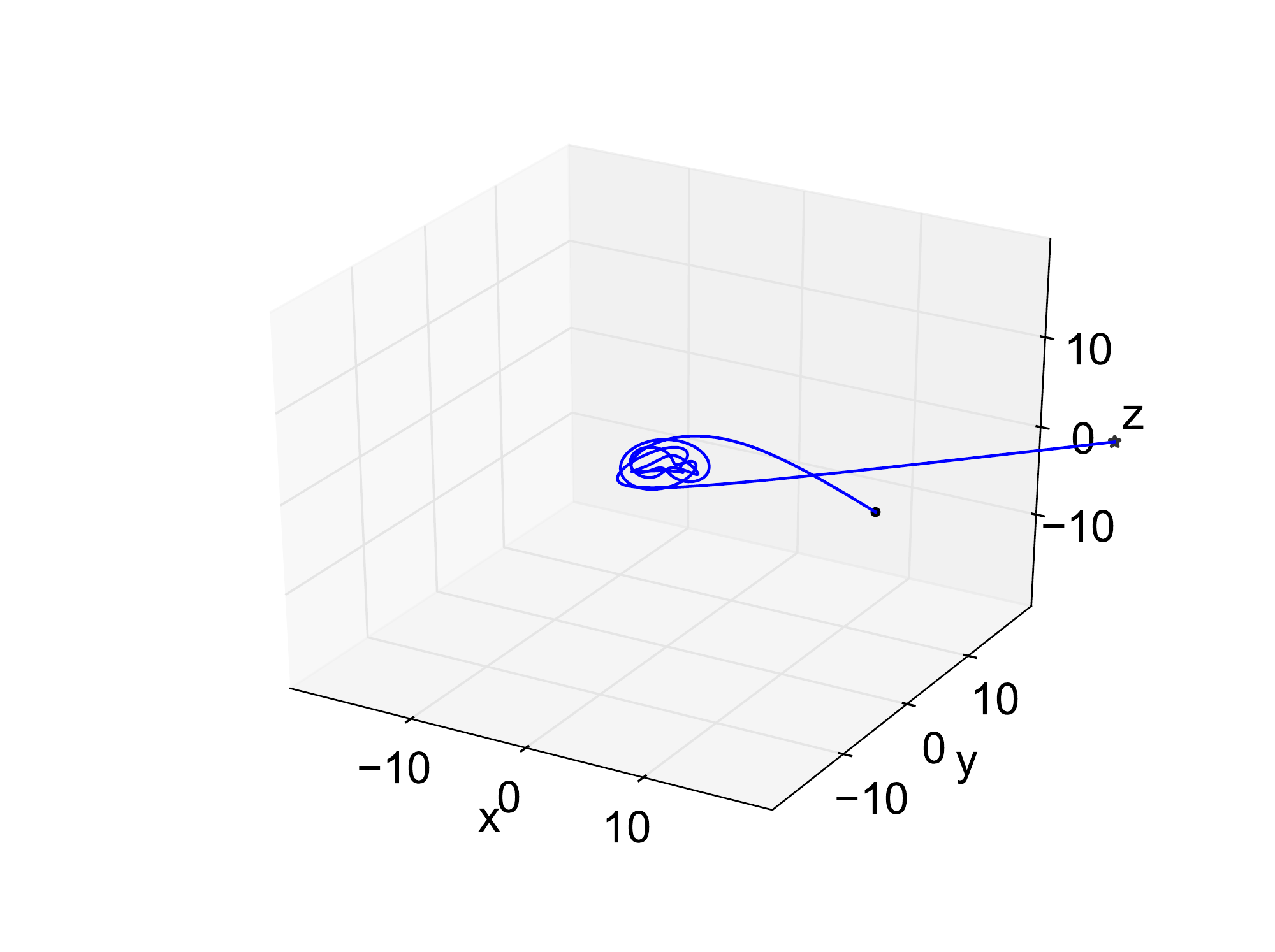}\
\put(0,50){${\bf 3}_{10}$}
\end{center}
\caption{(Top) Lensing of configuration 10 with three highlighted points. Corresponding scattering orbits in  the effective potential (left) and spacetime (right).}
\label{gallery10}
\end{figure}

As a first example, we show in Fig.~\ref{gallery10} the lensing of configuration 10, as seen from the equatorial plane. In order to identify points in the image, we introduced an image coordinate system $(X,Y)$ ranging from $(-1,-1)$ at the lower left corner of the image to $(1,1)$ at the upper right corner. We then selected three points in the lensing image, expressed as ${\bf 1}_{10}$,  ${\bf 2}_{10}$, and  ${\bf 3}_{10}$, where the subscript denotes the configuration these points belong to. The corresponding impact parameters and their location in $(X,Y)$ image coordinates are:

\begin{center}
\begin{tabular}{lcc}
\hline
\hline
Point  & $\eta$ & $(X,Y)$   \\
\hline
\hline
${\bf 1}_{10}$   & -9.00 & $(0.790, 0.289)$  \\
 \hline
${\bf 2}_{10}$ & -8.50 & $ (0.732, 0.026)$  \\
 \hline
${\bf 3}_{10}$  & -8.00 & $(0.690, 0.189)$  \\
 \hline
\end{tabular}
\end{center}

Point ${\bf 1}_{10}$ corresponds to an impact parameter for which the effective potential forms a pocket around a stable light ring (marked by a \textcolor{OliveGreen}{green} upright triangle in the potential plot) that does not connect with the exterior region. Hence this trajectory cannot get trapped and this point belongs to a non-chaotic region in the lensing. The corresponding spacetime trajectory exhibits only weak bending around the center. In order to ease the representation of the trajectory, Cartesian-like coordinates $(x,y,z)$ are used, defined from $(r,\theta,\varphi)$ as if these were standard spherical coordinates.\\
For point ${\bf 2}_{10}$, the effective potential has a pocket with a small opening (a ``throat'') around an unstable light ring (marked by a \textcolor{red}{red} inverted triangle in the potential plot), connecting it to the asymptotic region. But as the corresponding orbit has small $\theta$ motion, the photon enters and exits the pocket after a single bounce off the boundary of the pocket. The corresponding point in the lensing image is at the threshold between a chaotic and non-chaotic region. Finally, point ${\bf 3}_{10}$ corresponds to a trajectory that gets trapped for some time in the pocket,  bouncing off its boundary a few times before finding its way out. In the spacetime, the photon circles around the central region a few times, before being scattered off to infinity. In the lensing image this point appears inside a chaotic region.

\begin{figure}[!h]
\begin{center}
\includegraphics[trim={2.5cm 0cm 1.5cm 1.2cm}, clip, height=.28\textheight, angle =0]{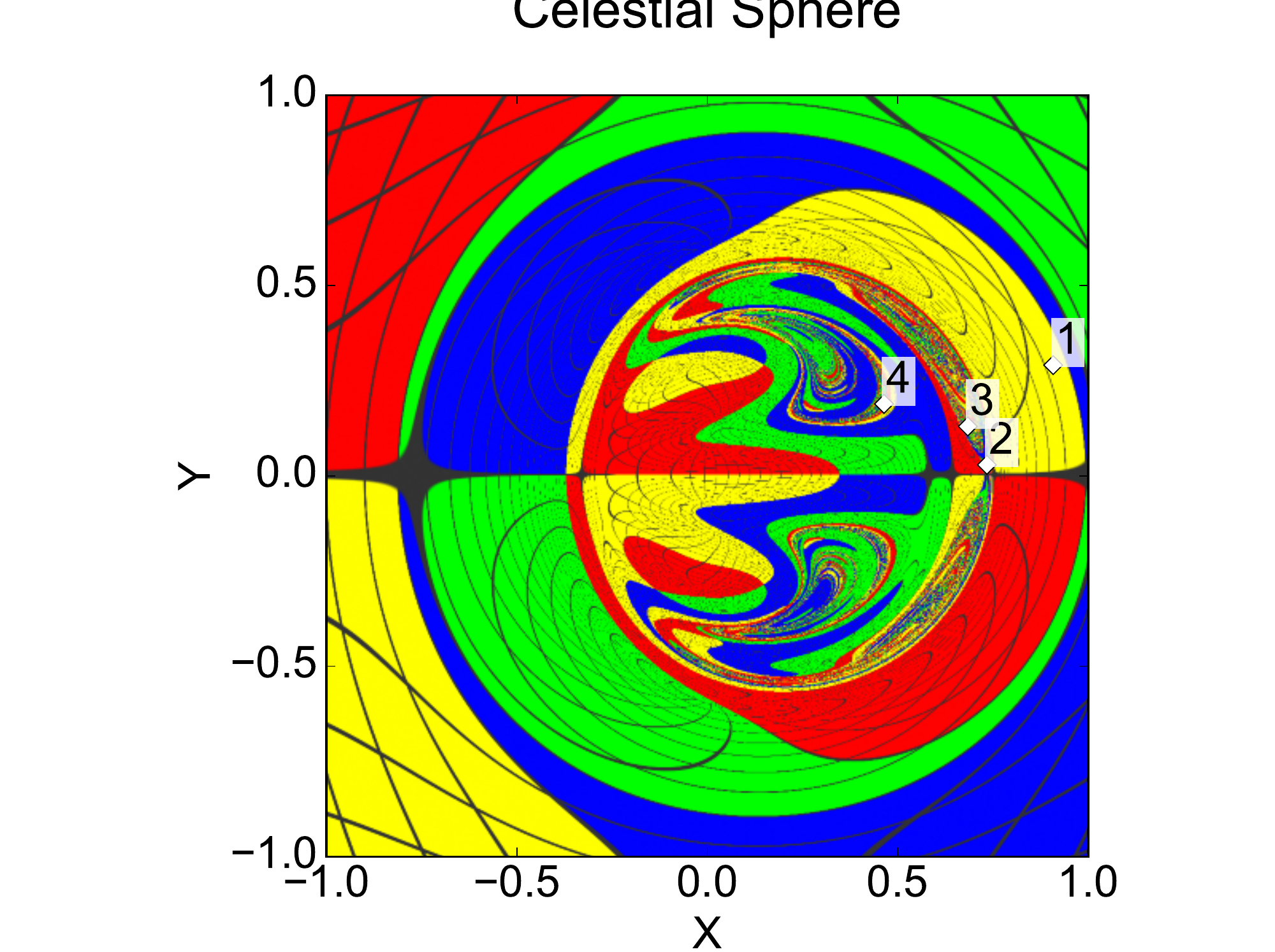}

\vspace{0.6cm}



\includegraphics[trim={0.2cm 0cm 1.5cm 1.6cm}, clip, height=.18\textheight, angle =0]{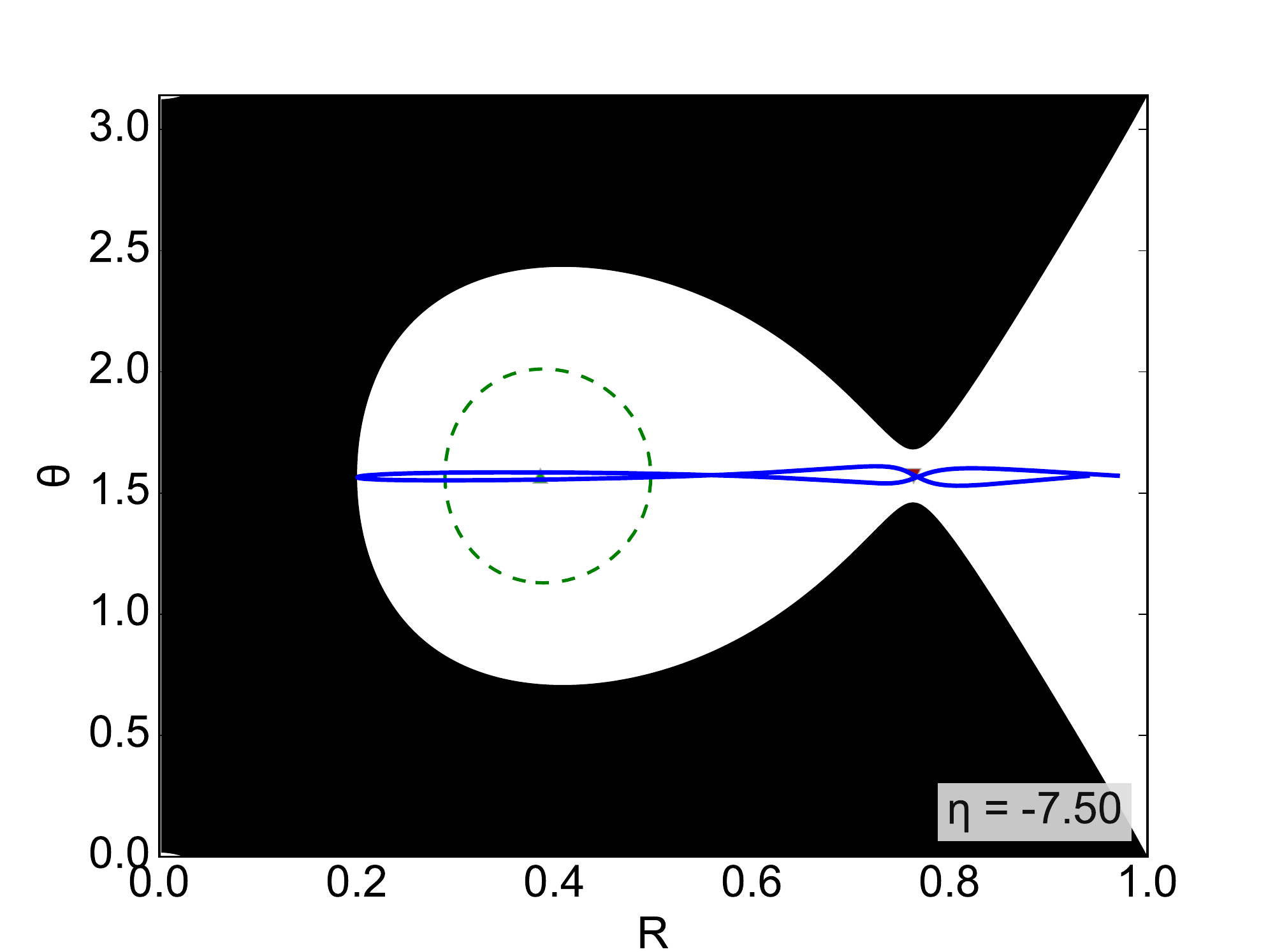}\
\includegraphics[trim={3.0cm 0.3cm 0.6cm 2.3cm}, clip, height=.16\textheight, angle =0]{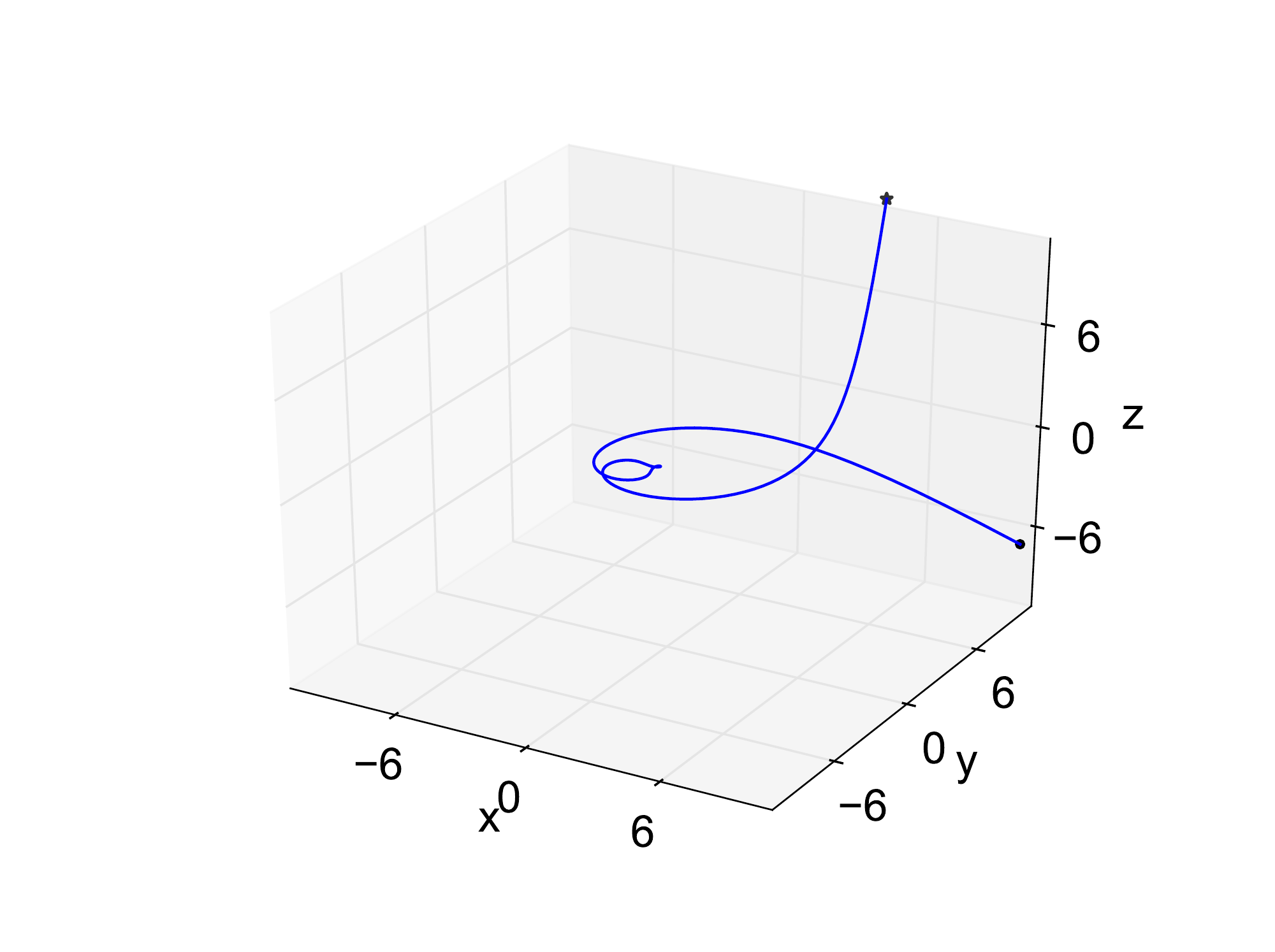}\
\put(0,50){${\bf 2}_{11}$}

\vspace{0.3cm}

\includegraphics[trim={0.2cm 0cm 1.5cm 1.6cm}, clip, height=.18\textheight, angle =0]{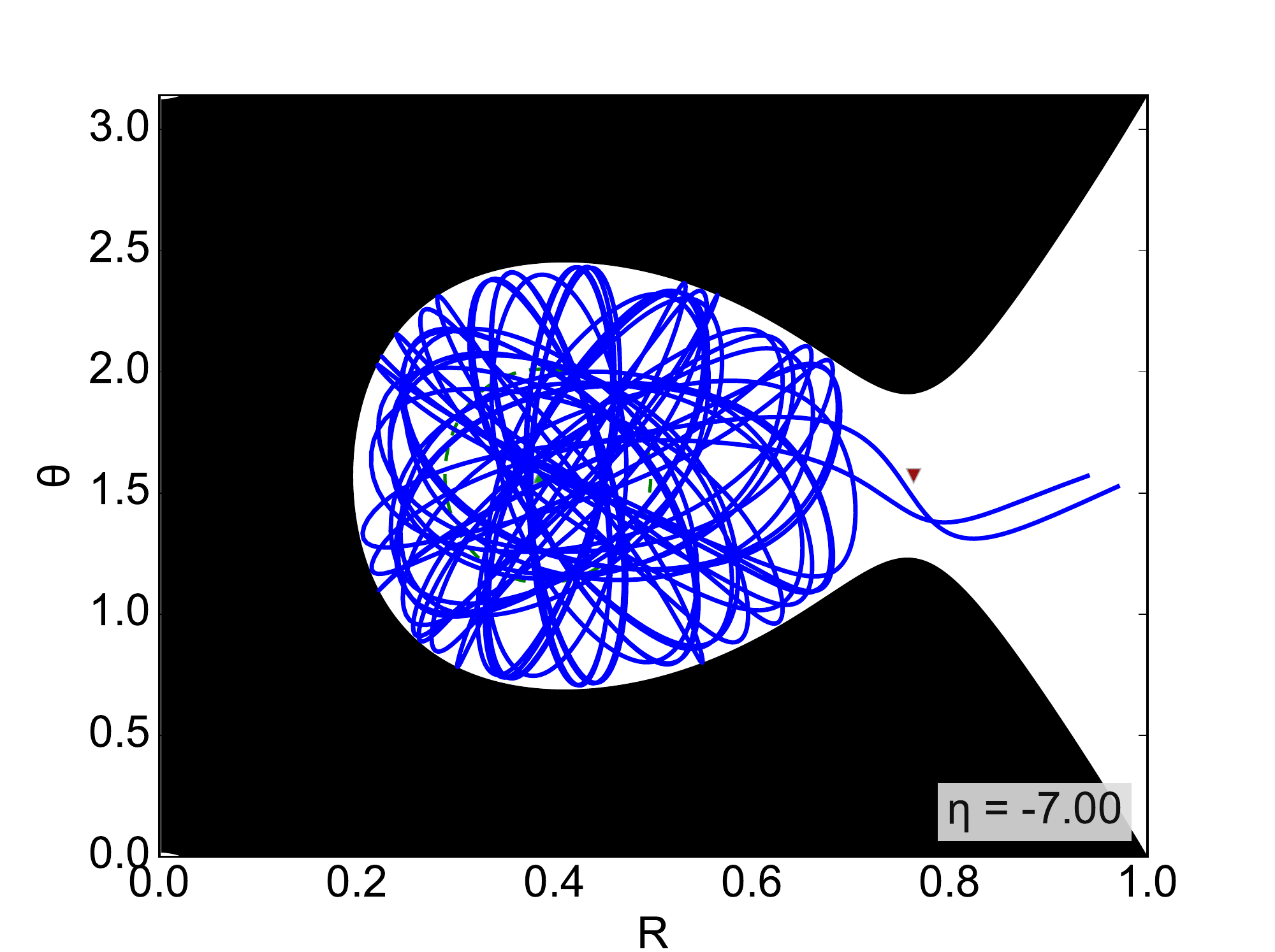}\
\includegraphics[trim={3.0cm 0.3cm 0.6cm 2.3cm}, clip, height=.16\textheight, angle =0]{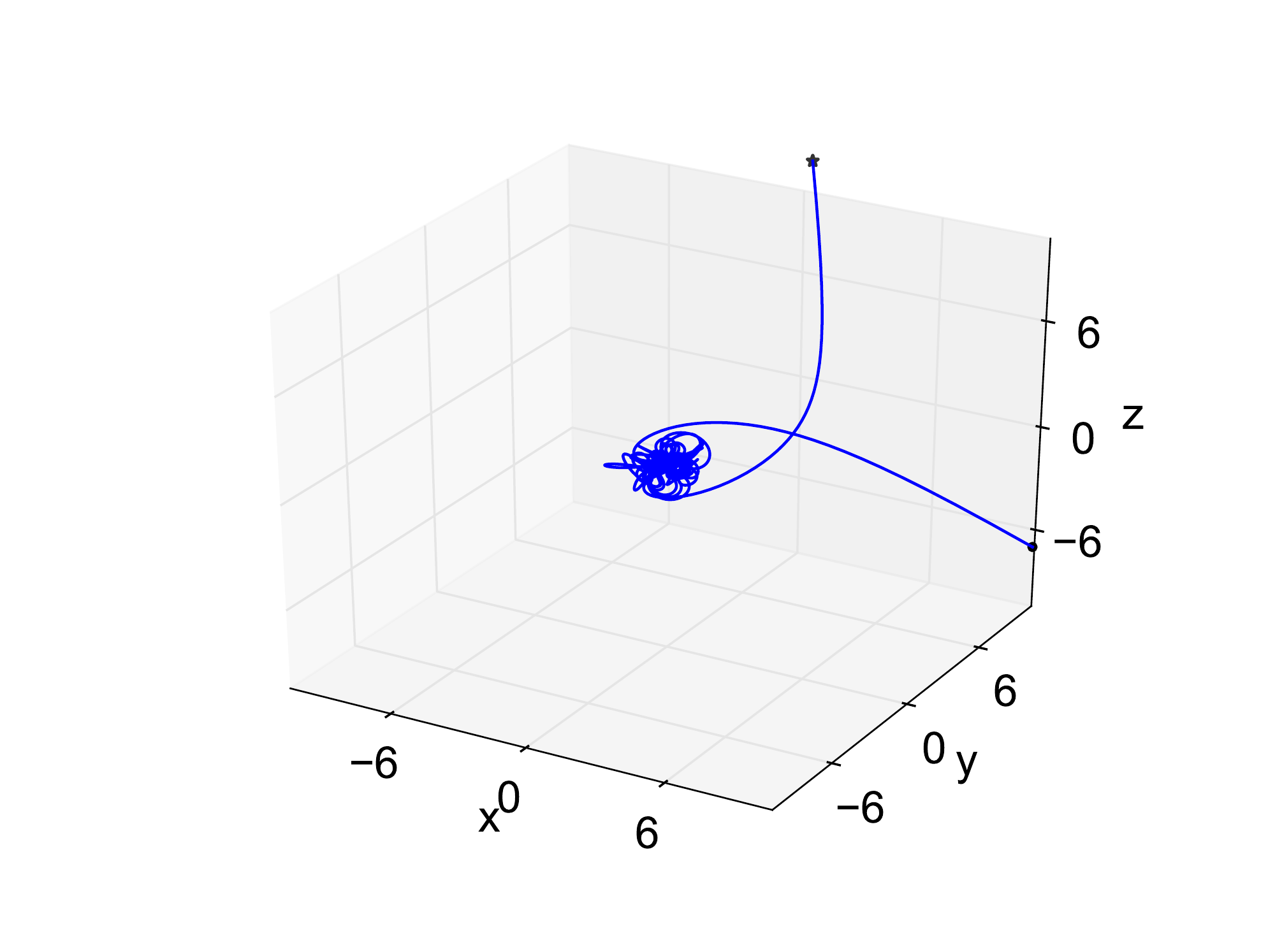}\
\put(0,50){${\bf 3}_{11}$}

\vspace{0.3cm}

\includegraphics[trim={0.2cm 0cm 1.5cm 1.6cm}, clip, height=.18\textheight, angle =0]{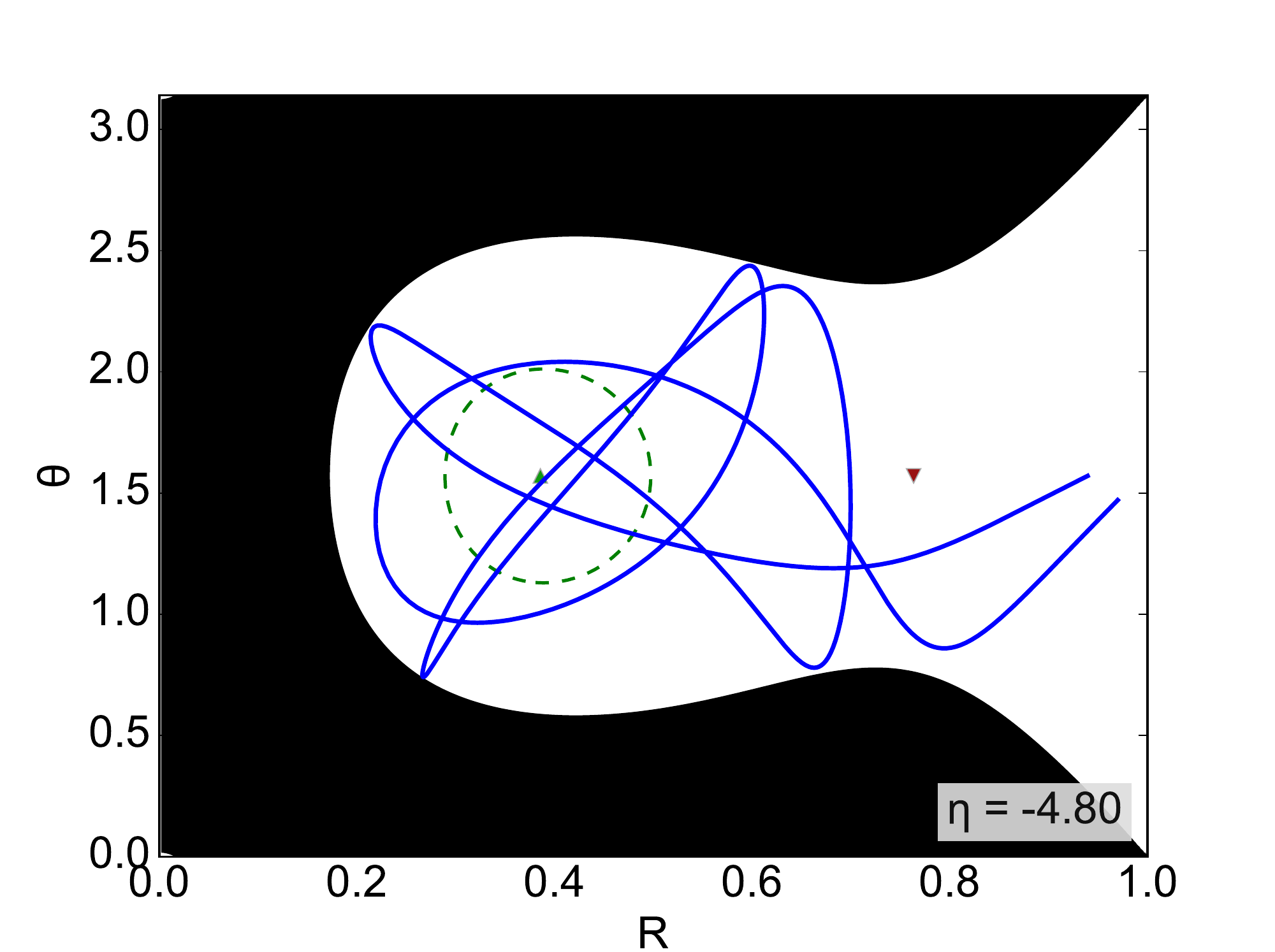}\
\includegraphics[trim={3.0cm 0.3cm 0.6cm 2.3cm}, clip, height=.16\textheight, angle =0]{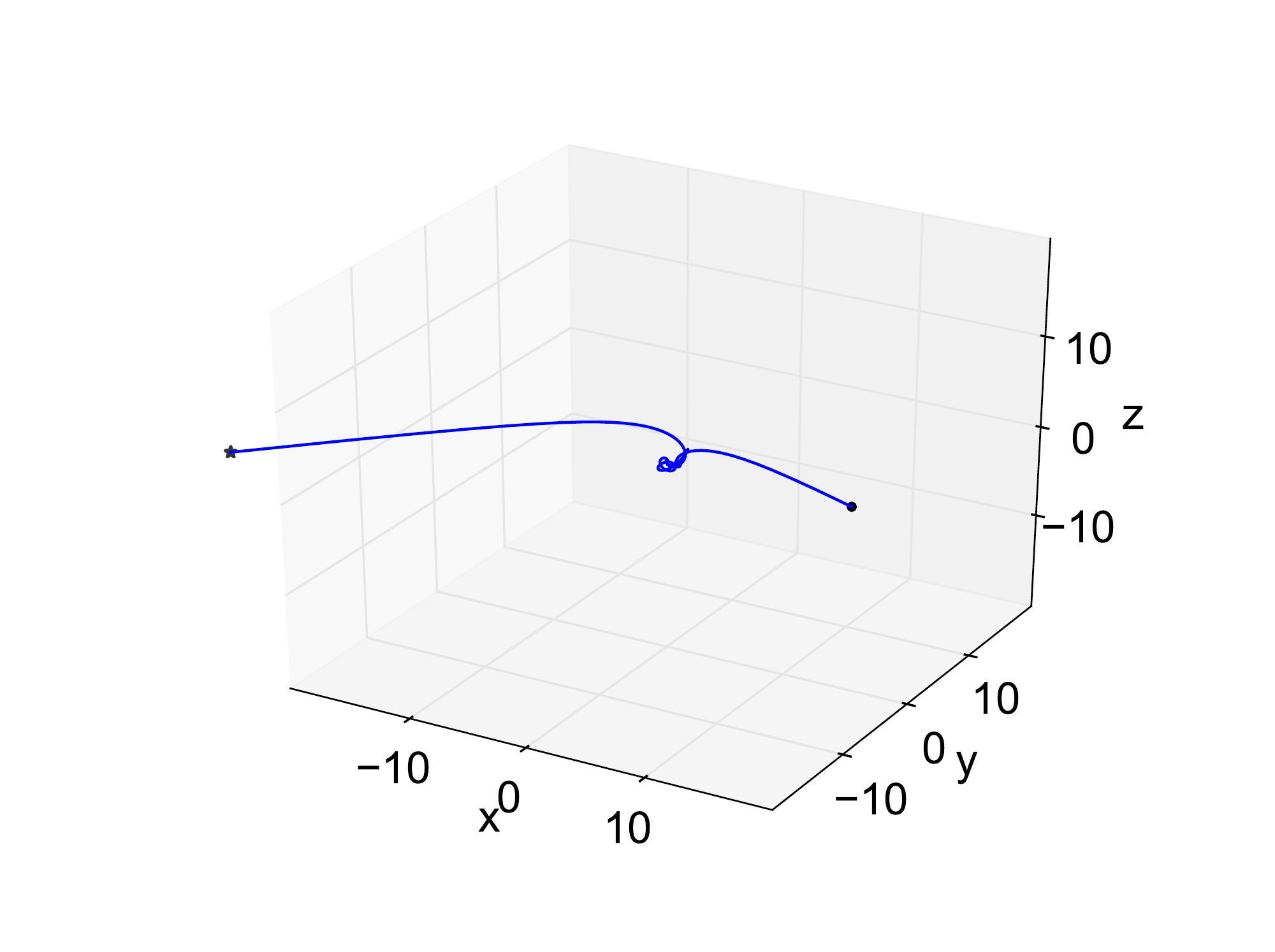}\
\put(0,50){${\bf 4}_{11}$}
\end{center}
\caption{(Top) Lensing of configuration 11 with four highlighted points. Corresponding scattering orbits (except point ${\bf 1}_{11}$) in  the effective potential (left) and  spacetime (right).}
\label{gallery11}
\end{figure}

In Fig.~\ref{gallery11}, we exhibit the lensing of configuration 11, again seen from the equatorial plane, and we have selected four points in the lensing image, denoted ${\bf 1}_{11}$ to  ${\bf 4}_{11}$. The corresponding impact parameters and their location in $(X,Y)$ image coordinates are:

\begin{center}
\begin{tabular}{lcc}
\hline
\hline
Point  & $\eta$ & $(X,Y)$   \\
\hline
\hline
${\bf 1}_{11}$   & -9.00 & $(0.908, 0.291)$  \\
 \hline
${\bf 2}_{11}$ & -7.50 & $ (0.734, 0.029)$  \\
 \hline
${\bf 3}_{11}$  & -7.00 & $(0.685, 0.130)$  \\
 \hline
 ${\bf 4}_{11}$  & -4.80 & $(0.464, 0.189)$  \\
 \hline
\end{tabular}
\end{center}

The new qualitative feature in this configuration, with respect to the previous one, is the existence of an ergo-region.  Its boundary is shown as a dashed \textcolor{OliveGreen}{green} line in the effective potentials. Point ${\bf 1}_{11}$ corresponds to an impact parameter for which the pocket does not connect with the asymptotic region. Its potential and spacetime orbit are similar to those of point Point ${\bf 1}_{10}$ (Fig.~\ref{gallery10}), and it is not shown here. Points ${\bf 2}_{11}$ and ${\bf 3}_{11}$ are also qualitatively similar to points ${\bf 2}_{10}$ and ${\bf 3}_{10}$ as shown in Fig.~\ref{gallery10}. But point ${\bf 4}_{11}$ is qualitatively new, in the sense that the chaotic region has now extended to other disjoint parts on the lensing image.

\begin{figure}[!hp]
\begin{center}
\includegraphics[trim={2.5cm 0cm 1.5cm 1.2cm}, clip, height=.28\textheight, angle =0]{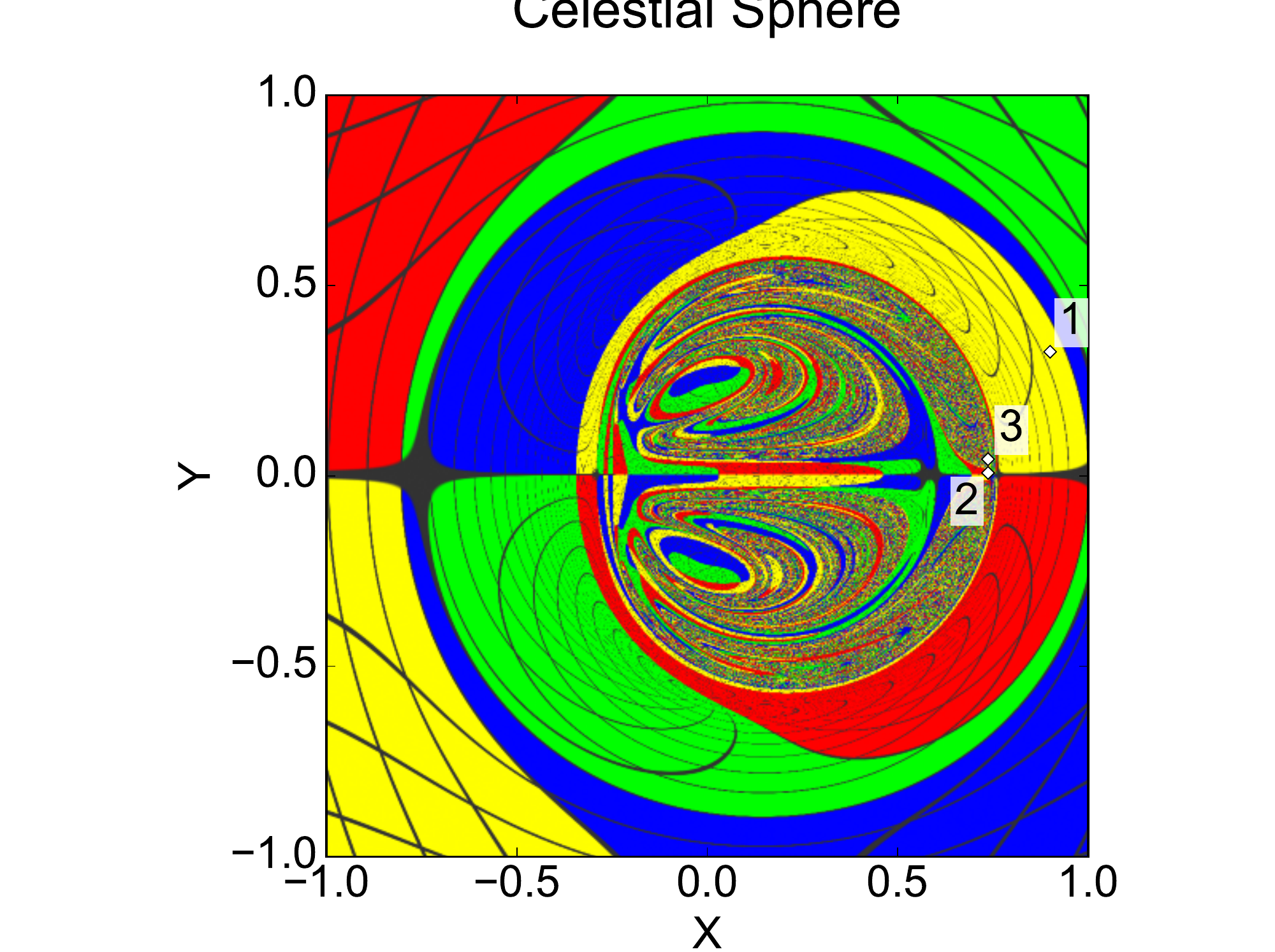}
\includegraphics[height=.28\textheight, angle =0]{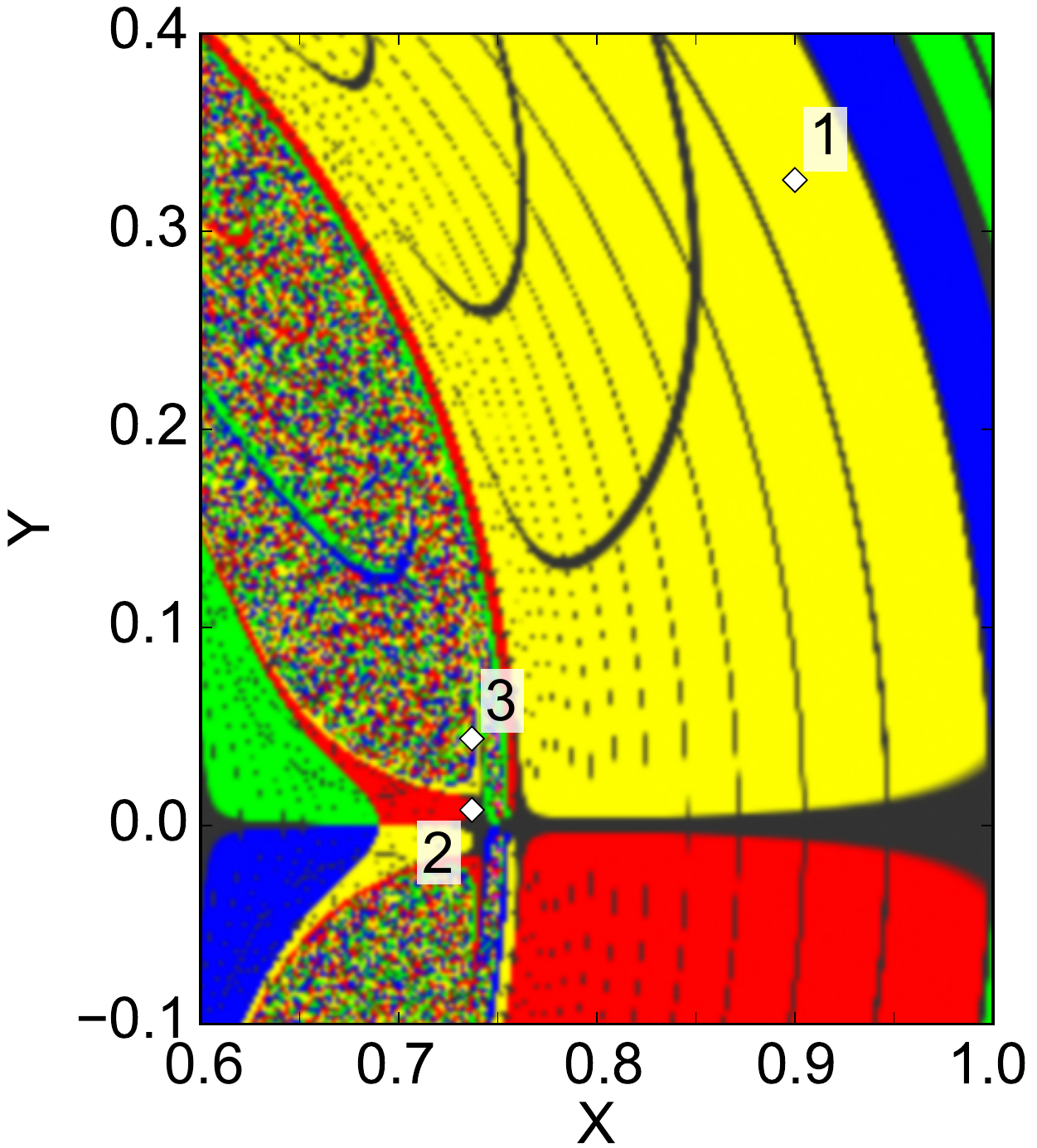}

\vspace{0.6cm}

\includegraphics[trim={0.2cm 0cm 1.5cm 1.6cm}, clip, height=.18\textheight, angle =0]{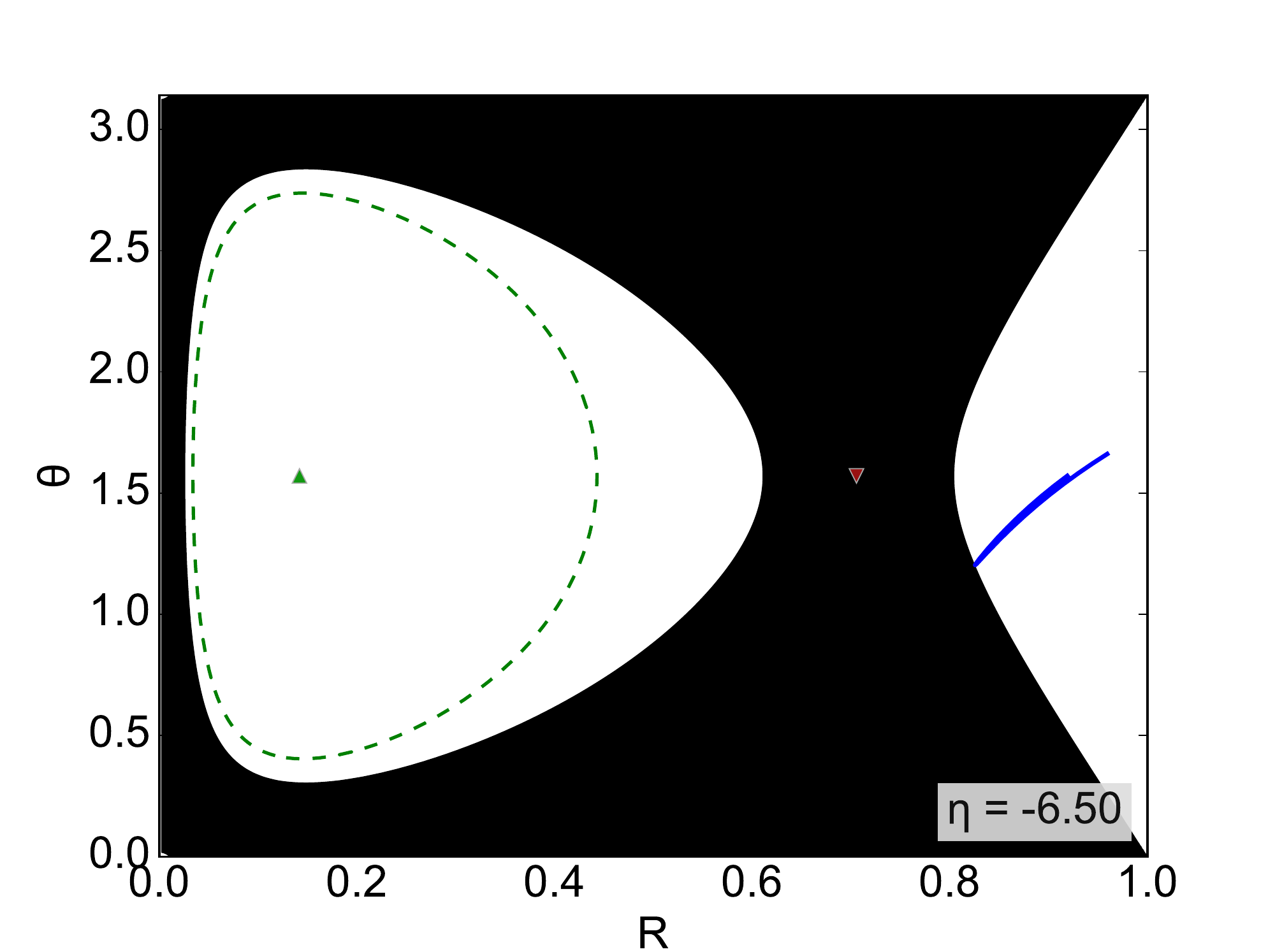}\
\includegraphics[trim={3.0cm 0.3cm 0.6cm 2.3cm}, clip, height=.16\textheight, angle =0]{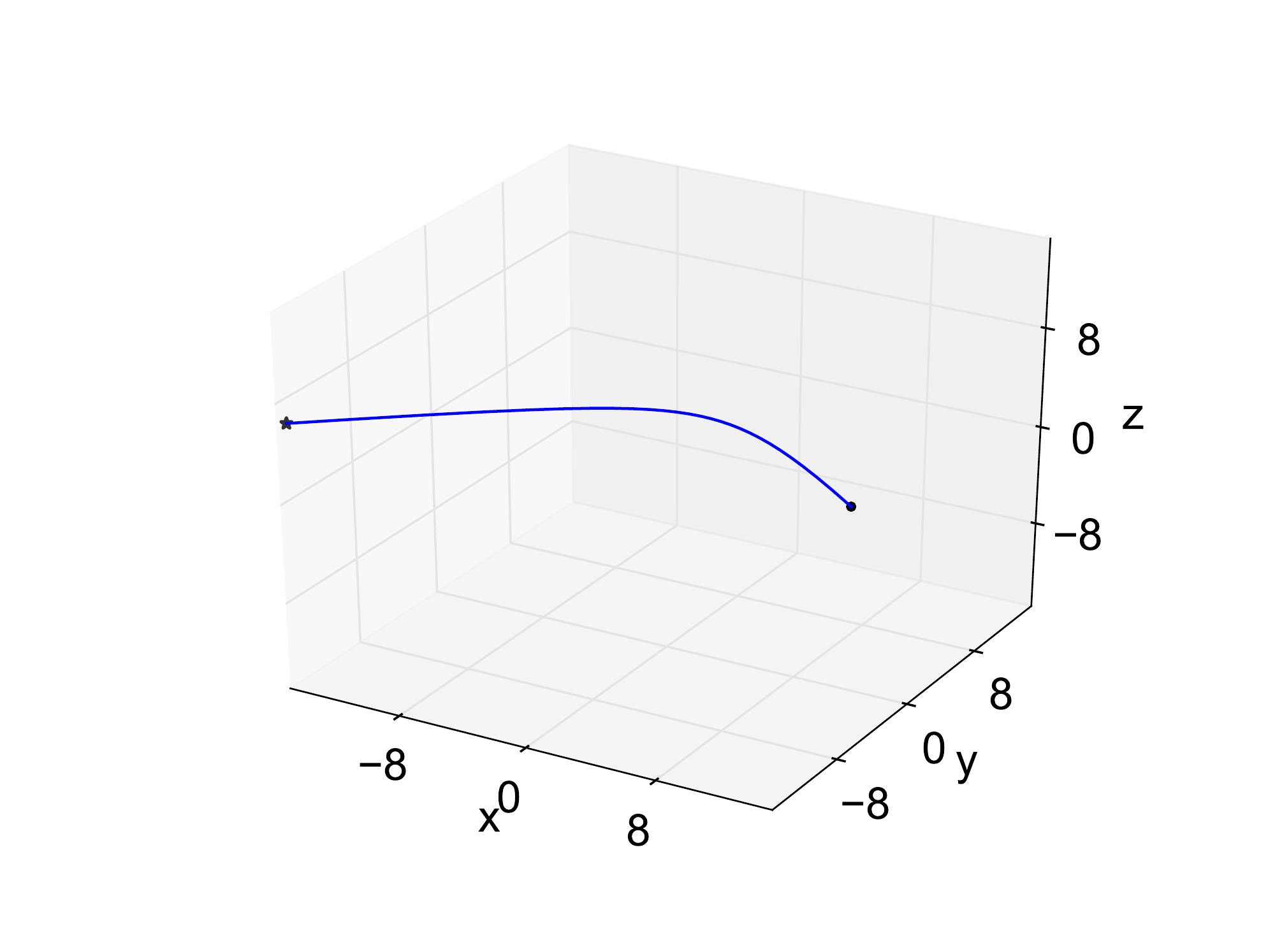}\
\put(0,50){${\bf 1}_{12}$}

\vspace{0.3cm}

\includegraphics[trim={0.2cm 0cm 1.5cm 1.6cm}, clip, height=.18\textheight, angle =0]{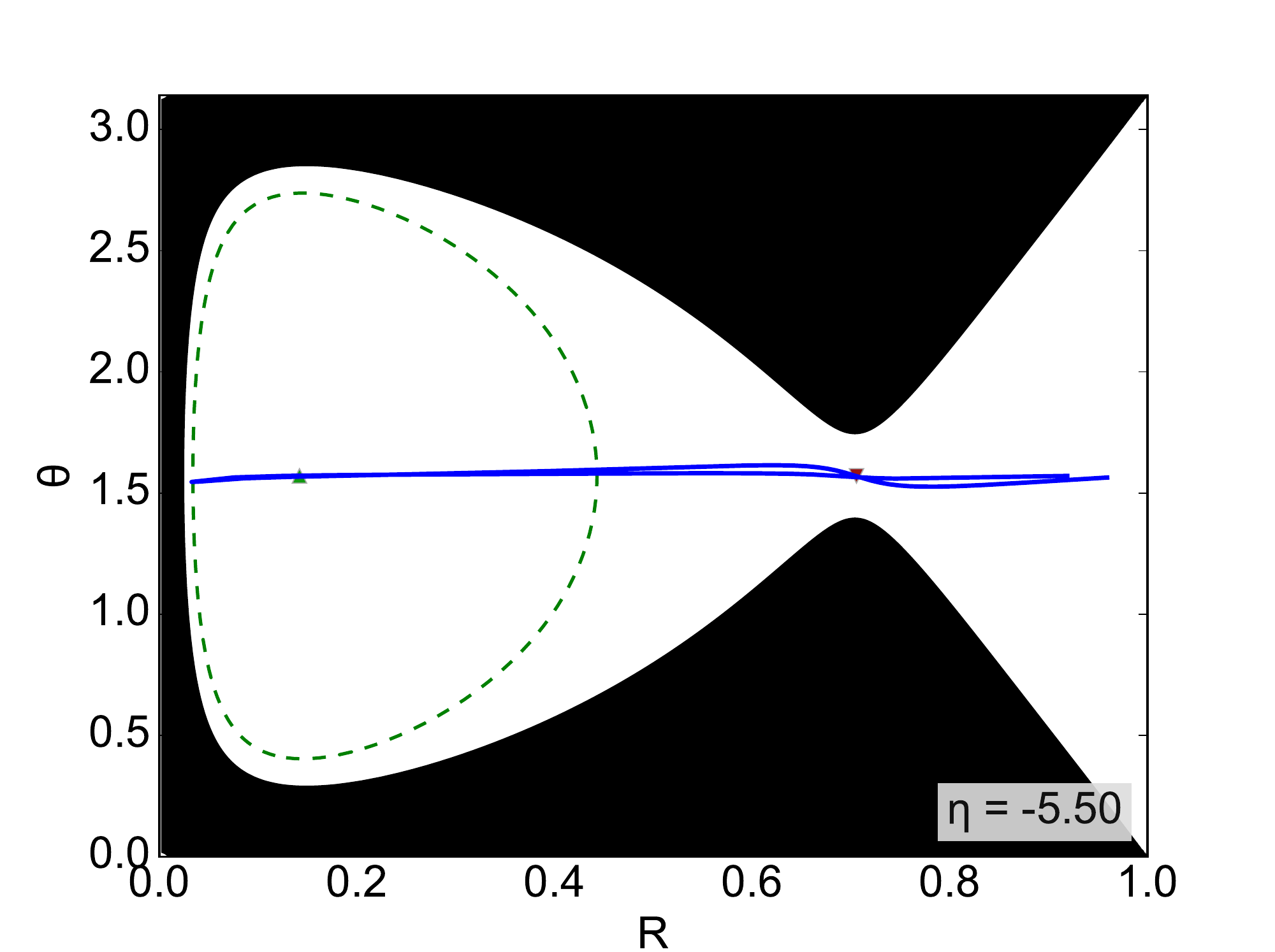}\
\includegraphics[trim={3.0cm 0.3cm 0.6cm 2.3cm}, clip, height=.16\textheight, angle =0]{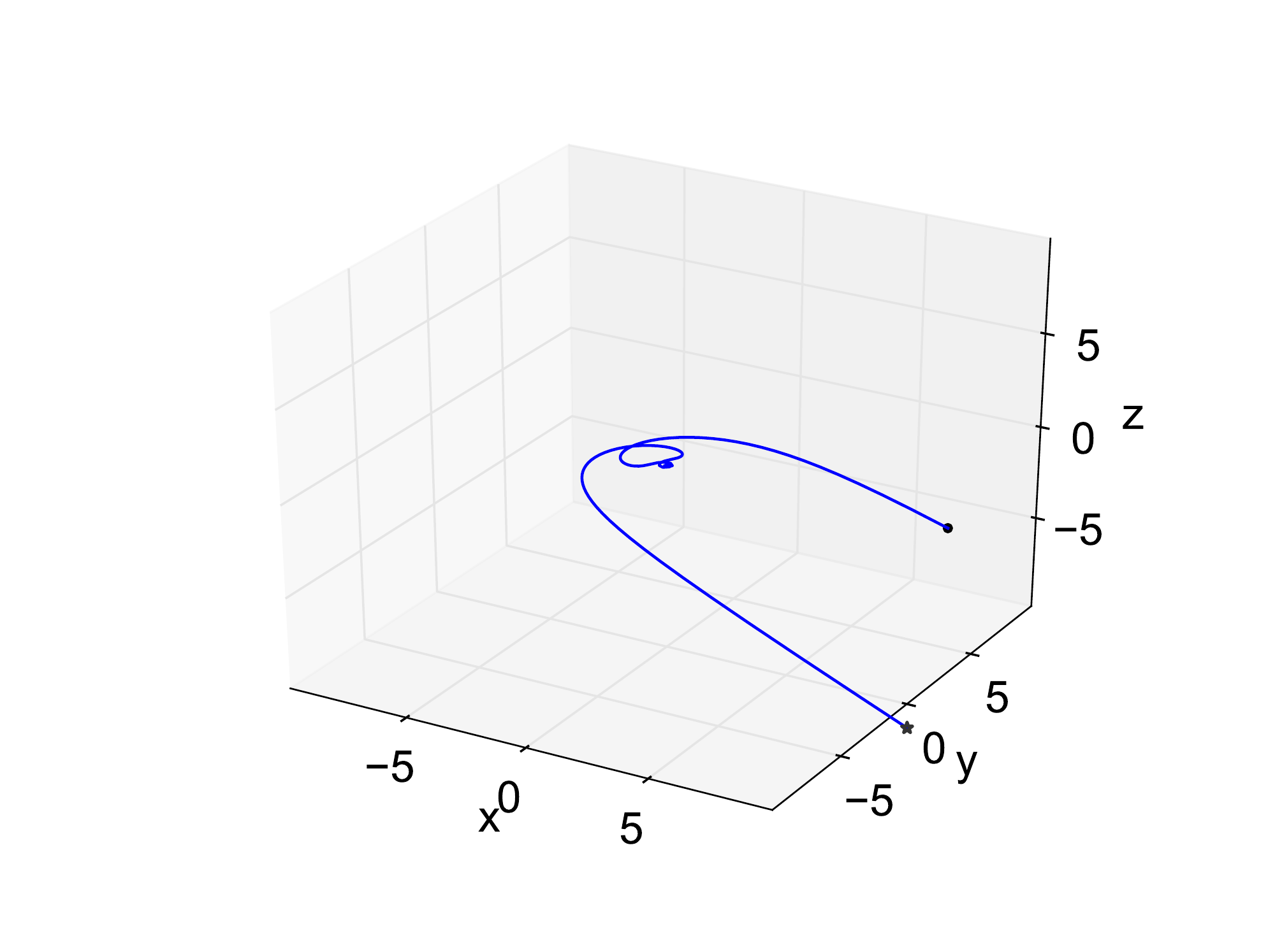}\
\put(0,50){${\bf 2}_{12}$}

\vspace{0.3cm}

\includegraphics[trim={0.2cm 0cm 1.5cm 1.6cm}, clip, height=.18\textheight, angle =0]{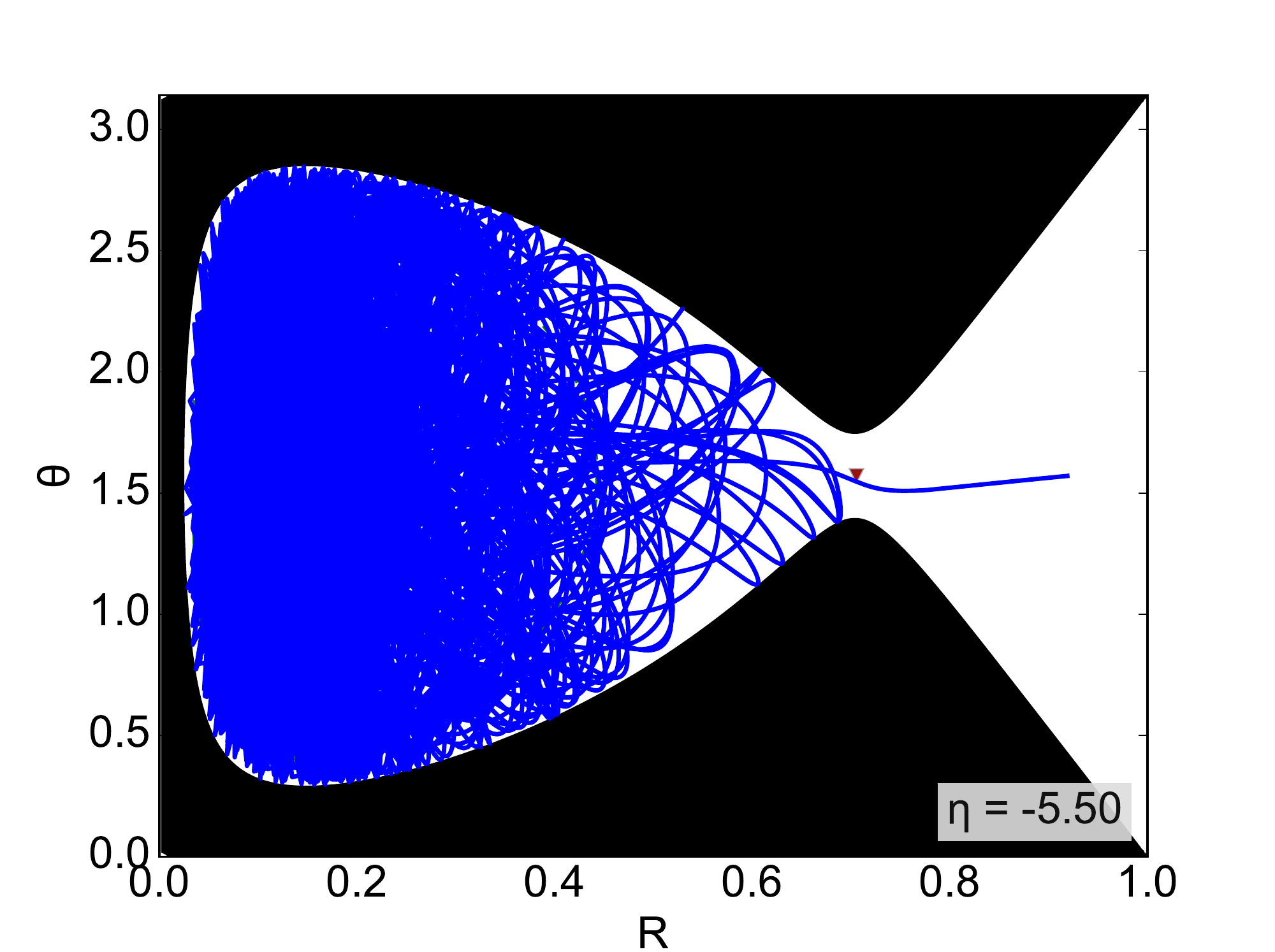}\
\includegraphics[trim={3.0cm 0.3cm 0.6cm 2.3cm}, clip, height=.16\textheight, angle =0]{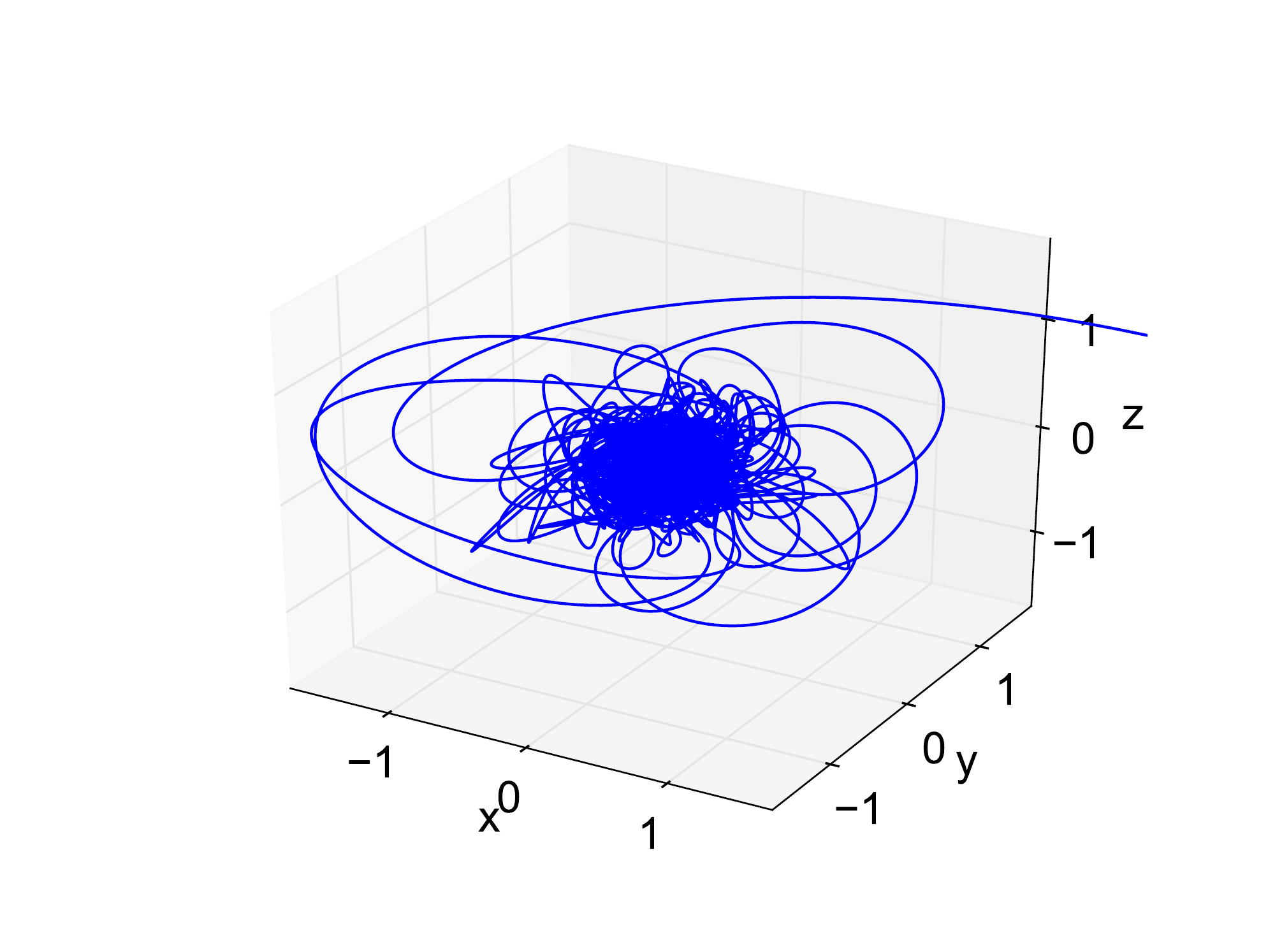}\
\put(0,50){${\bf 3}_{12}$}
\end{center}
\caption{(Top) Lensing of configuration 12 with three highlighted points and enlarged image of the selected points. Corresponding scattering orbits in  the effective potential (left) and spacetime (right).}
\label{gallery12}
\end{figure}

To close the gallery on RBSs, Fig.~\ref{gallery12} shows the lensing of configuration 12, seen from the equatorial plane, and  three highlighted points, denoted ${\bf 1}_{12}$,  ${\bf 2}_{12}$ and  ${\bf 3}_{12}$. The corresponding impact parameters and locations in $(X,Y)$ coordinates of the lensing image are:

\begin{center}
\begin{tabular}{lcc}
\hline
\hline
Point  & $\eta$ & $(X,Y)$   \\
\hline
\hline
${\bf 1}_{12}$   & -6.50 & $(0.900, 0.326)$  \\
 \hline
${\bf 2}_{12}$ & -5.50 & $ (0.737, 0.008)$  \\
 \hline
${\bf 3}_{12}$  & -5.50 & $(0.737, 0.044)$  \\
 \hline
\end{tabular}
\end{center}

The RBS 12 was not analysed in detail before, since the respective effective potential displays essentially the same features as the RBS 11. Nonetheless, the RBS configuration 12 exhibits one of the richest dynamical structures of the configurations presented here. In particular, large areas of the central region of the lensing image exhibit chaotic behaviour. The characteristics of points ${\bf 1}_{12}$ and ${\bf 2}_{12}$ are very similar to their counterparts in configurations 10 and 11. However, just a small perturbation of ${\bf 2}_{12}$ leads to point ${\bf 3}_{12}$. It is chosen such that its impact parameter allows the photon to enter a pocket with a very small opening. At the same time, it has sufficient $\theta$ momentum for it to get trapped in the pocket for a very long time. Its orbit fills out the pocket with an almost dense covering, as well as the central spacetime region, respectively. Given sufficiently long integration time, these types of orbits tend to escape eventually.

\begin{figure}[!htp]
\begin{center}
\includegraphics[trim={2.5cm 0cm 1.5cm 1.2cm}, clip, height=.28\textheight, angle =0]{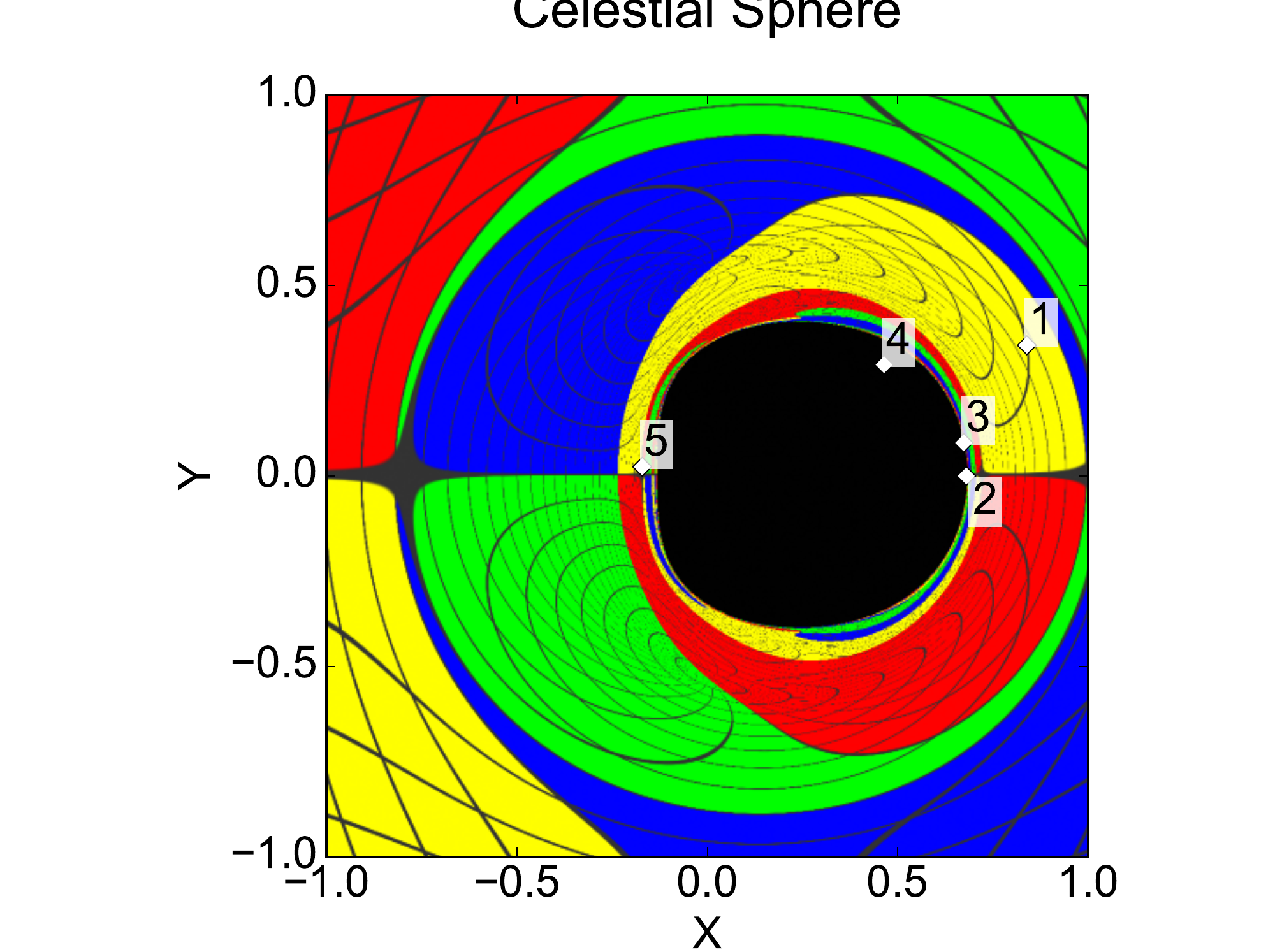}

\vspace{0.6cm}

\includegraphics[trim={0.2cm 0cm 1.5cm 1.6cm}, clip, height=.18\textheight, angle =0]{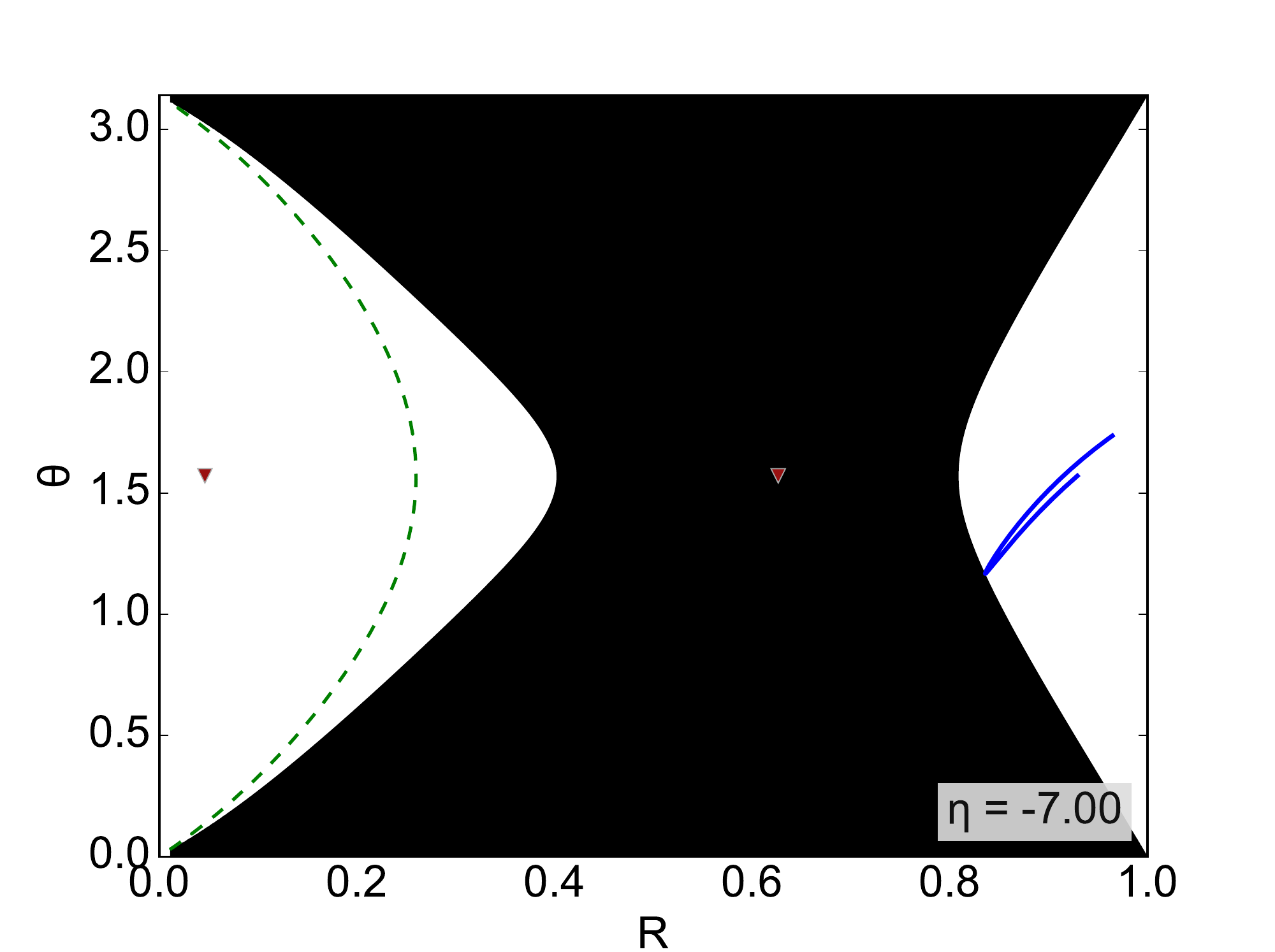}\
\includegraphics[trim={3.0cm 0.3cm 0.6cm 2.3cm}, clip, height=.16\textheight, angle =0]{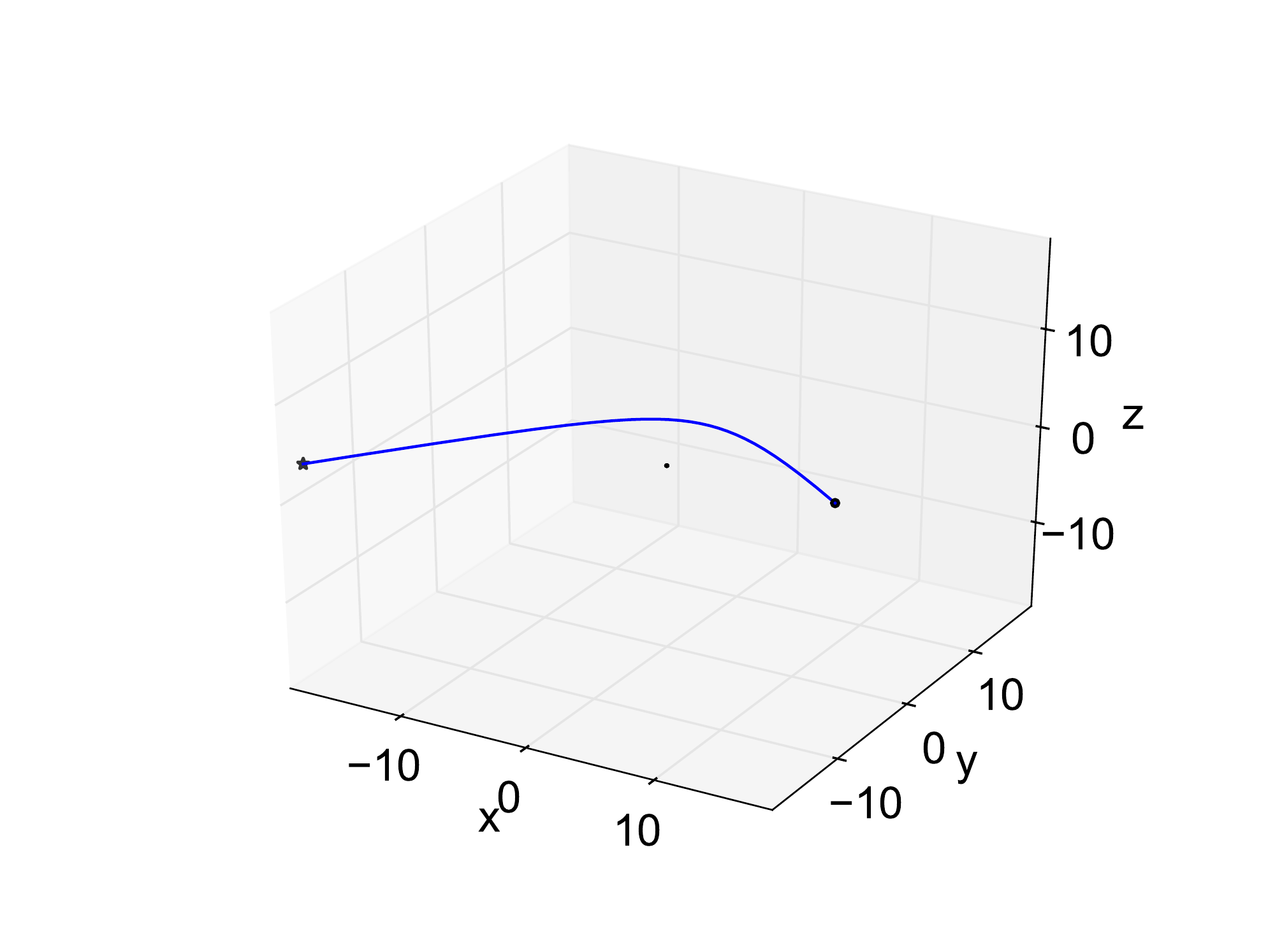}\
\put(0,50){${\bf 1}_{\rm II}$}

\vspace{0.3cm}

\includegraphics[trim={0.2cm 0cm 1.5cm 1.6cm}, clip, height=.18\textheight, angle =0]{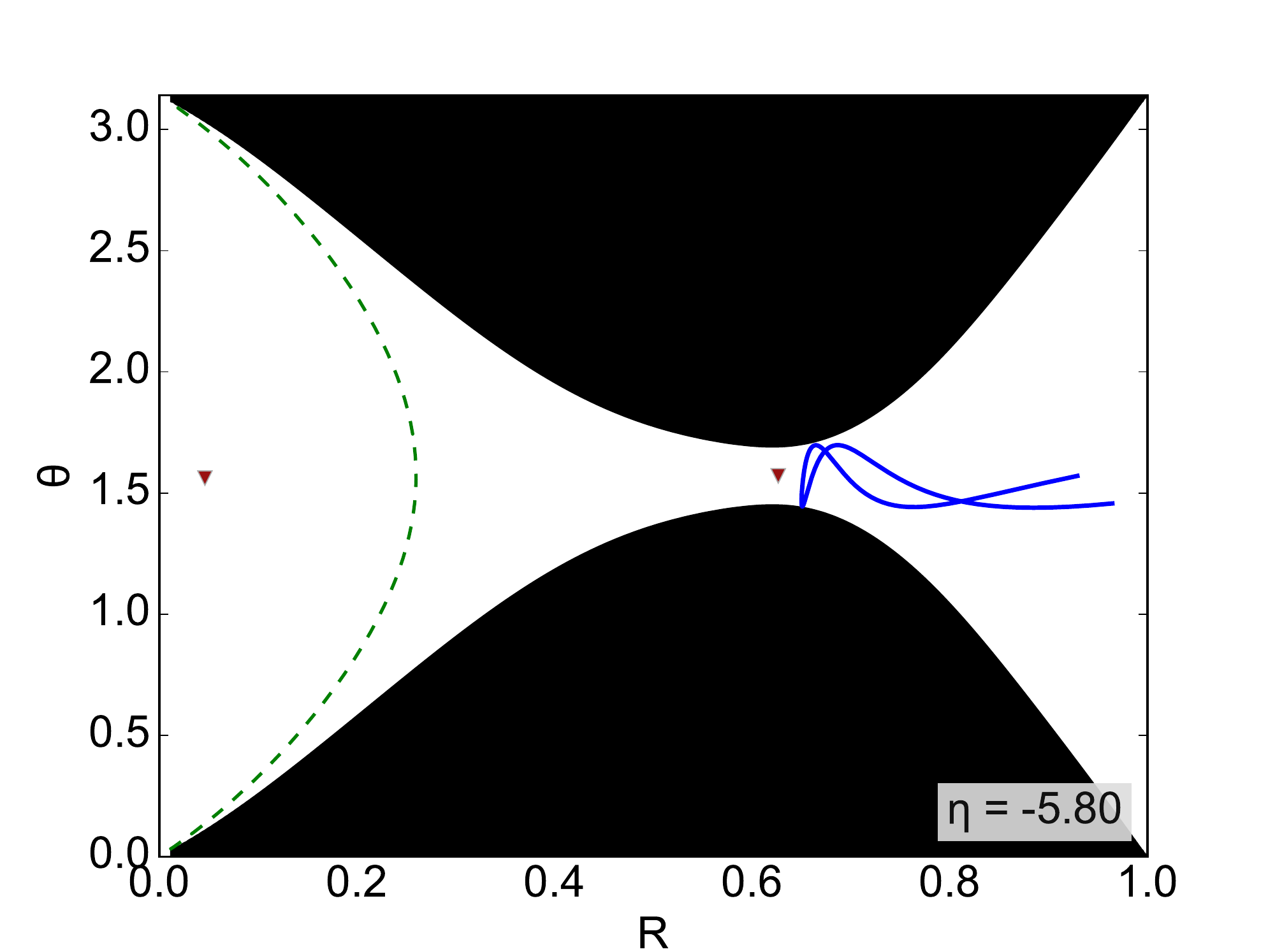}\
\includegraphics[trim={3.0cm 0.3cm 0.6cm 2.3cm}, clip, height=.16\textheight, angle =0]{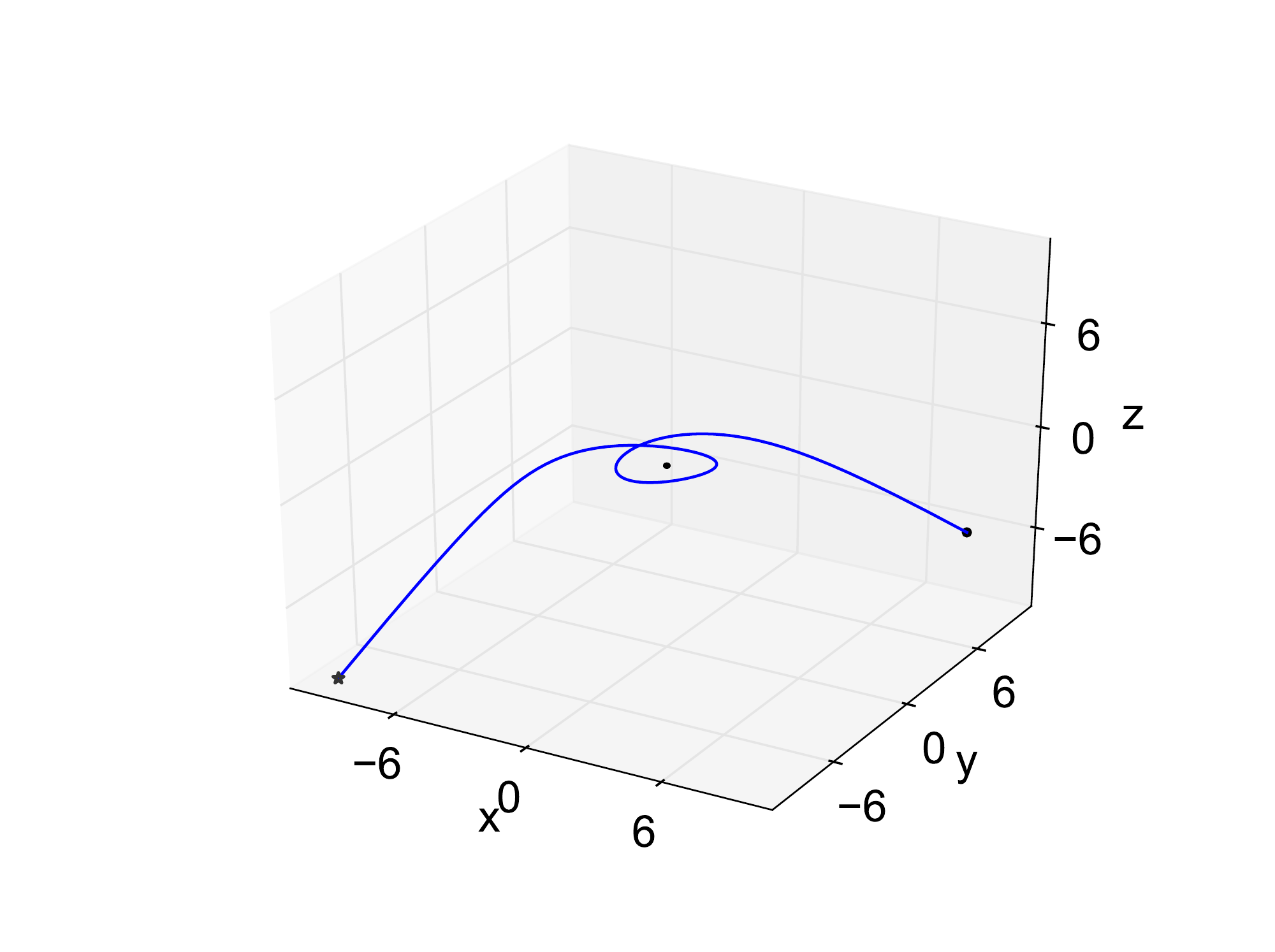}\
\put(0,50){${\bf 3}_{\rm II}$}

\vspace{0.3cm}

\includegraphics[trim={0.2cm 0cm 1.5cm 1.6cm}, clip, height=.18\textheight, angle =0]{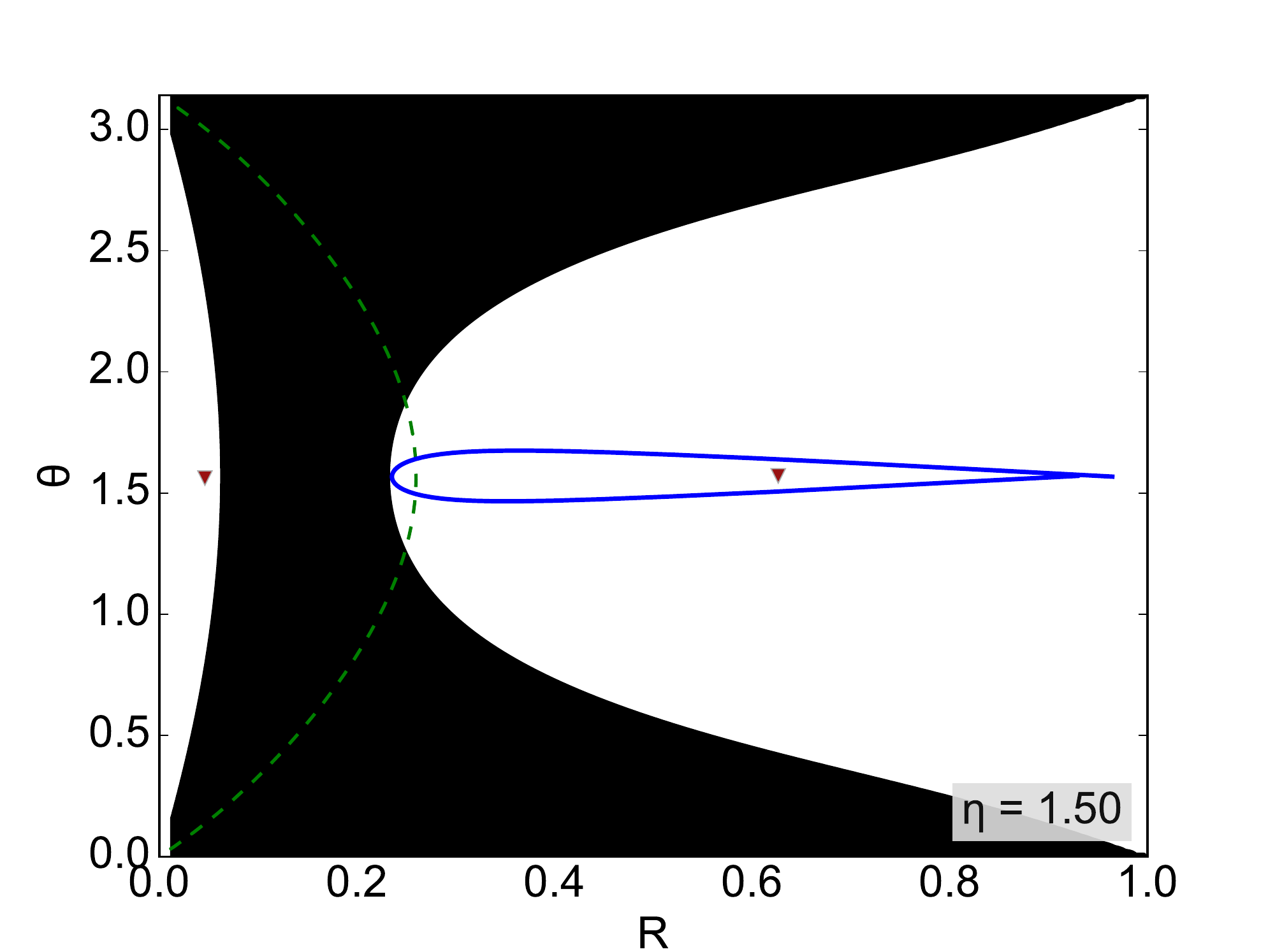}\
\includegraphics[trim={3.0cm 0.3cm 0.6cm 2.3cm}, clip, height=.16\textheight, angle =0]{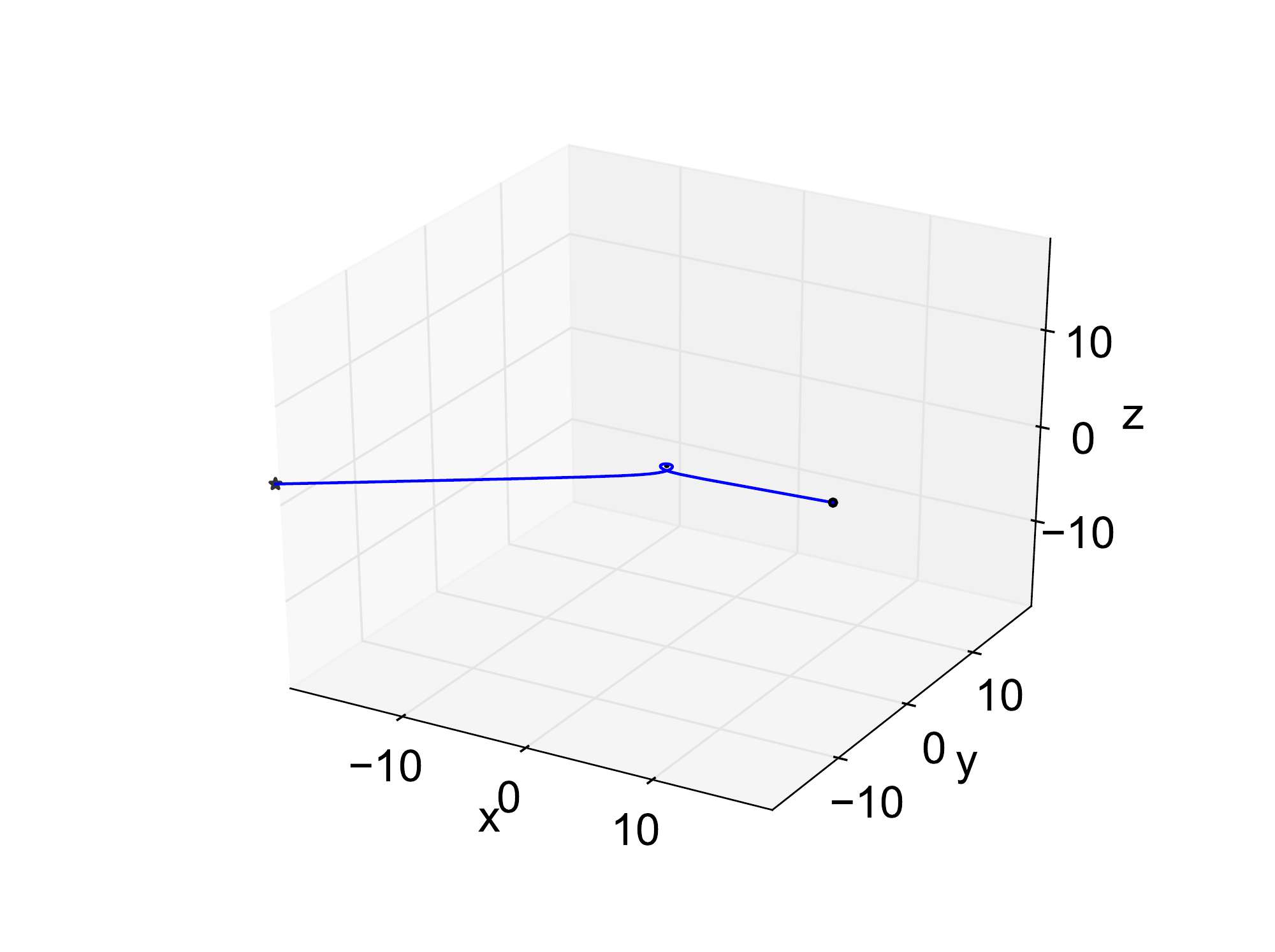}\
\put(0,50){${\bf 5}_{\rm II}$}
\end{center}
\caption{(Top) Lensing of configuration II with five highlighted points. Corresponding scattering orbits in  the effective potential (left) and  spacetime (right).  Fig.~\ref{galleryIIb} shows the absorption states.}
\label{galleryIIa}
\end{figure}

\subsection{Kerr BHs with scalar hair}

We now turn our attention to KBHsSH, in particular configurations II and III. Similarly to the RBS 12, the KBHSH II was not discussed before since its effective potential shares the same qualitative features as the Kerr case ($cf.$ Appendix~\ref{appendixa}). Fig.~\ref{galleryIIa} shows the lensing of configuration II, as before seen from the equatorial plane. We have selected five points in the lensing image, denoted ${\bf 1}_{\rm II}$ to  ${\bf 5}_{\rm II}$. The corresponding impact parameters, and the location in $(X,Y)$ image coordinates are:

\begin{center}
\begin{tabular}{lcc}
\hline
\hline
Point  & $\eta$ & $(X,Y)$   \\
\hline
\hline
${\bf 1}_{\rm II}$   & -7.00 & $(0.839, 0.343)$  \\
 \hline
${\bf 2}_{\rm II}$ & -5.87 & $ (0.680, 0.000)$  \\
 \hline
${\bf 3}_{\rm II}$  & -5.80 & $(0.673, 0.087)$  \\
 \hline
 ${\bf 4}_{\rm II}$  & -4.00 & $(0.464, 0.292)$  \\
 \hline
  ${\bf 5}_{\rm II}$  & +1.50 & $(-0.171, 0.024)$  \\
 \hline
\end{tabular}
\end{center}

\begin{figure}[!htp]
\begin{center}
\includegraphics[trim={0.2cm 0cm 1.5cm 1.6cm}, clip, height=.18\textheight, angle =0]{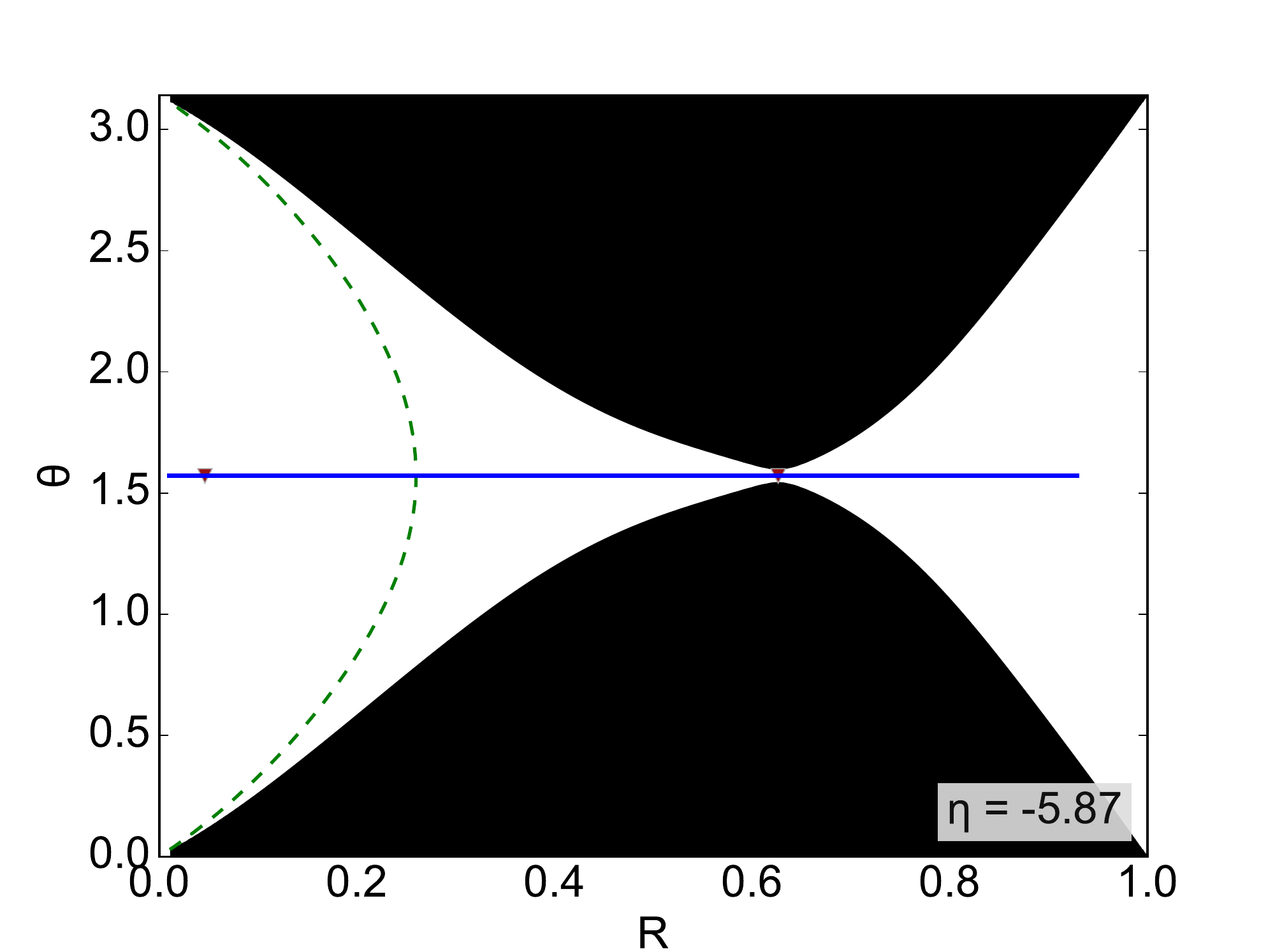}\
\includegraphics[trim={3.0cm 0.3cm 0.6cm 2.3cm}, clip, height=.16\textheight, angle =0]{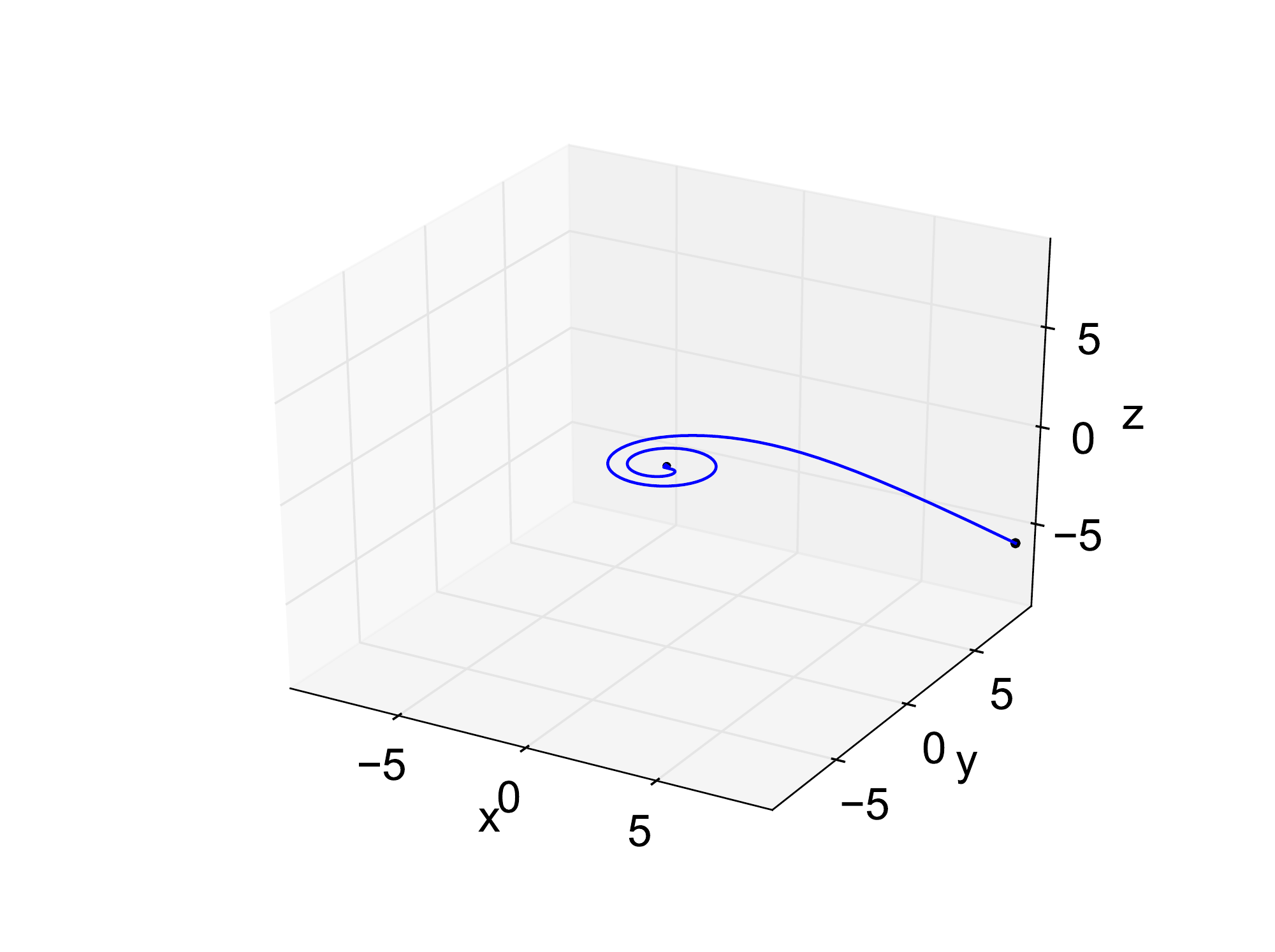}\
\put(0,50){${\bf 2}_{\rm II}$}

\vspace{0.3cm}

\includegraphics[trim={0.2cm 0cm 1.5cm 1.6cm}, clip, height=.18\textheight, angle =0]{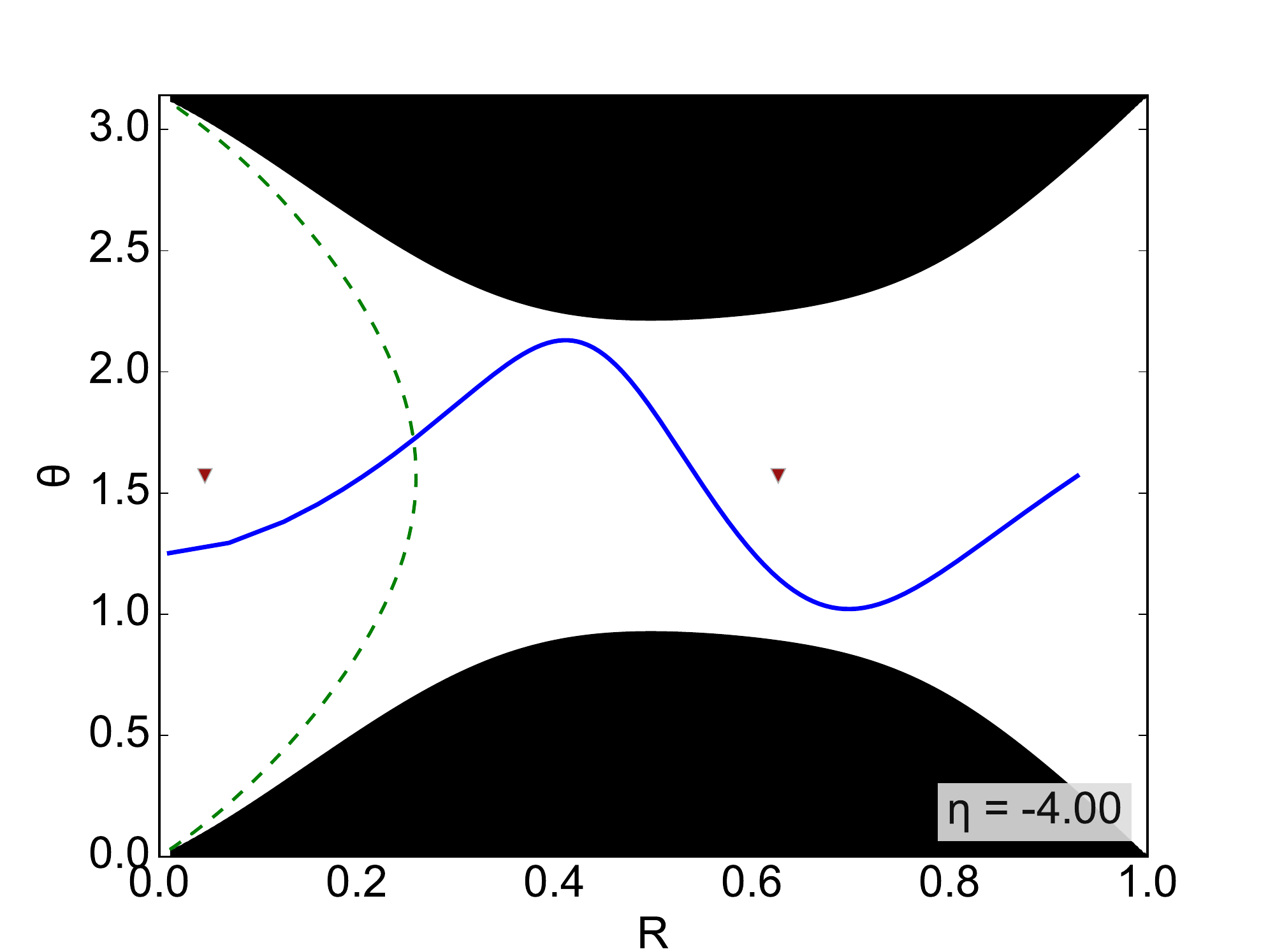}\
\includegraphics[trim={3.0cm 0.3cm 0.6cm 2.3cm}, clip, height=.16\textheight, angle =0]{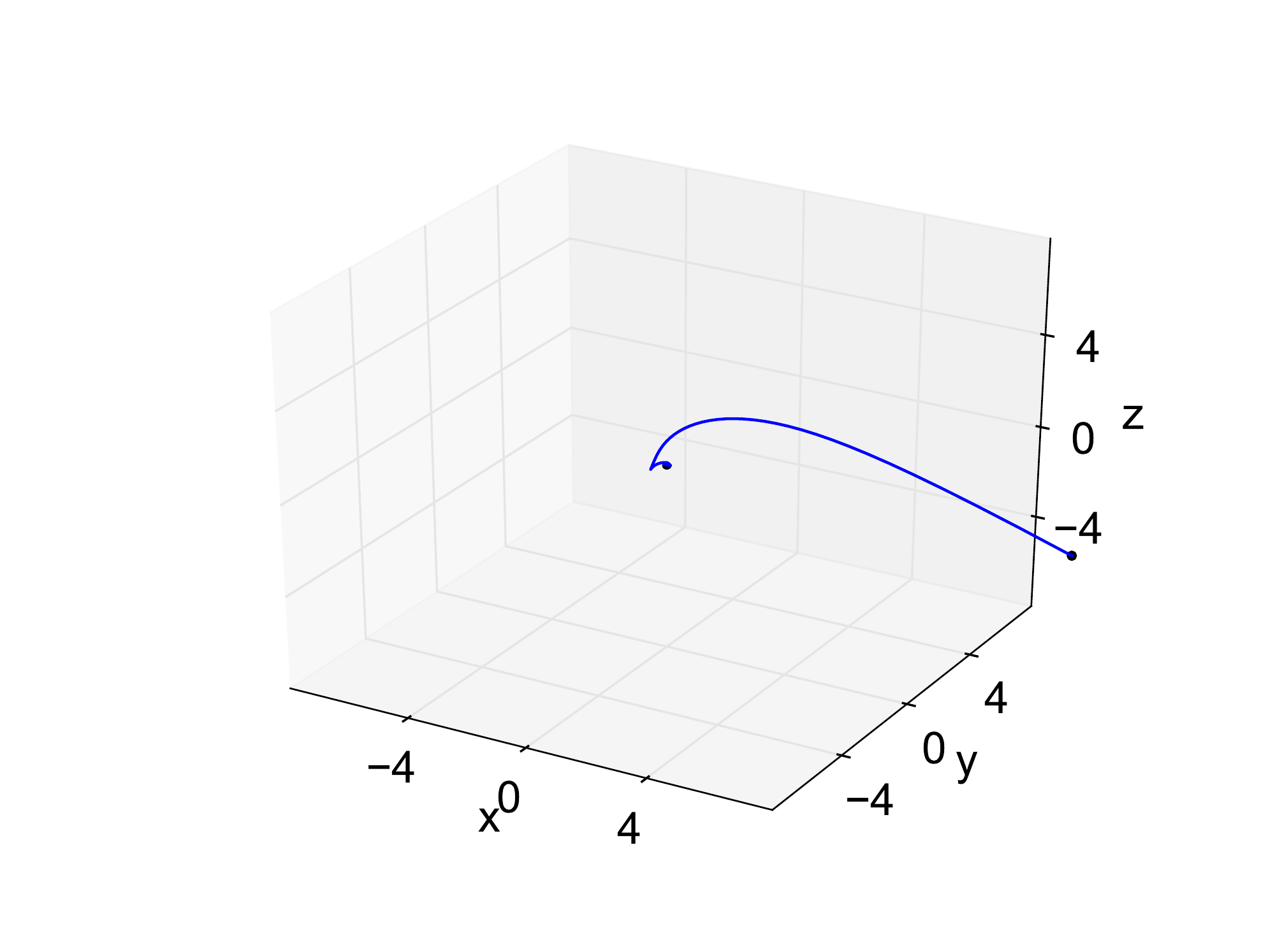}\
\put(0,50){${\bf 4}_{\rm II}$}
\end{center}
\caption{Absorption orbits in the effective potential (left) and  spacetime (right), corresponding to points  ${\bf 2}_{\rm II}$ and ${\bf 4}_{\rm II}$ in the lensing image of Fig.~\ref{galleryIIa}.}
\label{galleryIIb}
\end{figure}

In Fig.~\ref{galleryIIa} we show the effective potential and spacetime orbit for points ${\bf 1}_{\rm II}$, ${\bf 3}_{\rm II}$ and  ${\bf 5}_{\rm II}$, all of which are scattering states. Instead of an isolated pocket, the effective potential now has an inner allowed region connected to the BH horizon. Point ${\bf 1}_{\rm II}$ corresponds to a state for which this inner region is not accessible from infinity. The same holds for point  ${\bf 5}_{\rm II}$, which has an impact parameter with the opposite sign, and hence is located on the left side of the shadow. For point ${\bf 3}_{\rm II}$, the exterior and interior allowed regions are connected, but the orbit does not fall into the BH; it bounces off at the ``throat" of the potential and then escapes to infinity. In the lensing image this orbit corresponds to a region close to the shadow's edge, where smaller and smaller copies of the celestial sphere accumulate in an orderly fashion. In spacetime this orbit circles once around the BH before being scattered off to infinity. This circling occurs in the neighbourhood of the unstable light ring. In Fig.~\ref{galleryIIb} we instead show two orbits that are absorbed by the BH, corresponding to points ${\bf 2}_{\rm II}$ and ${\bf 4}_{\rm II}$ in the lensing image of Fig.~\ref{galleryIIa}. Point ${\bf 2}_{\rm II}$ lies just barely inside the shadow along the equatorial plane. The potential is just open for this impact parameter, allowing the photon to pass through to the inner region and fall into the shadow. Point ${\bf 4}_{\rm II}$ on the other hand lies well within the shadow and moves within a wide open effective potential.

\begin{figure}[!htp]
\begin{center}
\includegraphics[trim={2.5cm 0cm 1.5cm 1.2cm}, clip, height=.28\textheight, angle =0]{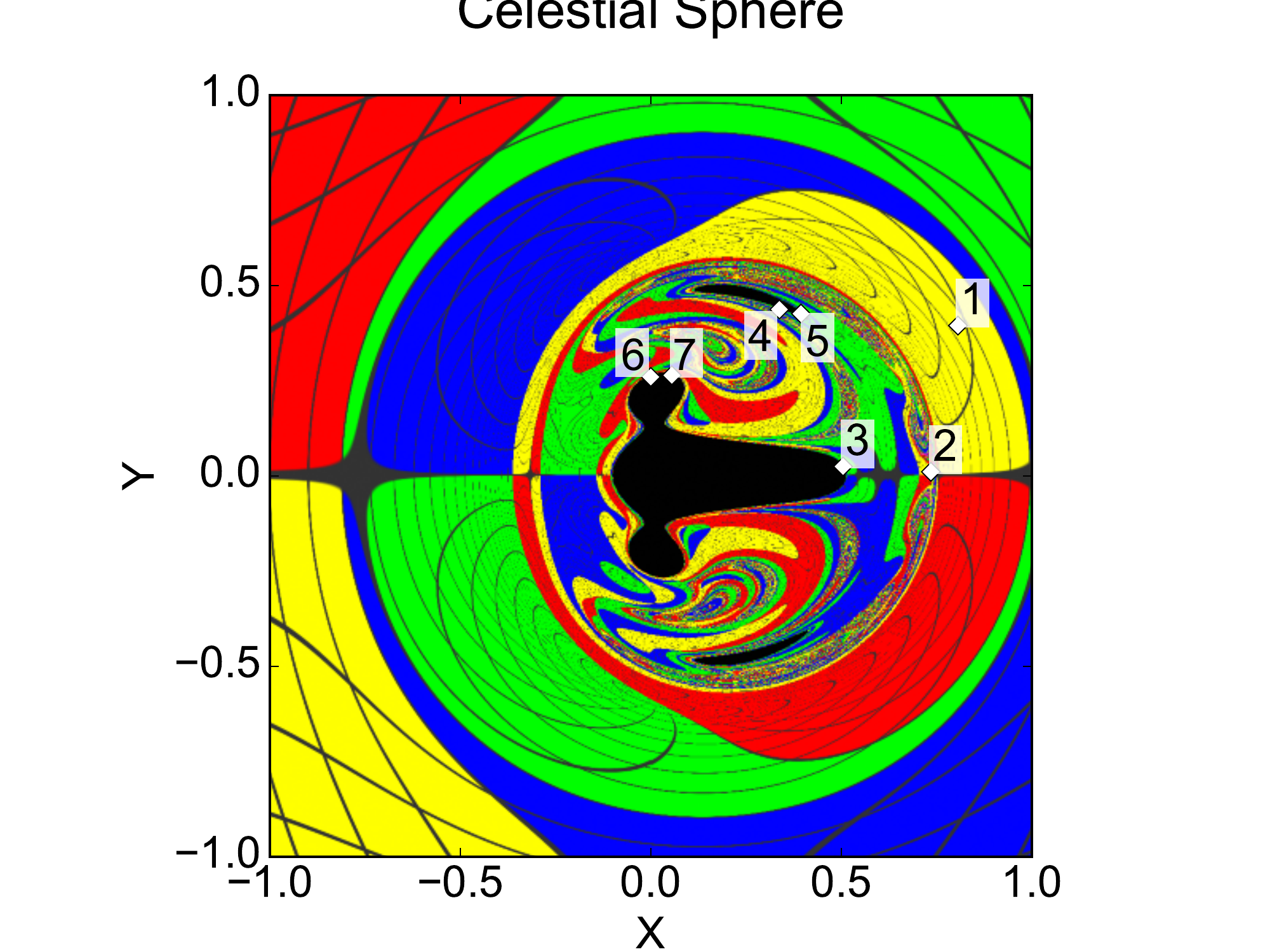}

\vspace{0.6cm}

\includegraphics[trim={0.2cm 0cm 1.5cm 1.6cm}, clip, height=.18\textheight, angle =0]{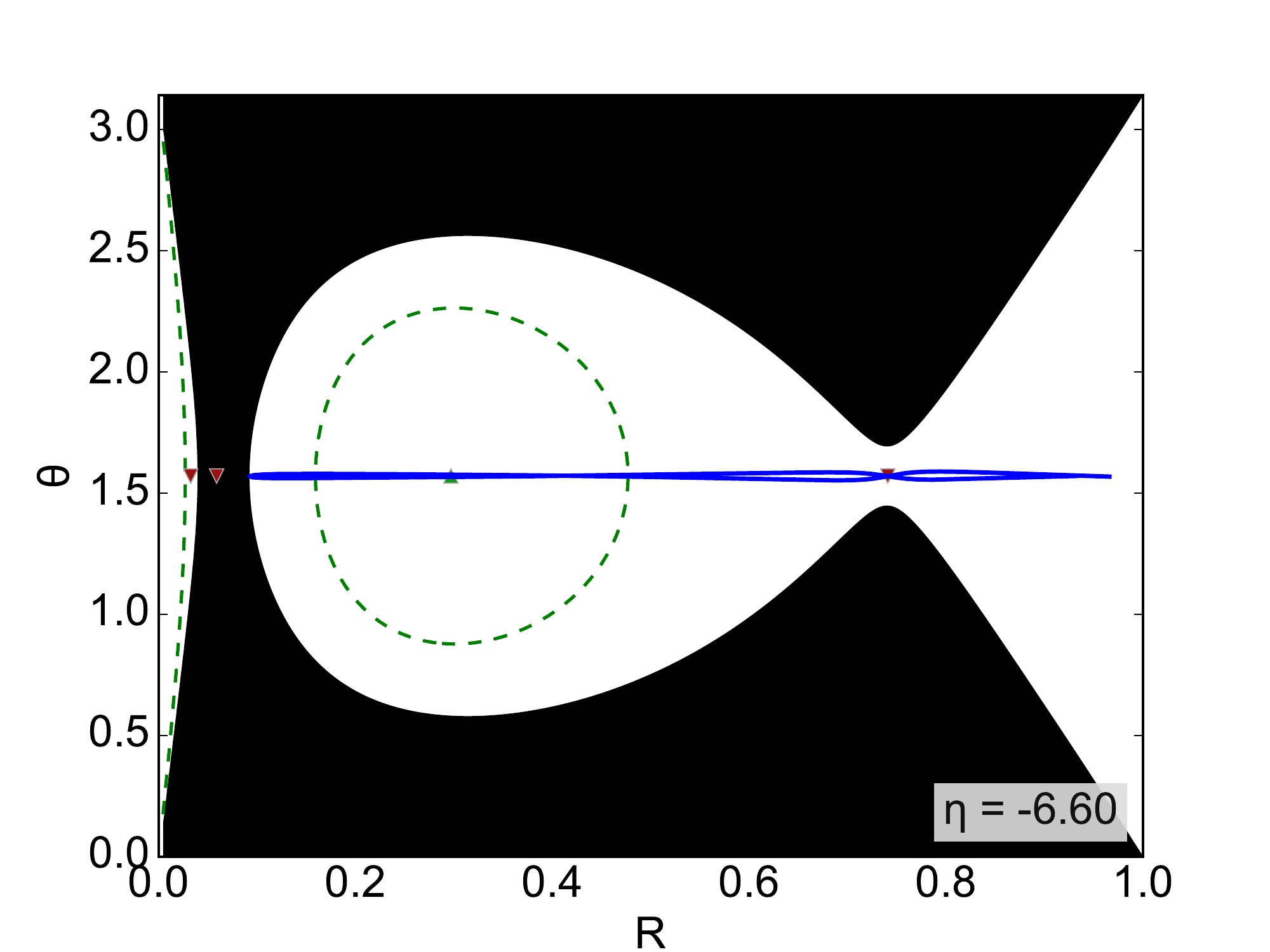}\
\includegraphics[trim={3.0cm 0.3cm 0.6cm 2.3cm}, clip, height=.16\textheight, angle =0]{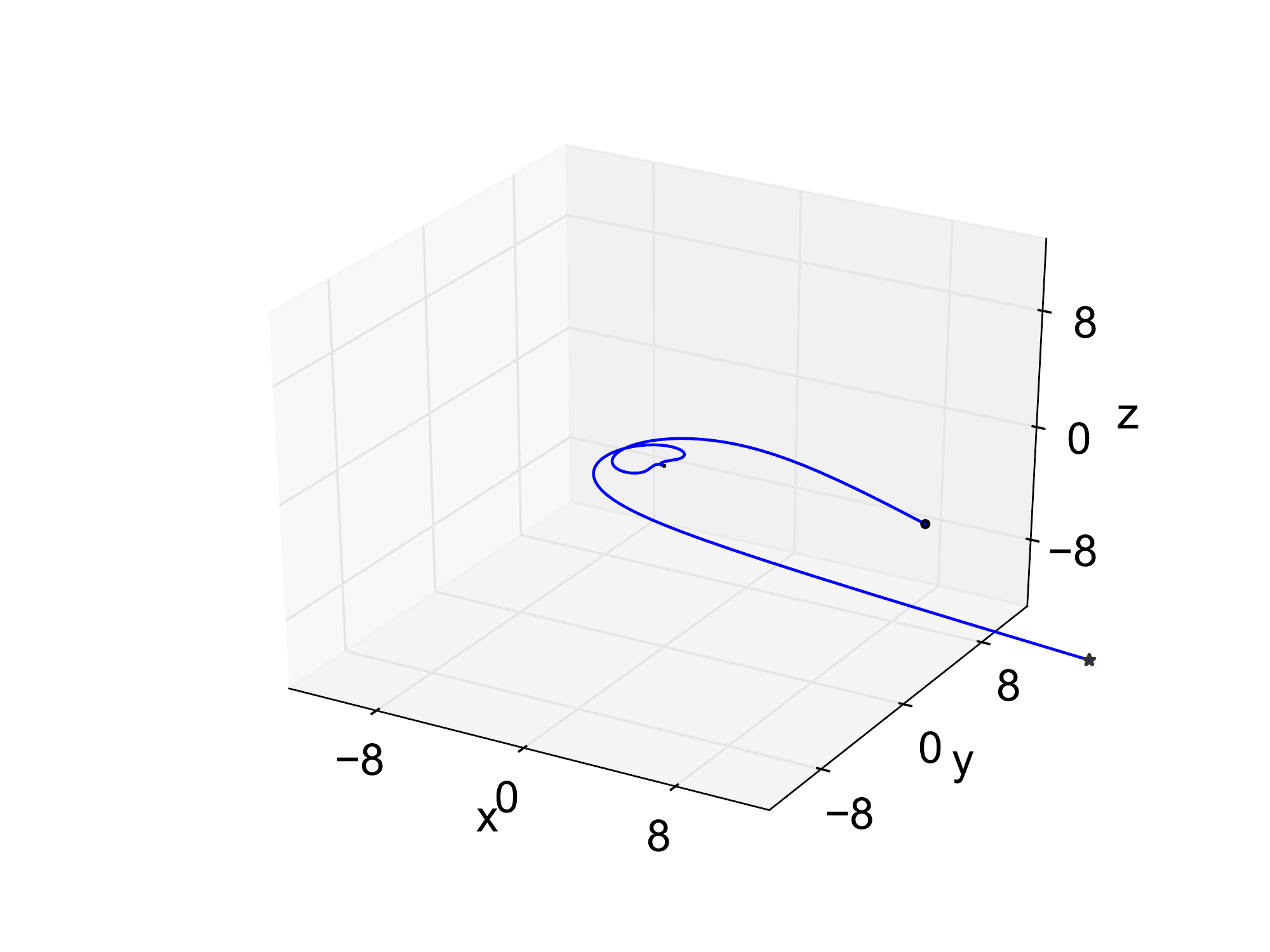}\
\put(0,50){${\bf 2}_{\rm III}$}

\vspace{0.3cm}

\includegraphics[trim={0.2cm 0cm 1.5cm 1.6cm}, clip, height=.18\textheight, angle =0]{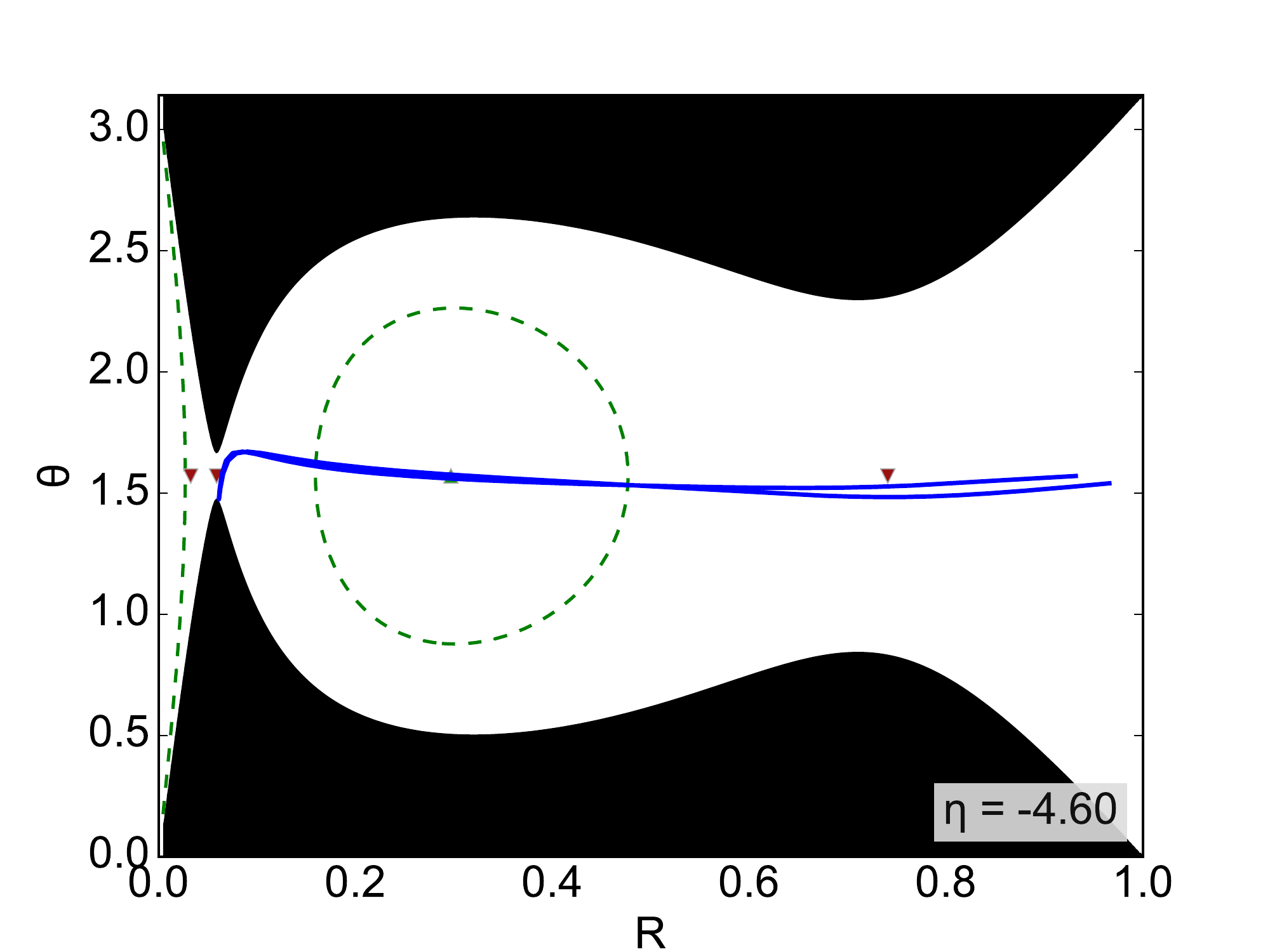}\
\includegraphics[trim={3.0cm 0.3cm 0.6cm 2.3cm}, clip, height=.16\textheight, angle =0]{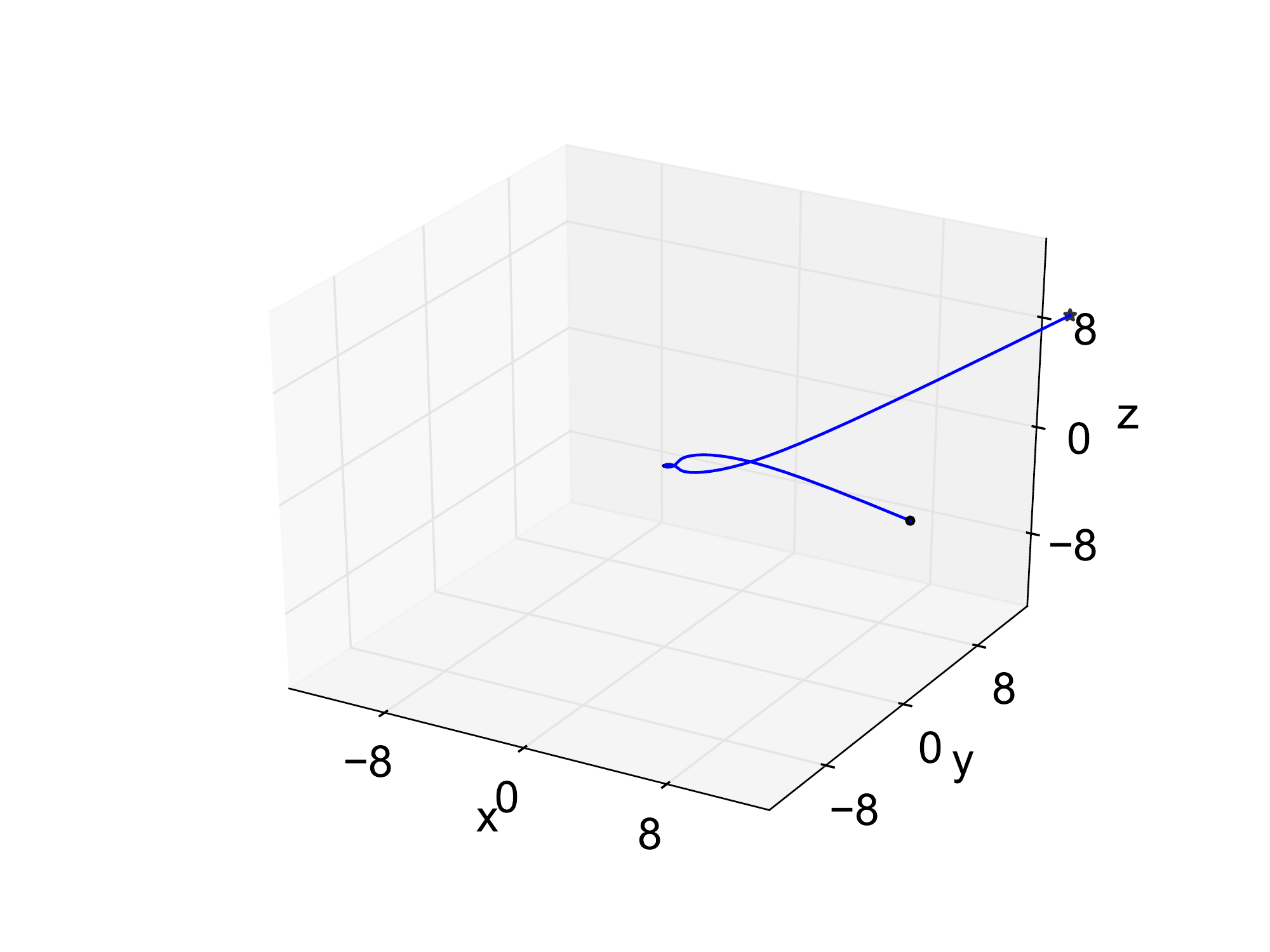}\
\put(0,50){${\bf 3}_{\rm III}$}

\vspace{0.3cm}

\includegraphics[trim={0.2cm 0cm 1.5cm 1.6cm}, clip, height=.18\textheight, angle =0]{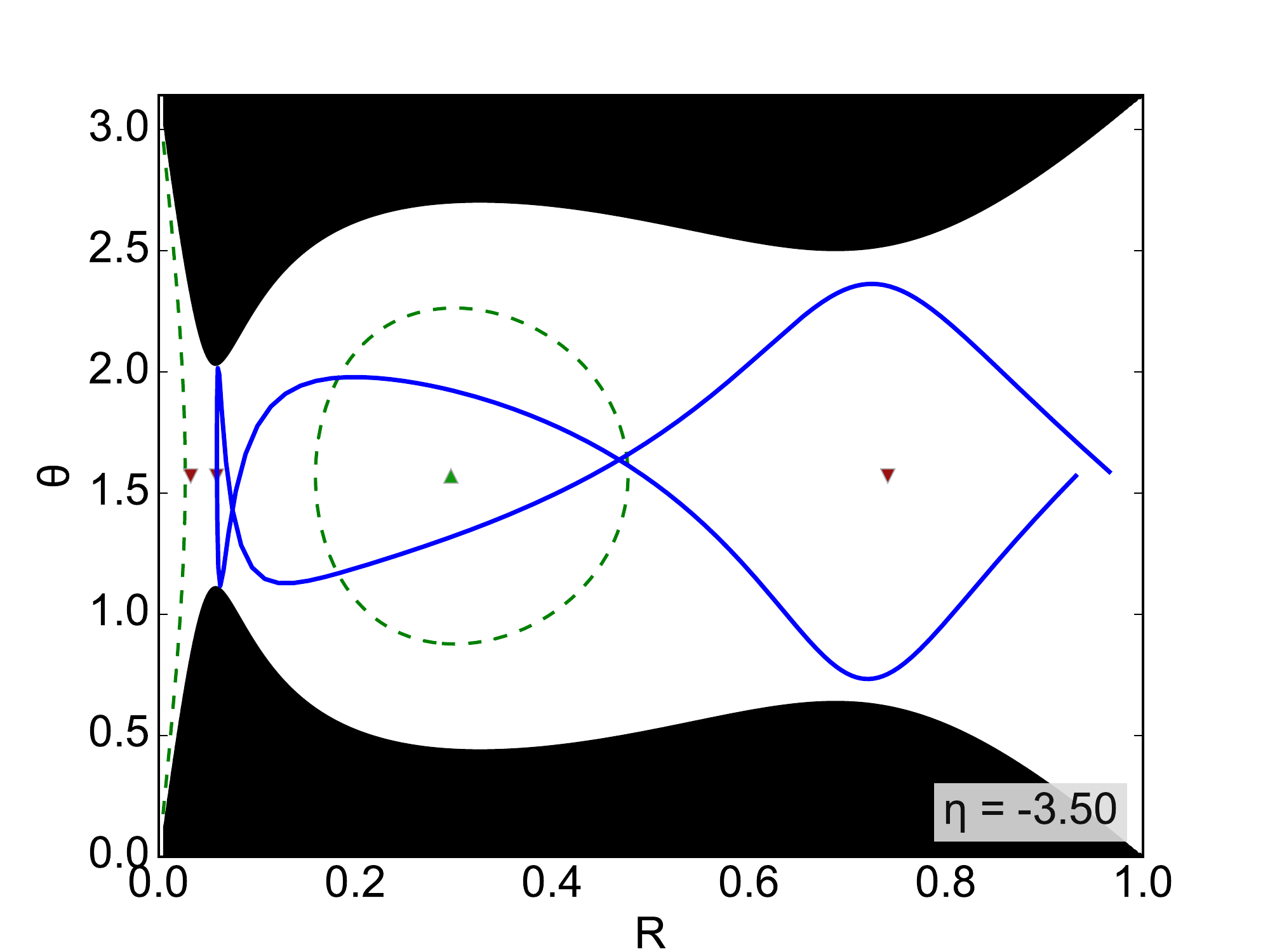}\
\includegraphics[trim={3.0cm 0.3cm 0.6cm 2.3cm}, clip, height=.16\textheight, angle =0]{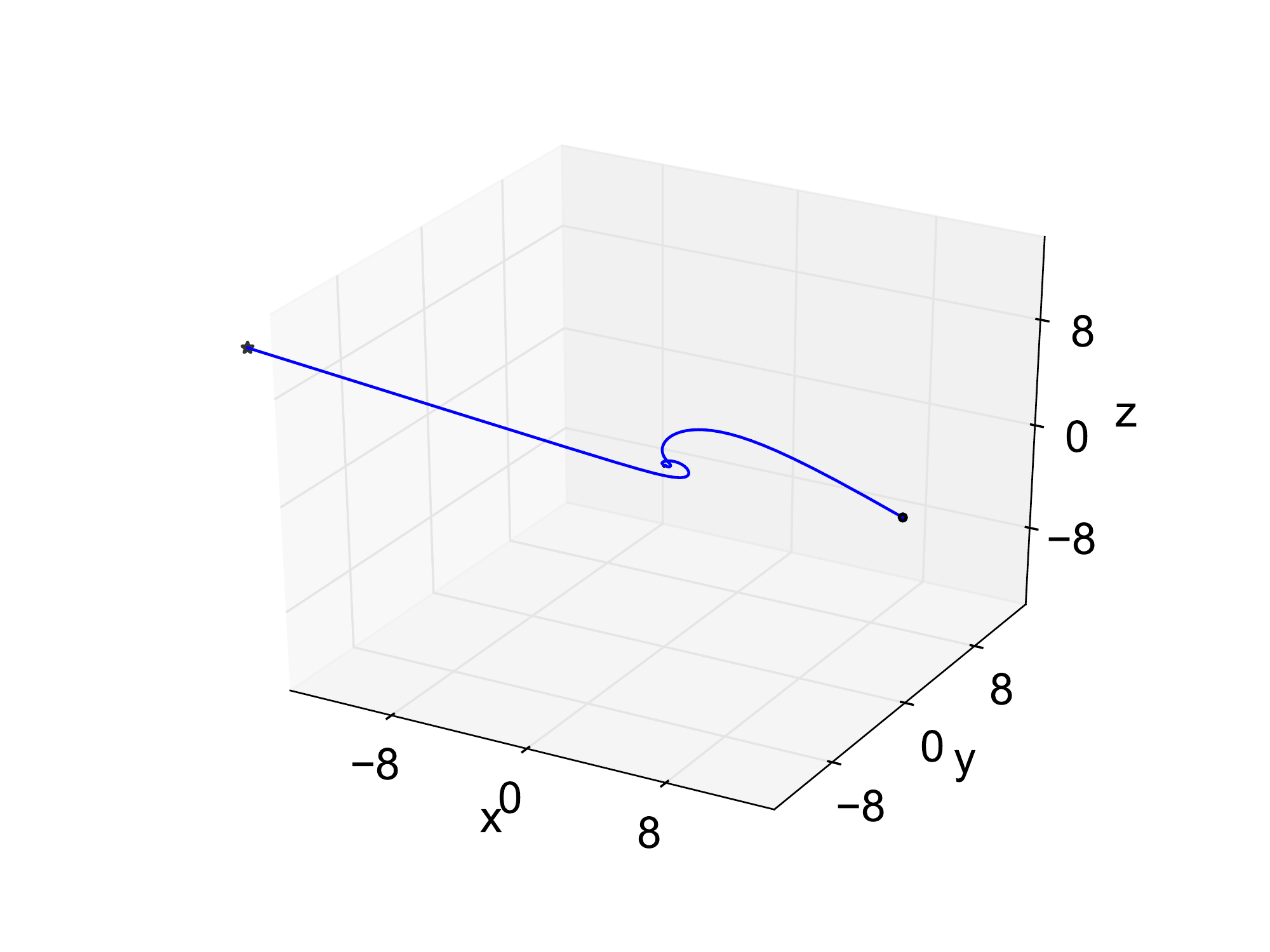}\  
\put(0,50){${\bf 5}_{\rm III}$}
\end{center}
\caption{(Top) Lensing of configuration III with seven highlighted points. Orbits for points ${\bf 2}_{\rm III}$, ${\bf 3}_{\rm III}$ and ${\bf 5}_{\rm III}$  in  the effective potential (left) and spacetime (right).}
\label{galleryIIIa}
\end{figure}

Finally, we consider the richest of our backgrounds, configuration III. In Fig.~\ref{galleryIIIa}, we exhibit the lensing of this configuration, seen from the equatorial plane. We selected seven points in the lensing image, denoted ${\bf 1}_{\rm III}$ to ${\bf 7}_{\rm III}$. The corresponding impact parameters and their locations in $(X,Y)$ image coordinates are:

\begin{center}
\begin{tabular}{lcc}
\hline
\hline
Point  & $\eta$ & $(X,Y)$   \\
\hline
\hline
${\bf 1}_{\rm III}$   & -7.00 & $(0.806, 0.395)$  \\
 \hline
${\bf 2}_{\rm III}$ & -6.60 & $ (0.735, 0.011)$  \\
 \hline
${\bf 3}_{\rm III}$  & -4.60 & $(0.504, 0.025)$  \\
 \hline
  ${\bf 4}_{\rm III}$  & -3.00 & $(0.337, 0.437)$  \\
 \hline
   ${\bf 5}_{\rm III}$  & -3.50 & $(0.394, 0.426)$  \\
 \hline
 ${\bf 6}_{\rm III}$  & 0.00 & $(0.000, 0.260)$  \\
 \hline
   ${\bf 7}_{\rm III}$  & -0.50 & $(0.055, 0.265)$  \\
 \hline
\end{tabular}
\end{center}

In Fig.~\ref{galleryIIIa} we show the effective potential and spacetime orbit for points ${\bf 2}_{\rm III}$, ${\bf 3}_{\rm III}$ and  ${\bf 5}_{\rm III}$, all of which are scattering states. The orbit of point ${\bf 1}_{\rm III}$ is similar to that of ${\bf 1}_{\rm II}$ (Fig.~\ref{galleryIIa}), except that there are three (rather than two) disconnected regions, one of which is connected to infinity, another one to the horizon and the third is an intermediate closed pocket (as in Fig.~\ref{galleryIIIc}, left panel). It is therefore not shown. The effective potential in this configuration exhibits features from both a more Kerr-like BH, such as configuration II, and a RBS with an ergoregion and light rings.

Observe the difference between points ${\bf 3}_{\rm III}$ and  ${\bf 5}_{\rm III}$; both scatter off the innermost throat, which connects the pocket with the near-horizon region of the effective potential. But whereas point ${\bf 3}_{\rm III}$ is the result of a single scattering, point ${\bf 5}_{\rm III}$ also scatters off the outermost throat (which is almost nonexistent). Recall that each throat satisfies $\partial_r h_+=0$ at the boundary of the allowed region and is likely connected to a fundamental orbit (an unstable spherical orbit) ($cf.$ Section \ref{seclr}, Appendix \ref{appendixao}, \ref{appendixa} and \ref{appendixa2}). Moreover, notice that point ${\bf 3}_{\rm III}$ is close to the edge of the main part of the shadow, whereas point ${\bf 5}_{\rm III}$ is close to the edge of one of the eyebrows\footnote{Secondary shadows, disconnected from a larger one are dubbed \textit{eyebrows} \cite{Bohn:2014xxa}.}.

To further extrapolate these results, recall that in the familiar Schwarzschild or Kerr case, the edge of the shadow connects to a self-similar structure with infinitely many copies of the whole celestial sphere. This is due to photons that \textit{approximately resonate} the unstable light ring. In a spacetime endowing photon orbits with an effective potential as that in Fig.~\ref{galleryIIIa}, there are several light rings. Thus, photons can approximately resonate with each of these, or, in principle any combination thereof. This creates a \textit{hierarchy} of resonances, wherein more excited ones resonate more times, with different light rings. The plausible scenario we have just described suggests that the photons approaching the edge of the main shadow and of the eyebrows approximately resonate with different combinations of fundamental orbits.

\begin{figure}[!htp]
\begin{center}
\includegraphics[trim={0.2cm 0cm 1.5cm 1.6cm}, clip, height=.18\textheight, angle =0]{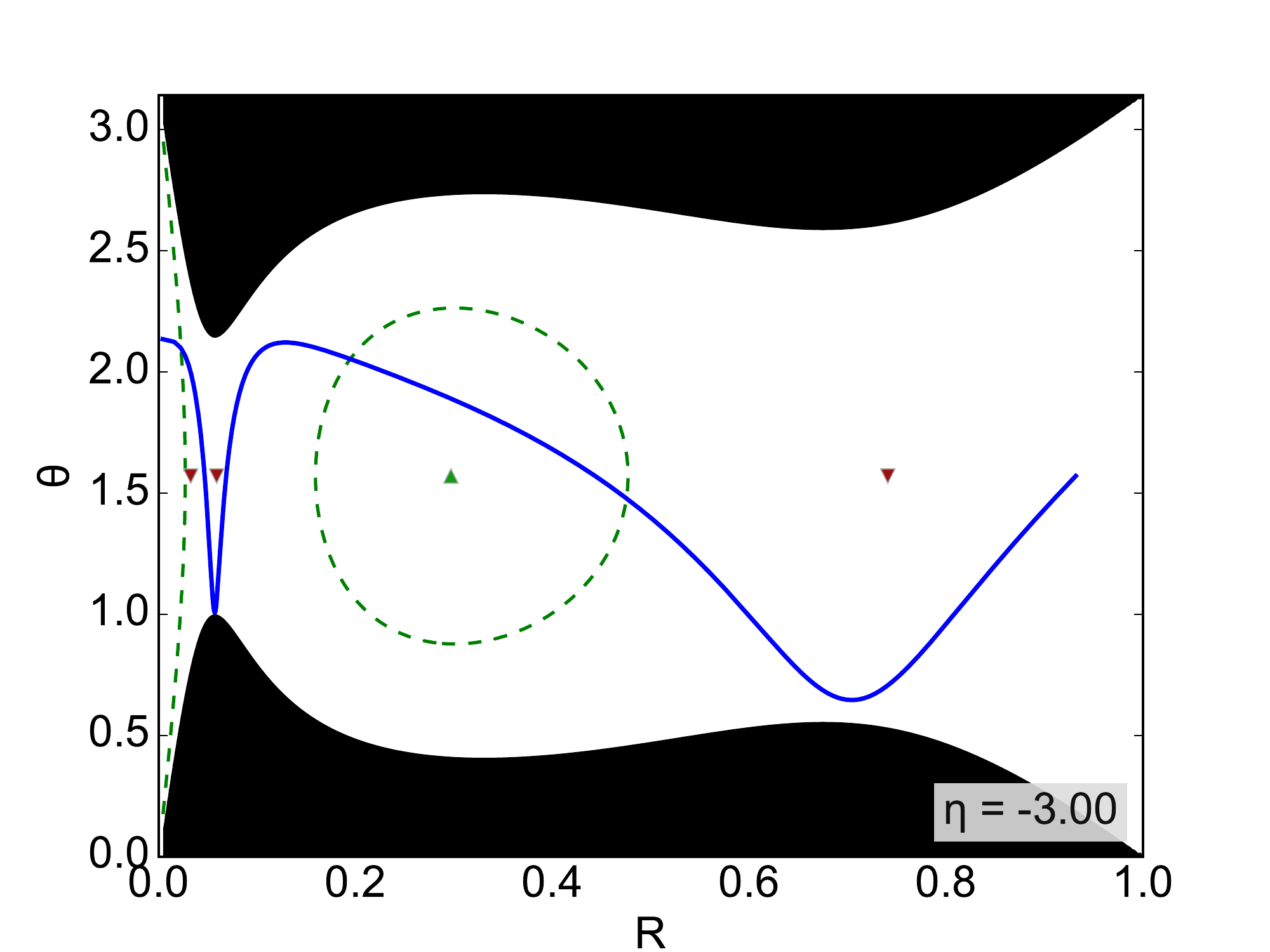}\ 
\includegraphics[trim={3.0cm 0.3cm 0.6cm 2.3cm}, clip, height=.16\textheight, angle =0]{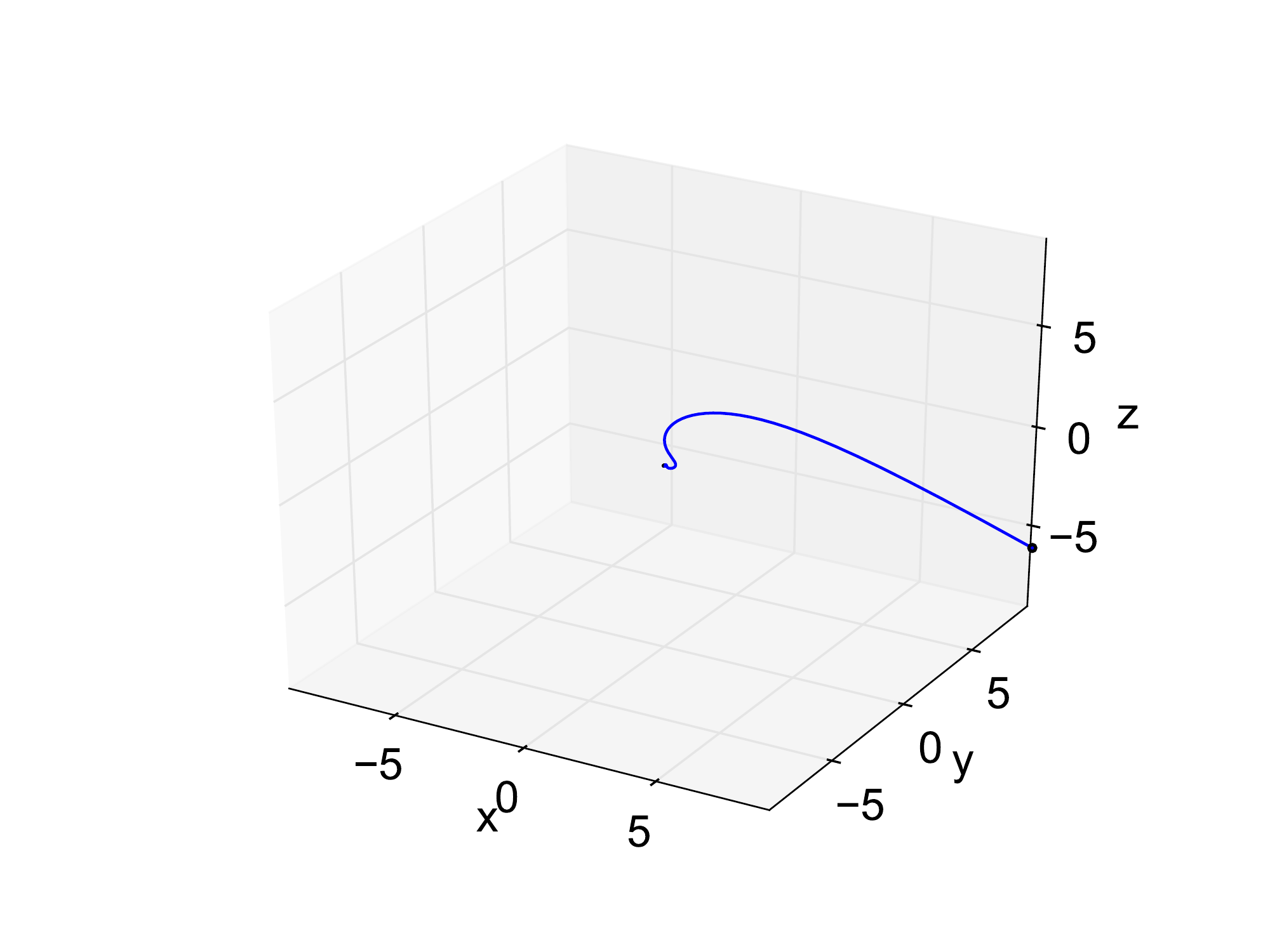}\
\put(0,50){${\bf 4}_{\rm III}$}

\vspace{0.3cm}

\includegraphics[trim={0.2cm 0cm 1.5cm 1.51cm}, clip, height=.18\textheight, angle =0]{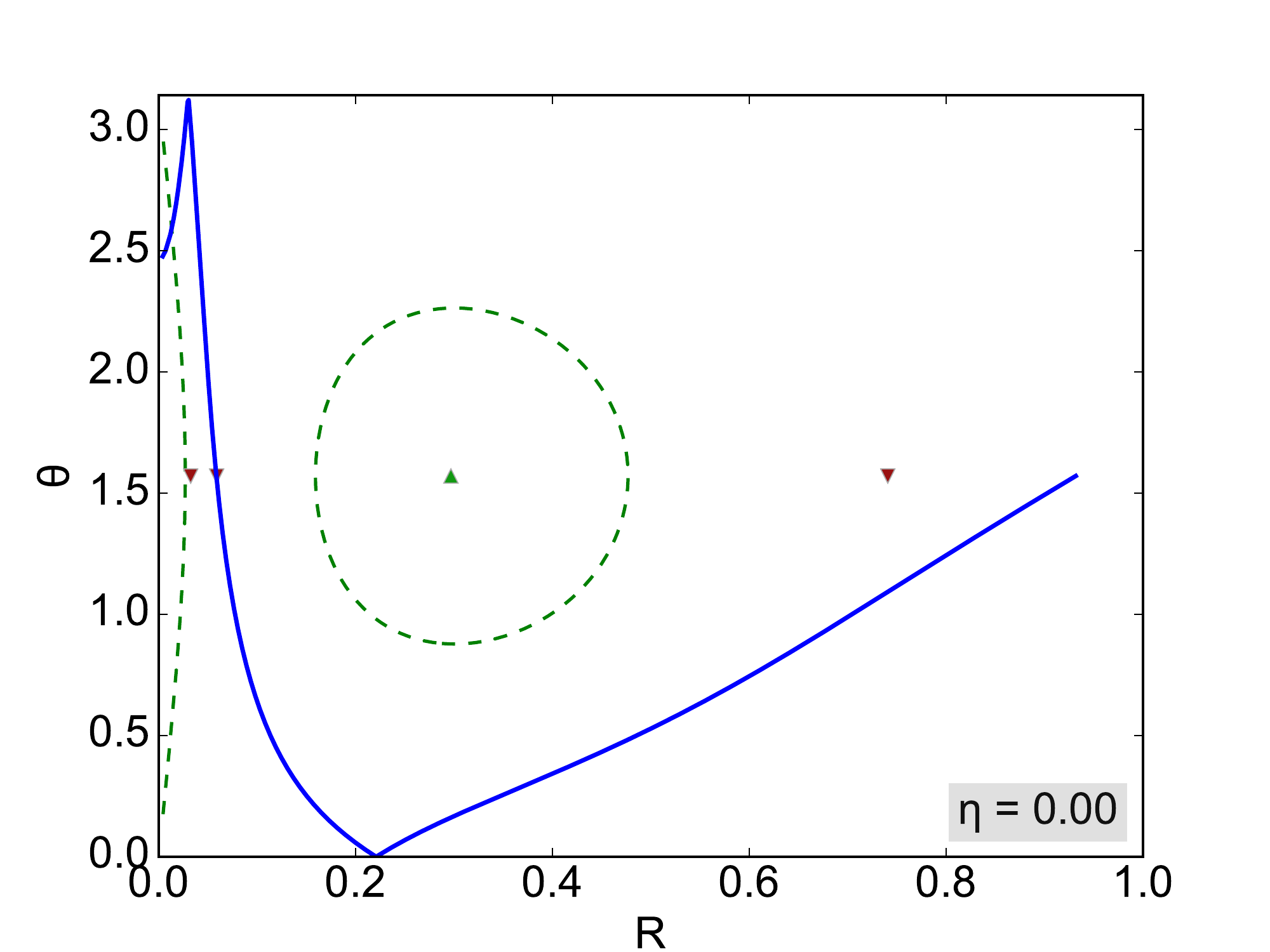}\ 
\includegraphics[trim={3.0cm 0.3cm 0.6cm 2.3cm}, clip, height=.16\textheight, angle =0]{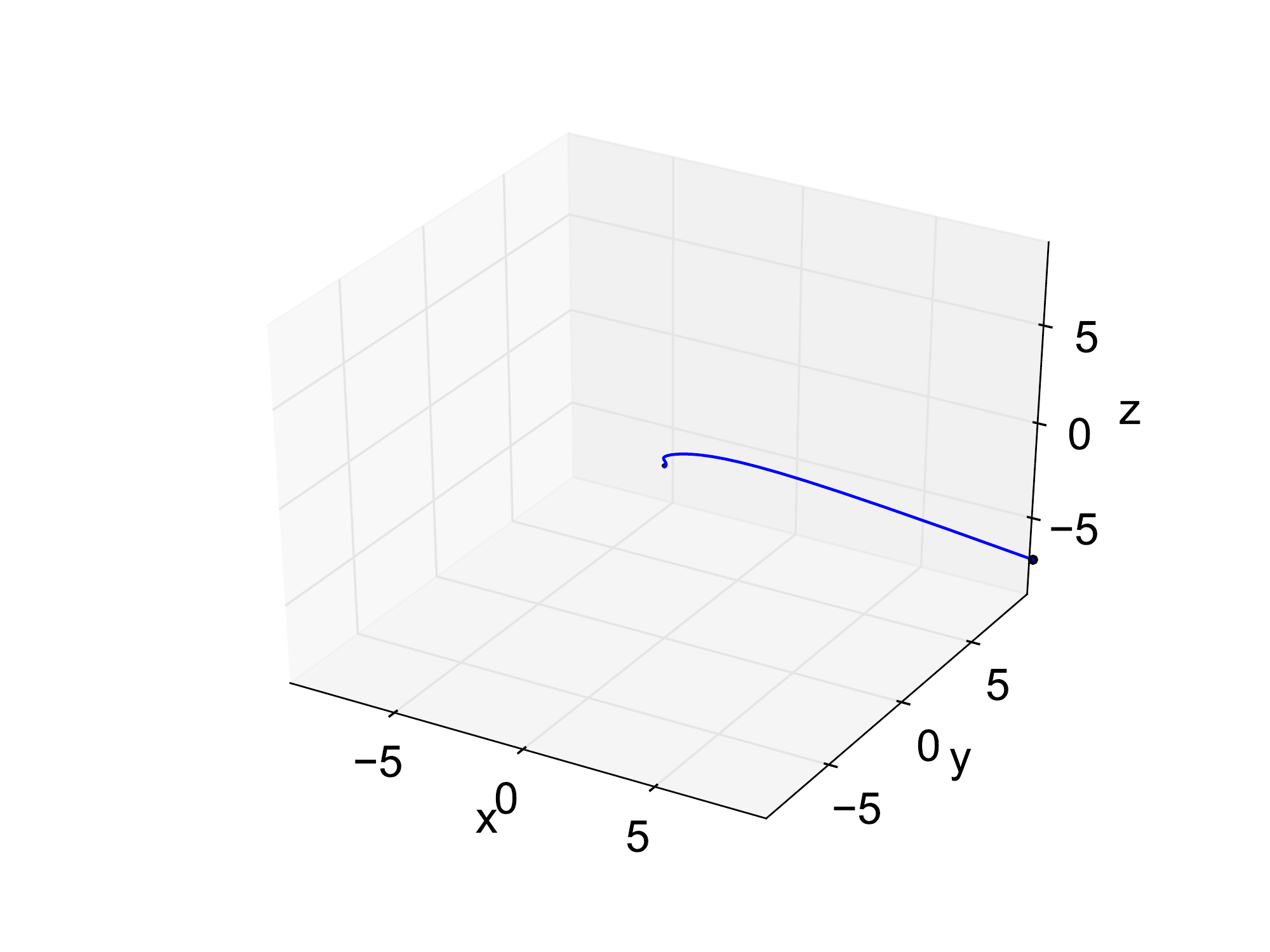}\
\put(0,50){${\bf 6}_{\rm III}$}

\vspace{0.3cm}

\includegraphics[trim={0.2cm 0cm 1.5cm 1.6cm}, clip, height=.18\textheight, angle =0]{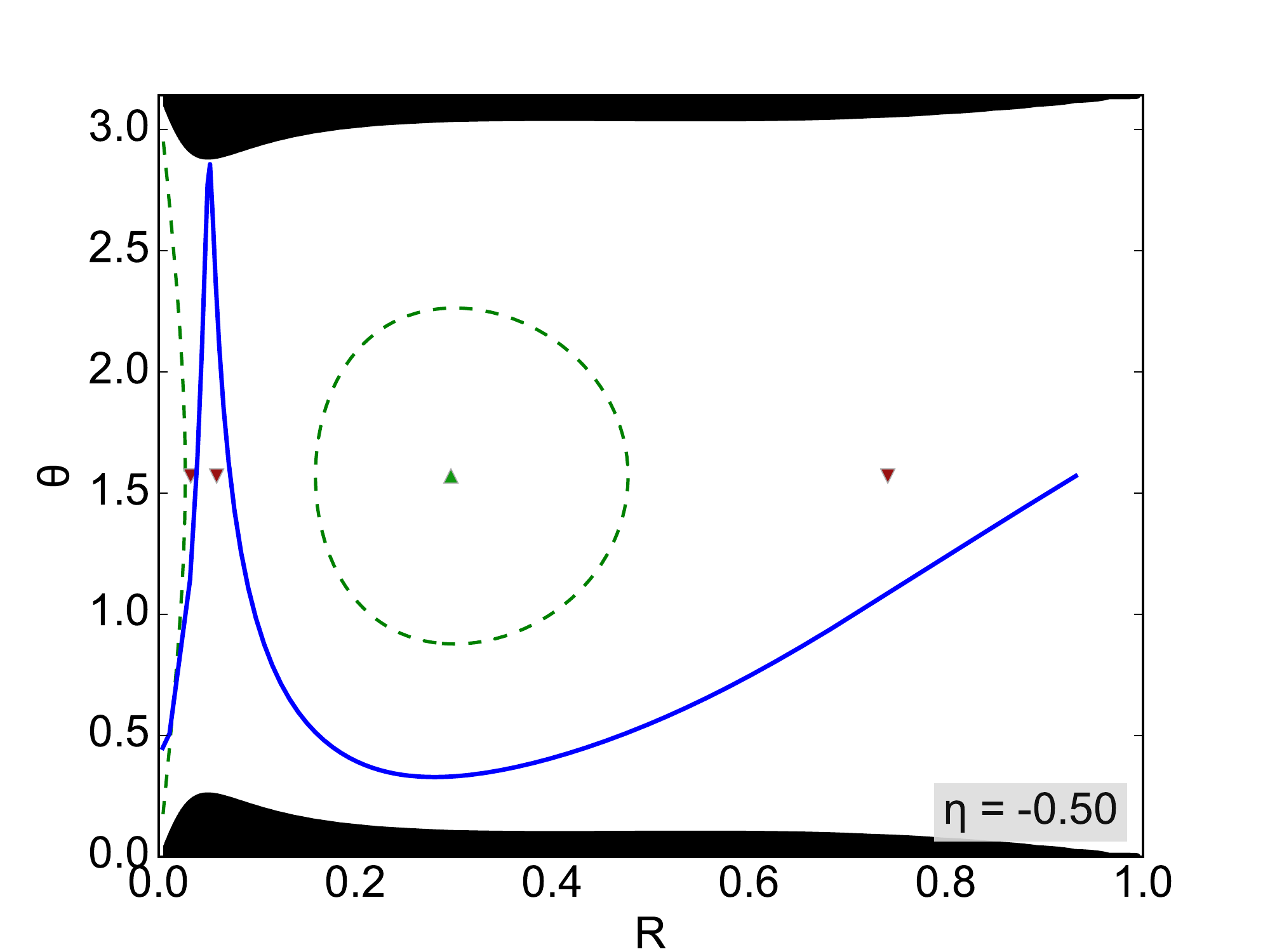}\ 
\includegraphics[trim={3.0cm 0.3cm 0.6cm 2.3cm}, clip, height=.16\textheight, angle =0]{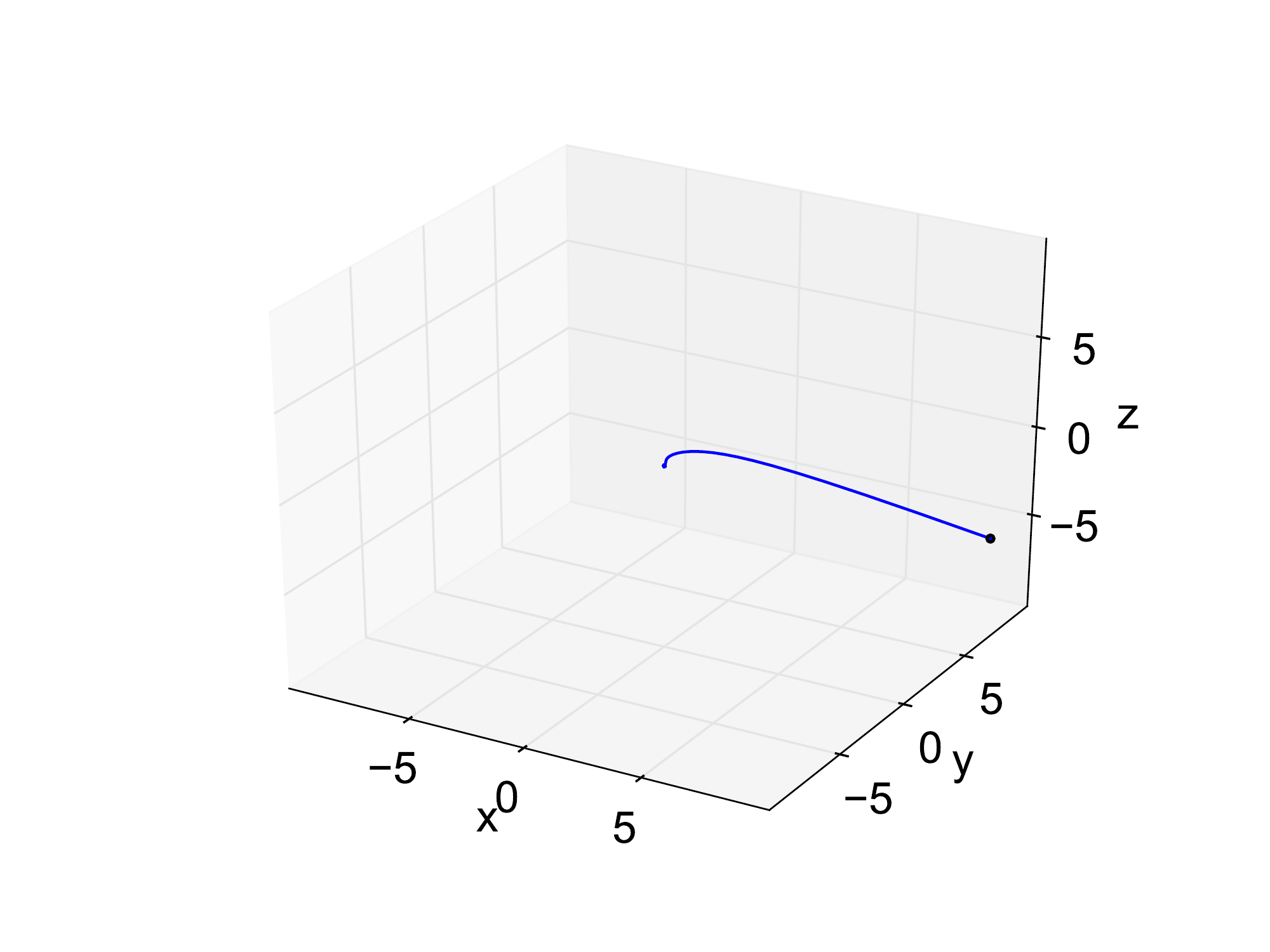}\
\put(0,50){${\bf 7}_{\rm III}$}

\end{center}
\caption{Absorption orbits in the effective potential (left) and  spacetime (right), corresponding to points ${\bf 4}_{\rm III}$, ${\bf 6}_{\rm III}$ and ${\bf 7}_{\rm III}$ in the lensing image of Fig.~\ref{galleryIIIa}.}
\label{galleryIIIb}
\end{figure}

\begin{figure}[!htp]
\begin{center}
\includegraphics[trim={0.2cm 0cm 1.5cm 1.6cm}, clip, height=.18\textheight, angle =0]{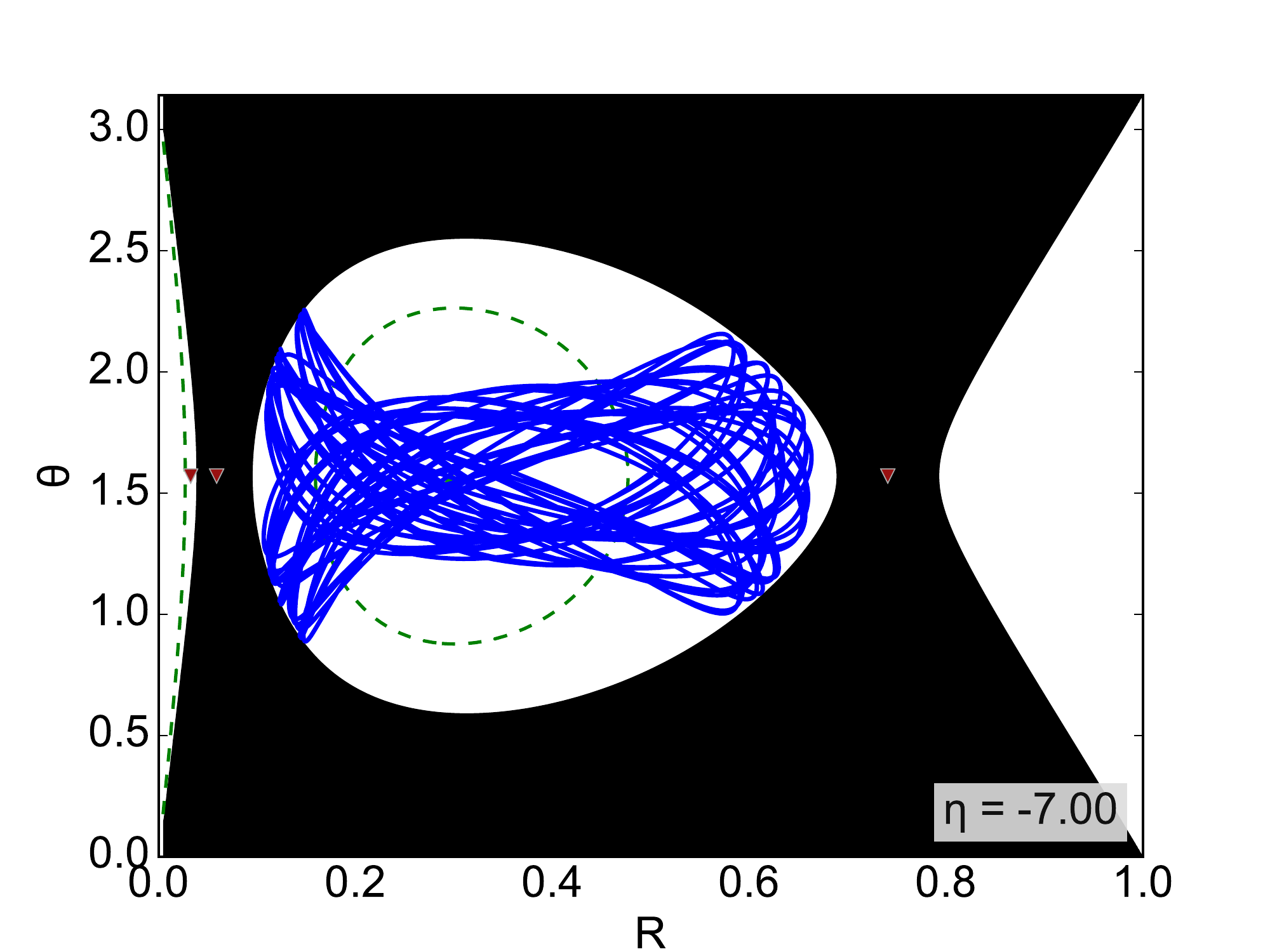}\
\includegraphics[trim={3.0cm 0.3cm 0.6cm 2.3cm}, clip, height=.16\textheight, angle =0]{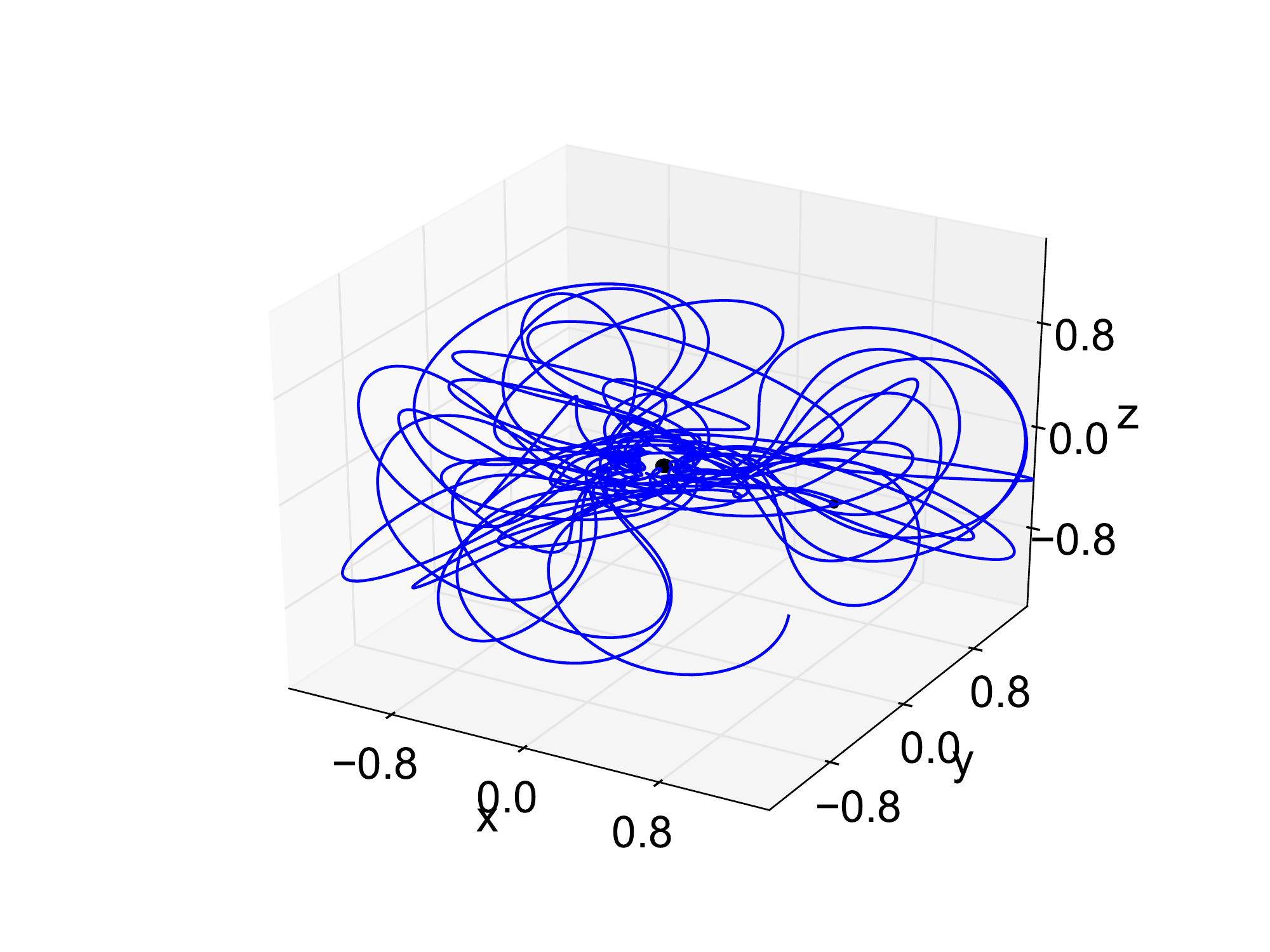}\

\vspace{0.3cm}

\includegraphics[trim={0.2cm 0cm 1.5cm 1.6cm}, clip, height=.18\textheight, angle =0]{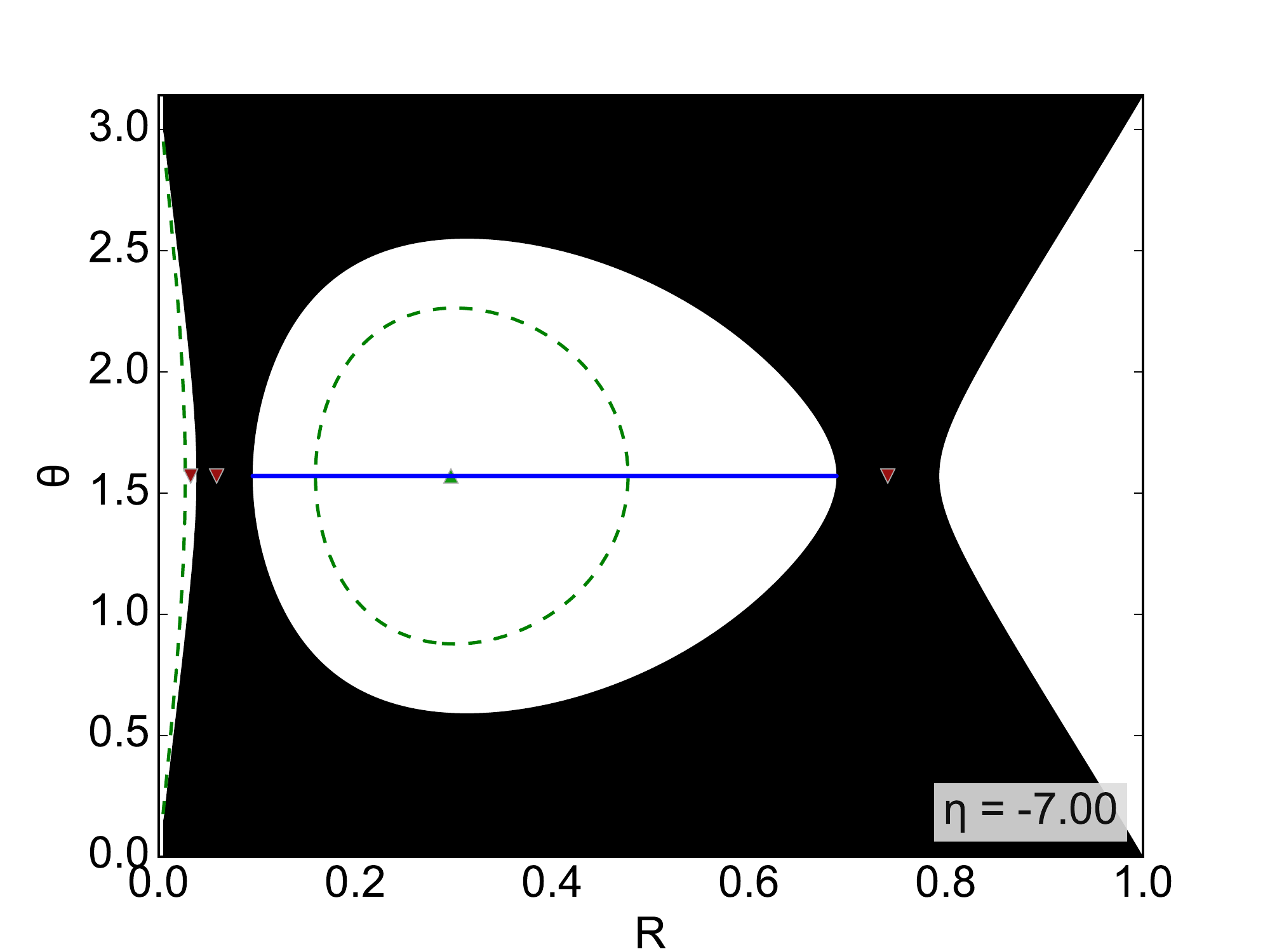}\
\includegraphics[trim={3.0cm 0.3cm 0.6cm 2.3cm}, clip, height=.16\textheight, angle =0]{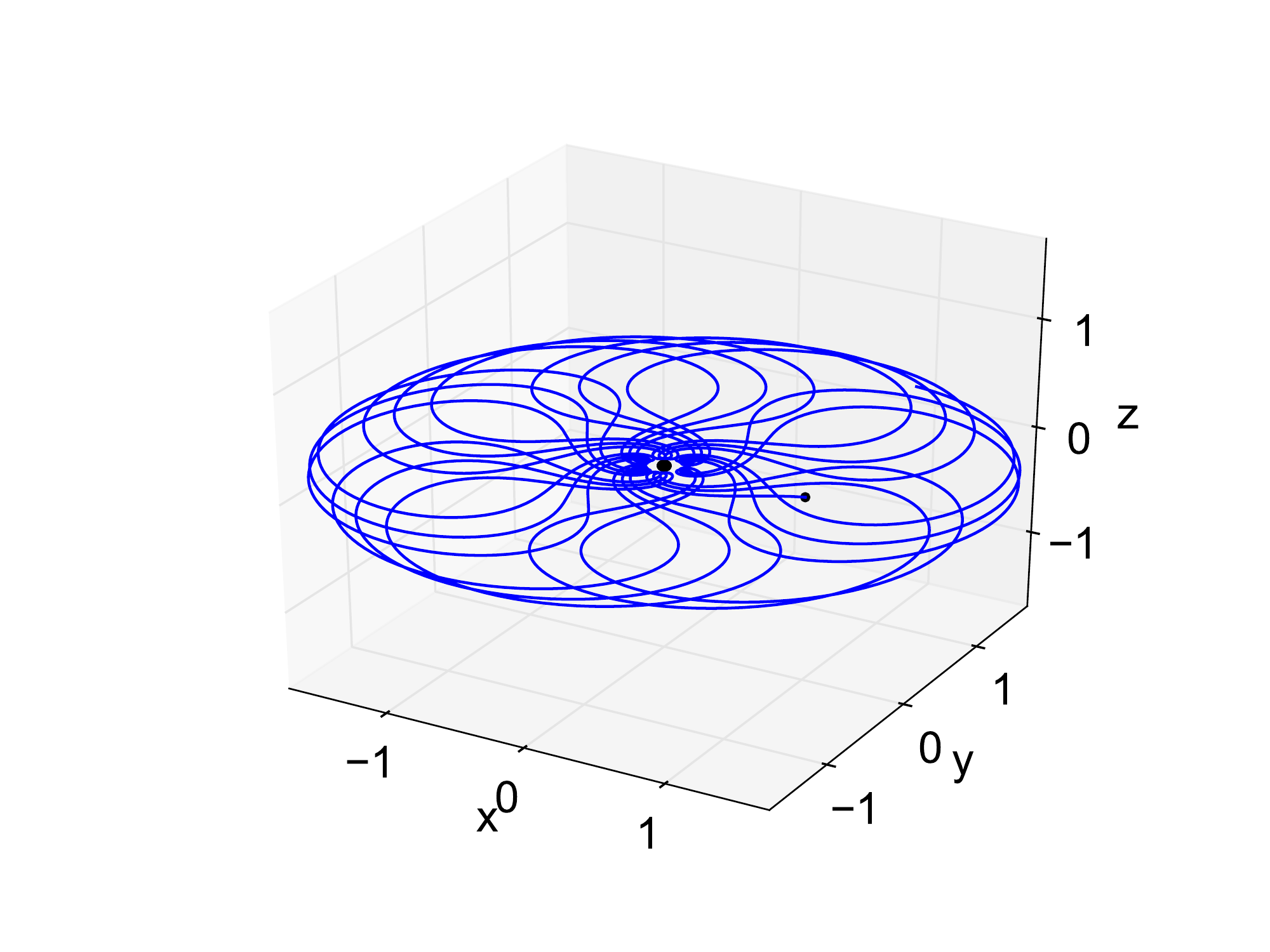}\
\end{center}
\caption{Effective potential (left) and spacetime orbits (right), of two bound orbits in configuration III ,with $\eta=-7.00$, one with $\theta$ motion and the other without. Observe that it is possible to have regular orbits even inside the pocket (bottom row).}
\label{galleryIIIc}
\end{figure}

This possibility is supported by Fig.~\ref{galleryIIIb} where we show three orbits that fall into the BH, two close to the edge of the main part of the shadow (points ${\bf 6}_{\rm III}$ and ${\bf 7}_{\rm III}$), and the other one close to the edge of one of the eyebrows (point ${\bf 4}_{\rm III}$); the latter can be seen to scatter off both throats of the potential. Point ${\bf 6}_{\rm III}$ in particular illustrates a case with zero impact parameter, that nevertheless displays a non-trivial trajectory. Clearly the effective potentials cannot describe all the dynamics.

Finally, in Fig.~\ref{galleryIIIc} we show two bound states around configuration III, for the same impact parameter as for point ${\bf 1}_{\rm III}$. One of these bound states has non-zero $\theta$ momentum and the other one is purely planar. This illustrates that for the same values of the impact parameter there can be many different orbits, including both scattering and non-scattering states.

\FloatBarrier

\bigskip
 \section{Overview and discussion}
 \label{sec4}

In this paper we have performed a detailed study of photon orbits in the background of KBHsSH and RBHs, extending and complementing the results in~\cite{Cunha:2015yba}. We now summarize some of our main results:


\begin{description}
\item[$\bullet$] For null geodesics, the Hamiltonian $\mathcal{H}=0$ restricts the motion of the light ray and sets a forbidden region in the phase space ($r,\theta$). The boundary of the latter can be studied in a systematic way by defining two potentials $h_\pm$, such that their countour lines delimit the boundary of the forbidden region for each value of the impact parameter $\eta$.
\item[$\bullet$] For some configurations, this boundary forms a \textit{pocket} that can be closed for some interval of $\eta$, giving rise to \textit{bound} orbits. However, there is a open interval of $\eta$ values that can leave an arbitrarily small entrance to the pocket, leading to \textit{trapped} or \textit{quasi-bound} orbits. The formation of such pockets can be traced back to the presence of a \textit{stable light ring}, combined with at least one unstable light ring. The latter is associated to a ``throat'' (a pocket entrance) that connects the interior of the pocket with a different region of the allowed phase space.
\item[$\bullet$]  The existence of a pocket is strongly correlated to the existence of chaos in the motion of the light ray, leading to turbulent patterns in the gravitational lensed image of the configuration. However, despite inducing chaos, pockets are neither a necessary nor sufficient condition for a particular trajectory to lie in such a chaotic pattern.
\item[$\bullet$] A common feature of chaotic orbits appears to be having \textit{more than one radial turning point}, a feature which embodies a deviation from Kerr spacetime \cite{Wilkins1972}. Nevertheless, it is still possible to have several turning points for a regular scattering, and hence this is not a sufficient condition for chaos.
\item[$\bullet$] The ergoregion does not appear to play a major role in this context, despite enhancing the chaotic patterns in the image. 
\item[$\bullet$] If an event horizon and a pocket are both present, the existence of a two throat system may be the origin of the formation of disconnected shadows, first reported in \cite{Cunha:2015yba} for KBHsSH. 
\end{description}

To conclude, and following the above observations, we would like to emphasize that:

-  not all KBHsSH display chaotic lensing. For instance, configuration II in~\cite{Cunha:2015yba} exhibits effective potentials very similar to those of Kerr ($cf.$ Appendix~\ref{appendixa}), even though the corresponding shadow is quite distinct. This also provides an example for which lack of integrability, in the sense of Liouville\footnote{Except for the corresponding Kerr boundary line (see Fig. \ref{fig_overview}), it is unlikely that any KBHsSH has a hidden constant of the motion (which exists in Kerr), and hence geodesic motion is almost certainly non-integrable in (almost) all the domain of existence.}, does not imply chaos;

- an important part of our analysis in this paper relied on numerical ray tracing. The results obtained using different ray tracing codes agree, lending them credibility. Such numerical methods, however, have issues for very long term integrations. Thus, our discussion of the chaotic patterns is mostly focused on their emergence, rather than on their precise quantitative properties, for which numerical errors may become important;

- finally, a similar analysis to that performed herein can certainly be pursued for other similar types of backgrounds, as, $e.g.$ the ones discussed in~\cite{Kleihaus:2005me,Brito:2015pxa,Kleihaus:2015iea,Herdeiro:2015tia,Herdeiro:2016tmi,Delgado:2016jxq}.

\vspace{0.5cm}

\section*{Acknowledgements}
P.C. is supported by Grant No. PD/BD/114071/2015 under the FCT-IDPASC Portugal Ph.D. program and by the Calouste Gulbenkian Foundation under the Stimulus for Research Program 2015.  C.H. and E. R. acknowledge funding from the FCT-IF programme. H.R. is supported by Grant No. PD/BD/109532/2015 under the MAP-Fis Ph.D. program.  This project has received funding from the European Union's Horizon
2020 research and innovation programme under the Marie Sklodowska-Curie
grant agreement No 690904 and by the CIDMA project UID/MAT/04106/2013.

\begin{appendix}

\section{Quasi-bound orbit in $(r,\dot{r})$ phase space}
\label{appendixao}
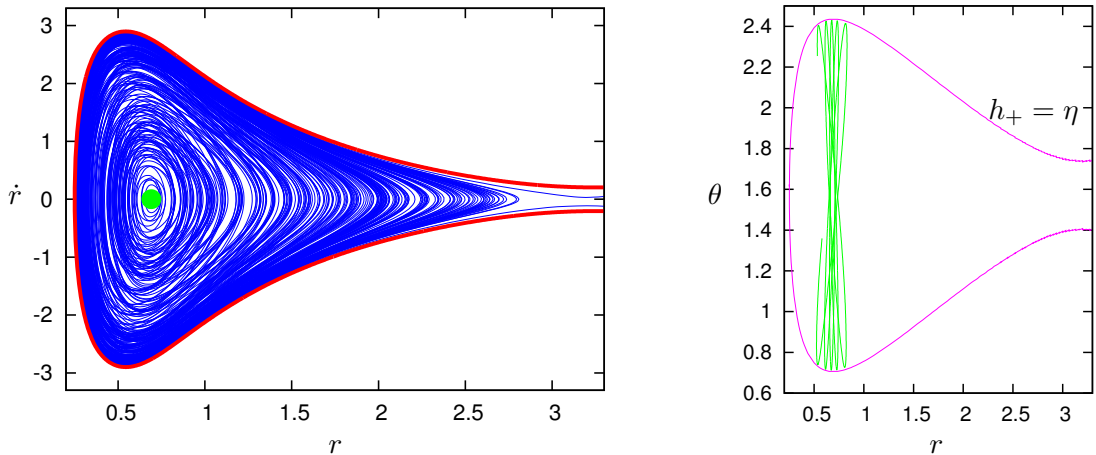
\begin{figure}[h!]
\begin{center}
\includegraphics[height=6cm]{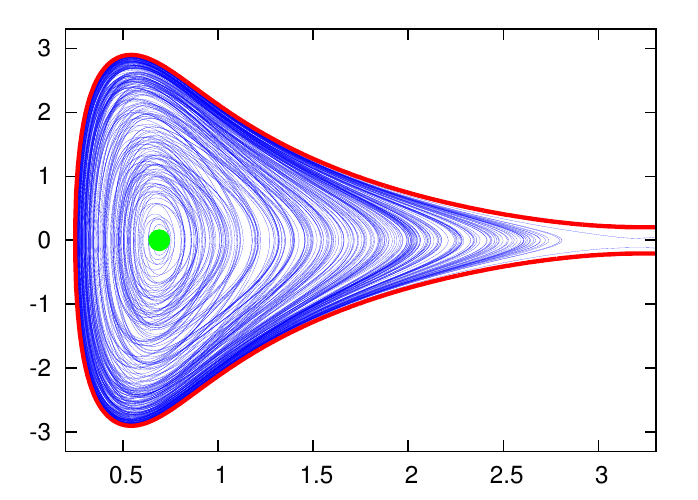}\hspace*{0.8cm}\includegraphics[height=6cm]{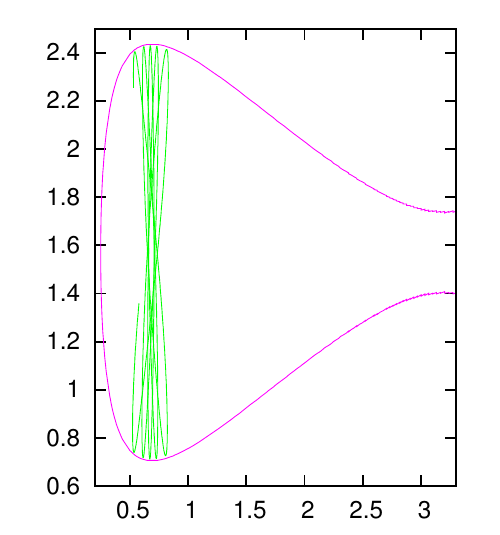}
\put(-75,-7){$r$}
\put(-158,87){$\theta$}
\put(-53,119){$h_+=\eta$}
\put(-300,-7){$r$}
\put(-420,87){$\dot{r}$}
\end{center}
\caption{\small (Left panel) Phase diagram $(r,\dot{r})$ of a photon trajectory in the RBS configuration 11 with $\eta\simeq-7.46$. The \textcolor{red}{red} line is given by the function $\pm\sqrt{-V/g_{rr}}$ and the \textcolor{OliveGreen}{green} dot satisfies $\partial_r h_+=0$ and $h_+=\eta$; (right panel) segment of the previous trajectory equivalent to the central \textcolor{OliveGreen}{green} dot, in $(r,\theta)$-space. The connection to a spherical orbit is apparent. The purple line represents the boundary of the allowed region.}
\label{fig_diagram}
\end{figure}

A quasi-bound orbit displays an interesting dynamics. The motion is 2-dimensional, in $r,\theta$. Thus, focusing, on the ($r,\dot{r}$) phase space, one can anticipate effective energy losses (gains) due to the coupling to the $\theta$ motion. This is exactly what can be observed in a neat way for some trajectories. As an example, the plot in Fig. \ref{fig_diagram} (left panel) displays a trajectory in phase space ($r,\dot{r}$) for a photon that enters a trapping region in the RBS configuration 11. The orbit spans a ``pear-like" curve which decreases in size, resembling the well known picture for a harmonic oscillator with friction (wherein the curve is an ellipse). Here, however, the energy is not being lost, rather it is being transferred into the $\theta$-motion.

The envelope curve in Fig.~\ref{fig_diagram}\, left (the \textcolor{red}{red} solid line) can be computed as follows. The maximum possible value of $\dot{r}$ can be obtained on the equatorial plane\footnote{One can convince oneself of this by looking at the potential $V$ and realising that the minimum is on the equator.} ($\theta=\pi/2$) with $\dot{\theta}=0$. This implies
\be
\dot{r}=\sqrt{-\frac{V(r,\pi/2)}{g_{rr}}}\ .
\ee
This function (\textcolor{red}{red} solid line in Fig. \ref{fig_diagram} -- left) describes perfectly the envelope shape.

For a given value $\eta$, the conditions $h_+=\eta$ and $\partial_r h_+=0$ are satisfied in phase space by the \textcolor{OliveGreen}{green} dot in the figure. The trajectory of the photon near that central dot is represented in Fig. \ref{fig_diagram} (right), displaying multiple reflections on the contour line $h_+=\eta$. The reflection points are outside the equatorial plane and close to the condition $\partial_r h_+=0$, leading to little motion along the $r$ coordinate. By analogy with the Kerr analysis ($cf.$ Appendix~\ref{appendixa}), it seems plausible that these points are connected to a \textit{spherical orbit}, with motion along $\theta$ only, and constant $r$ ($cf.$ Appendix \ref{appendixa2}).

\medskip

\section{The Kerr case}
\label{appendixa}

In this appendix, we will implement the $h_\pm$ framework for the Kerr spacetime. This is a well known case, but it is typically treated by separating variables. As such, we provide here a treatment parallel to that used in the main text for solutions for which no separation of variables is known (or likely to exist).

\begin{figure}[th!]
\begin{center}
\includegraphics[width=10cm]{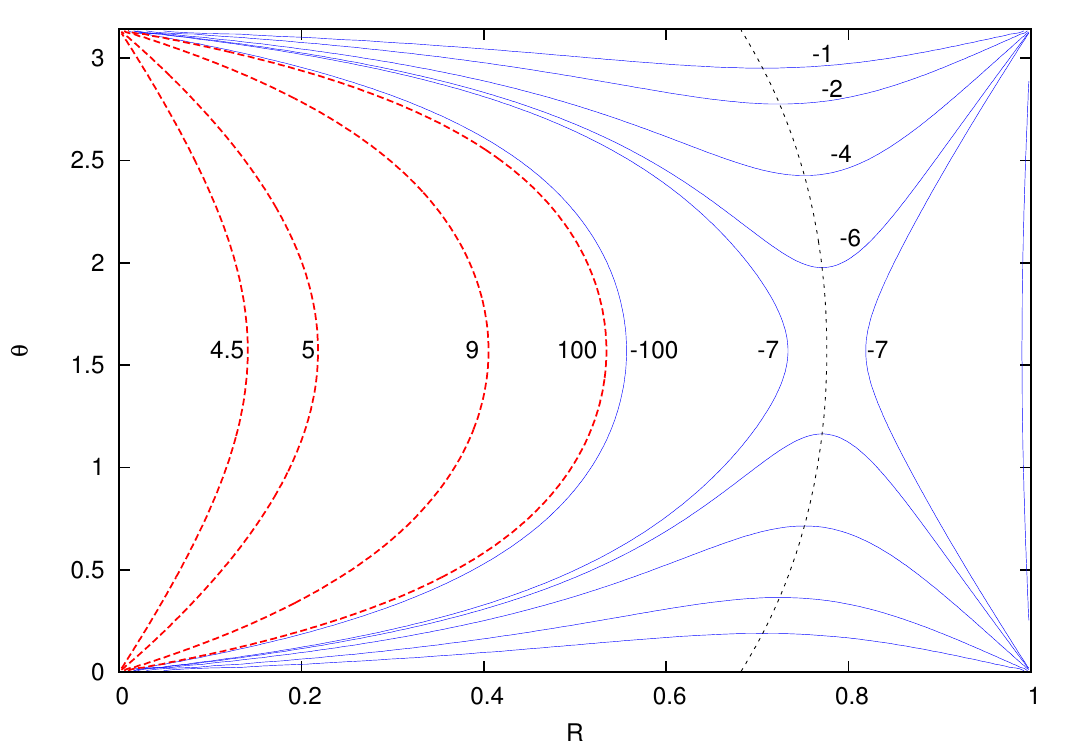} 
\put(0,110){$h_+$}\\
\includegraphics[width=10cm]{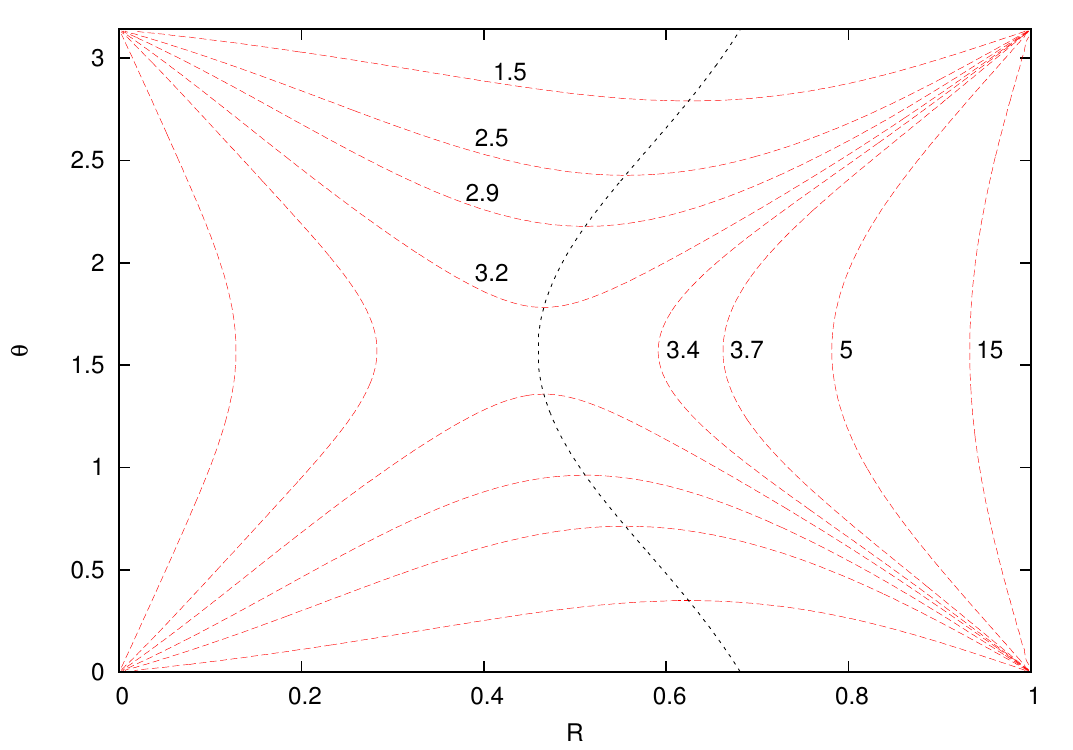}
\put(0,110){$h_-$}\\
\end{center}
\caption{\small Contour plots of $h_+$ (top panel) and $h_-$ (bottom panel), for the Kerr BH solution. The solid lines (\textcolor{blue}{blue}) represent negative $\eta$ values, whereas dashed lines (\textcolor{red}{red}) represent positive values. The dotted line (black) is given by both $\pi/2+\psi(r)$ and $\pi/2-\psi(r)$, coinciding with the condition $\partial_r h_\pm=0$. The saddle points of $h_\pm$ are consistent with the position of the light rings, as expected. The coordinate $R$ is computed with the same expression as before (\ref{R_def}), despite $r$ being a Boyer-Lindquist coordinate now.}
\label{fig_Kerr}
\end{figure}

In the Kerr case, we have two unstable light rings on the equatorial plane at radial coordinates $r_1$ (for co-rotating photons) and $r_2$ (for counter-rotating photons), in Boyer-Lindquist coordinates, given a value of the rotation parameter $a$ such that $a/M\in[-1;1]$ \cite{Bardeen1973,Teo2003}:
\[r_1\equiv2M\left\{1+\cos\left(\frac{2}{3}\arccos\left[-\frac{|a|}{M}\right]\right)\right\},\]
\[r_2\equiv2M\left\{1+\cos\left(\frac{2}{3}\arccos\left[\frac{|a|}{M}\right]\right)\right\},\]
where $M$ is the Kerr ADM mass. Moreover, we have that $r_1\leqslant r_2$. Between these radii we can have \textit{unstable spherical orbits}, which are not restricted to the equatorial plane and for which $\theta$ oscillates between $\pi/2\pm\psi$, where $\psi\in[0,\pi/2]$. In particular, given a radial coordinate $r$ such that $r_1\leqslant r \leqslant r_2$, we can have a spherical\footnote{\textit{Spherical} photon orbits denote null geodesics with a constant radial coordinate $r$; the latter is not related (in general) to the standard spherical coordinates.} photon orbit at that position as long as we have the correct restrictions on the constants of geodesic motion. For Kerr these constants are $E,\Phi$ and $Q$, the latter being the Carter constant \cite{Bardeen1973,Carter1968}. Specifically, the relations that must be satisfied are:
\begin{align*}
&\eta\equiv\frac{\Phi}{E}=-\frac{r^3-3Mr^2+a^2r+Ma^2}{a(r-M)},\\
&\\
&\chi\equiv\frac{Q}{E^2}=-\frac{r^3(r^3-6Mr^2+9M^2r-4a^2M)}{a^2(r-M)^2}.
\end{align*}
The first equation establishes a connection between our impact parameter $\eta$ and the radial coordinate of a spherical orbit. From it, it is possible to conclude that the $\eta$ required is positive for $r_1$ and negative for $r_2$ (given $a>0$), with the physical interpretation that $r_1$ is connected to a co-rotating light ring, whereas $r_2$ is related to a counter-rotating one ($cf.$ Section \ref{seclr}) \cite{Teo2003}.\\
As mentioned, $\theta$ oscillates between $\pi/2\pm\psi$, where $\psi$ can be computed as:
\[\psi(r)=\arcsin\left\{-\sqrt{\frac{[a^2-\eta^2-\chi]+ \sqrt{[a^2-\eta^2-\chi]^2+4a^2\chi}}{2a^2}}\right\}.\]
Hence given a value of $r$  (with $r_1\leqslant r\leqslant r_2$) one can compute $\eta(r)$ and $\chi(r)$ and obtain the respective $\psi(r)$. The curve $\pi/2\pm\psi(r)$ in ($r,\theta$) space is represented in Fig. \ref{fig_Kerr} as a dotted line (black). In this figure are also represented the contour lines of the functions $h_\pm$, each with a saddle point that coincides with the position of a light ring. This is consistent with the previous discussion since the saddle point for $h_+$ ($h_-$) occurs for a negative (positive) value of $\eta$ and thus corresponds to a light ring which is counter(co)-rotating. Moreover, the $h_-$ saddle point (connected to co-rotation) occurs for a smaller radial coordinate that the $h_+$ saddle point (connected to counter-rotation), as expected.\\
Interestingly, it is clear that the curve given by $\pi/2+\psi(r)$ and $\pi/2-\psi(r)$ also satisfies the condition $\partial_r h_\pm=0$. As such, the latter also yields spherical orbits ($cf.$ Appendix \ref{appendixao} and \ref{appendixa2}). In particular, for $\eta=0$ we have $\psi=\pi/2$ for both $h_\pm$. Hence there is a continuous connection between spherical orbits as we go from $h_+$ to $h_-$ (or vice-versa). As a final observation, $h_+$ can diverge due to the existence of an ergoregion, in this case with spherical topology. As before, inside this region the $h_+$ contour lines are for positive $\eta$ ($cf.$ Section \ref{poch}).

%

\section{Acceleration field $\mathcal{F}_r$}
\label{appendixa2}

From one of Hamilton's equations \eqref{eq_Hamilton} we obtain:
\[\dot{p}_r=-\frac{1}{2}\left(\partial_r g^{rr} p_r^2 +\partial_r g^{\theta\theta} p_\theta^2 +\partial_r V\right).\]
Setting $p_r=0$ and solving for $p_\theta$ from $\mathcal{H}=0$, it leads to:
\[\dot{p_r}_{[p_r=0]}=-\frac{1}{2}\left(-\frac{V}{g^{\theta\theta}} \partial_r g^{\theta\theta} +\partial_r V\right).\]
Dividing by the photon's energy at spatial infinity $E$, we obtain a function which only depends on ($r,\theta$) and on the impact parameter $\eta=\Phi/E$:
\begin{equation}\mathcal{F}_r(r,\theta)\equiv -\frac{1}{2E^2}\Big(V\,\partial_r\log [g_{\theta\theta}] + \partial_rV\Big).\label{eq-Fr}\end{equation}
Hence, this function dubbed radial \textit{acceleration field} returns the value of $\dot{p}_r$ of the photon when $p_r=0$, divided by a scale factor. We remark that $g_{\theta\theta}$ is positive definite, and hence the logarithm is well defined. Now we will consider applications of the $\mathcal{F}_r$ function to some of the spacetimes.

\begin{figure}[th!]
\begin{center}
\includegraphics[width=8cm]{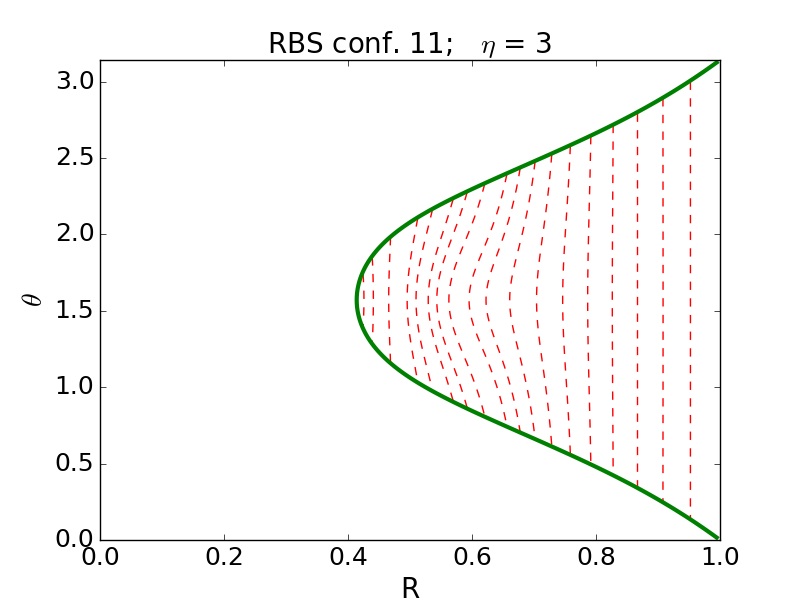}\includegraphics[width=8cm]{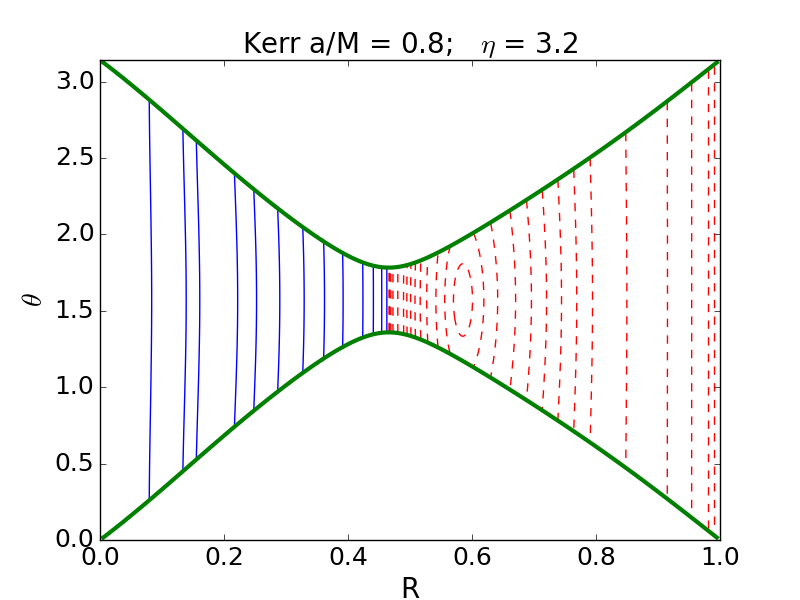}\\
\includegraphics[width=8cm]{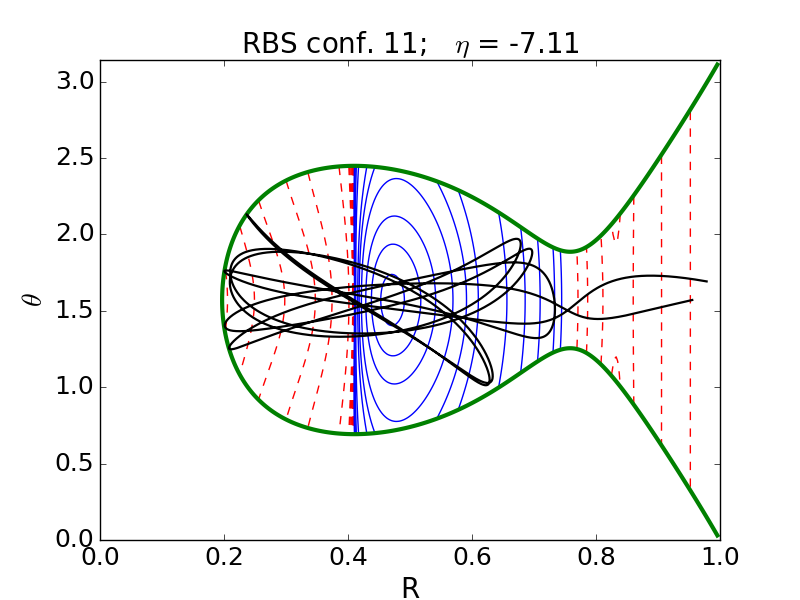} \includegraphics[width=8cm]{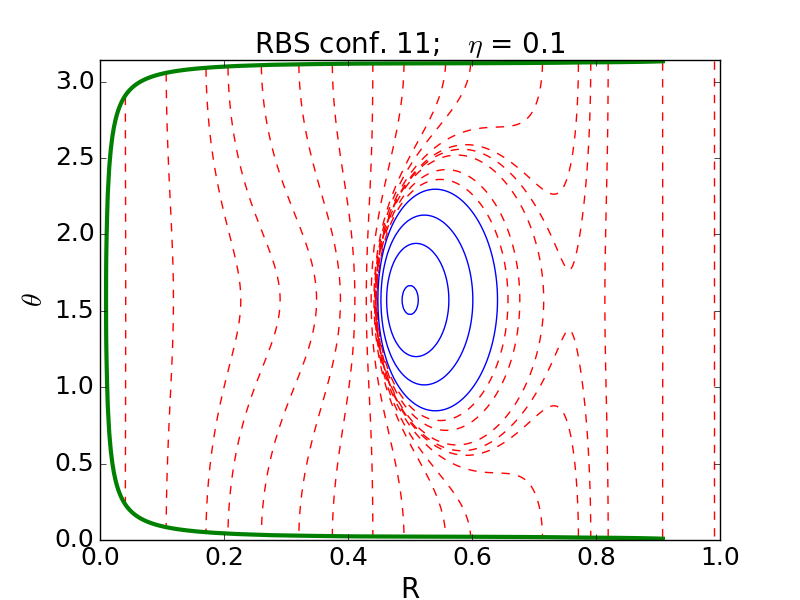}
\end{center}
\caption{\small Contour lines of the acceleration field $\mathcal{F}_r$, with dashed \textcolor{red}{red} (solid \textcolor{blue}{blue}) lines for positive (negative) values. All figures correspond to the RBS 11, except the top right panel which corresponds to Kerr. The thick \textcolor{OliveGreen}{green} line sets the boundary of the allowed region. The black line on the bottom left image represents a single photon trajectory.}
\label{fig_Fr}
\end{figure}

In Fig. \ref{fig_Fr} are displayed the contour plots of $\mathcal{F}_r$ for the RBS configuration 11 and the Kerr case. The dashed \textcolor{red}{red} (solid \textcolor{blue}{blue}) lines represent positive (negative) values of $\mathcal{F}_r$. Starting from the top left figure, for the RBS 11 with $\eta=3$, the acceleration field only has positive values (dashed \textcolor{red}{red} lines) inside the allowed region. This actually implies that in this case the motion can have at most one radial turning point. For instance, if the light ray has at any given point $p_r=0$, then $\mathcal{F}_r>0$ implies that $\dot{p}_r>0$, and $p_r$ cannot become negative afterwards since there is no negative $\mathcal{F}_r$ region.

Going now to Fig. \ref{fig_Fr} top right we have the Kerr case with $\eta=3.2$. The transition line from positive to negative values are the set of points such that $\mathcal{F}_r=0$, which implies that if $p_r=0$ at those points then $\dot{p}_r=0$ (but $p_\theta\neq 0$ in general). This appears to be  connected to a \textit{spherical orbit} at that location ($cf.$ Appendix~\ref{appendixa}). Moreover, notice that if $\partial_r V=0$ at the boundary of the allowed region $(V=0)$, then $\mathcal{F}_r=0$. Since $\partial_rV=0 \iff \partial_r h_\pm=0$ if $V=0$ ($cf.$ Section \ref{seclr}) then it may be possible to establish a connection between $\partial_r h_\pm=0$ and spherical orbits ($cf.$ Appendix \ref{appendixao} and \ref{appendixa}).

Curiously, it is also possible in this case to conclude that there is at most one radial turning point. For instance, if during the motion $p_r=0$ inside the \textcolor{red}{red} (\textcolor{blue}{blue}) region, then the value of $\dot{p}_r$ is positive (negative) and $r$ will start to increase (decrease). Since the sign of $\mathcal{F}_r$ will not change after this point, then the photon cannot have $p_r<0$ ($p_r>0$) afterwards and so the photon eventually escapes (falls into) the BH. This is consistent with the literature, as it is known that null geodesics in a Kerr spacetime have at most one radial turning point \cite{Wilkins1972}. However, we note that the present approach is valid even when the geodesic equations are not fully separable and integrable, despite that not being the case for Kerr.

Continuing to Fig. \ref{fig_Fr} bottom left, we have the RBS configuration 11 again, now with the different impact parameter $\eta=-7.11$. In this case there are two disconnected lines for which $\mathcal{F}_r=0$, each connected to a spherical orbit, by analogy with Kerr. Contrary to the previous cases, in this situation it is possible to have more than one turning point, since after having $p_r=0$ the sign of $\mathcal{F}_r$ can still change. This can be traced back to the transition line $\mathcal{F}_r=0$ that goes from positive to negative values of $\mathcal{F}_r$ as $r$ increases, and ultimately to the existence of a stable light ring. Thus, it is possible to have a photon wobbling around that line, yielding several radial turning points (see Fig. \ref{fig_Fr} bottom left for an example).

Advancing to Fig. \ref{fig_Fr} bottom right we again have the RBS 11, now with $\eta=0.1$. The transition line $\mathcal{F}_r=0$ has now become closed in a loop, leading to an isolated region with $\mathcal{F}_r<0$. As in the case before, it is possible to have more than one radial turning point. However notice that there is no pocket in this case: $\partial_r h_-$ is never zero at the boundary of the allowed region (\textcolor{OliveGreen}{green} line). This case illustrates a situation for which chaos is possible even without a pocket ($cf.$ Section \ref{poch}).

%

\section{PYHOLE}
\label{appendixb}

\textsc{pyhole} is a ray tracing code written in Python 3 using the NumPy \cite{NumPy:2011} and SciPy \cite{SciPy:2007} extensions for scientific computing. This choice allows rapid and flexible design of the code and simple addition of features. Since Python is an interpreted language, the resulting code is platform independent and can be run on any system with a Python 3 interpreter. This flexibility comes at the cost of computational speed as Python code for numerical applications is typically slower than optimised native implementations in languages such as C++ or Fortran. However, by optimizing the code specifically for NumPy, modern Python implementations can reduce the overhead significantly. We further address the issue by parallelizing the code and providing native implementations of the crucial code paths (see Appendix~\ref{pyper}).

The design of the code is fully modular to allow for different metrics (numerical as well as analytical), propagators, and projections. The code consists of two major components:
\begin{enumerate}
    \item the computational component that performs the ray tracing for each pixel in the image
    \item the interactive component that renders a picture from the ray tracing data and allows analysis of the results by the user.
\end{enumerate}

The computational component of the code consists of a ``scene" object which collects all the settings of a particular setup to be computed. This includes settings for the observer position, the background metric and the display styles to be used. Within each scene, a ``camera'' object encapsulates the observers reference frame information as well as the projection of incoming rays onto the image plane. Besides the camera, the user specifies a ``metric" object, which contains all information about the metric that is used for ray propagation. Internally, the camera passes the metric on to a ``propagator" object along with each initial condition to be propagated. The propagator then performs the actual numerical integration backwards along the light ray, returning the position of the source of the light ray on the celestial sphere, or event horizon, to the camera. With this setup it is possible to interchange the various objects for different implementations of each functionality.

At the core of all this sits the numerical propagator. In the Python code, we use the built in Adams/BDF multi-step method with automatic step size control provided by SciPy, which is based on the Netlib VODE implementation\cite{Brown:1989}. We have performed tests with other Runge-Kutta based integration schemes \cite{Verner:2002} but the results did not change while the runtime of the code increased significantly. This is expected as the ODE near the BH becomes stiff, which is a situation not handled well by Runge-Kutta based integrators.

The mapping between a pixel on the projected image and the direction of the source of the corresponding light ray is the output of the computational module of the code. Computing this mapping for high resolution images (typically 1024$\times$1024 pixels) is a time consuming operation, which is why this mapping is pre-computed and stored in a file on disk. The analysis code is then capable of loading a pre-computed direction mapping instead of recomputing it.

The interactive component is a unique feature of \textsc{pyhole}. After rendering the image, it is presented to the user in a graphical user interface. It is then possible to select points within the image and show their trajectories, along with other relevant information such as plots of the effective potential associated with a given trajectory (see Figures \ref{gallery10} - \ref{galleryIIIc}). The trajectories are computed on the fly using the same propagator described above. Apart from single trajectories, this component also allows the analysis of the resulting image using other techniques including the visualization of the time delay function (see Fig. \ref{timedelay}), as well as shadow fragmentation analysis.

\subsection{Metric, interpolation and image}

The metric is specified as an object, which allows for the simple addition of different metrics by the user. For the RBSs and KBHsSH solutions analysed in this paper, we work with an interpolated metric constructed using a second order 2D spline interpolation of the functions $F_0$, $F_1$, $F_2$, $W$ over $R$ and $\theta$ as defined in \cite{Cunha:2015yba}. The derivatives of these functions with respect to the interpolation variables are obtained by taking derivatives of the interpolant. The interpolation is performed using the \texttt{RectBivariateSpline} functionality of SciPy, which is based on the FITPACK implementation in Netlib\cite{Dierckx:1993}. The resulting values are then inserted in the analytical expressions for the actual metric components and their derivatives with respect to $r$ and $\theta$. For reference, a third order spline interpolation has also been tried but it has been determined that this does not change the accuracy of the result.

Besides the metric, some further information is needed to produce a camera image. Typical user input to construct this image consists of:
\begin{itemize}
    \item the initial observer position,
    \item the field of view,
    \item the image resolution, and
    \item the radius of the celestial sphere.
\end{itemize}

The observer position is given by $(t=0, r, \theta, \phi=0)$ where without loss of generality $t=0$ and $\phi=0$ is assumed. Thus the user only supplies an initial $r$ and $\theta$ position. We are specifying the observer's radial position $r$ for each choice of metric through a fixed circumferential radius $R$, as in \cite{Cunha:2015yba}, and the same for the radial position of the celestial sphere ($cf.$ Section~\ref{sec3}).

The projections currently implemented are stereographic and gnomonic projection, as well as the equirectangular projection used in \cite{Cunha:2015yba}. Gnomonic projection is the projection resulting from a physical pinhole camera, while the very similar stereographic projection is a variation thereof commonly used in computer graphics. The equirectangular projection directly maps the two observer angles onto the coordinate axes without further geometric projection.

For all of these projections, the user input is a field of view, which specifies the angle between the central direction and the boundaries of the image plane. Together with the resolution of the image, this fixes the scaling factor used to convert pixels to degrees.

\begin{figure}[!htp]
\begin{center}
\includegraphics[height=.14\textheight, angle =0]{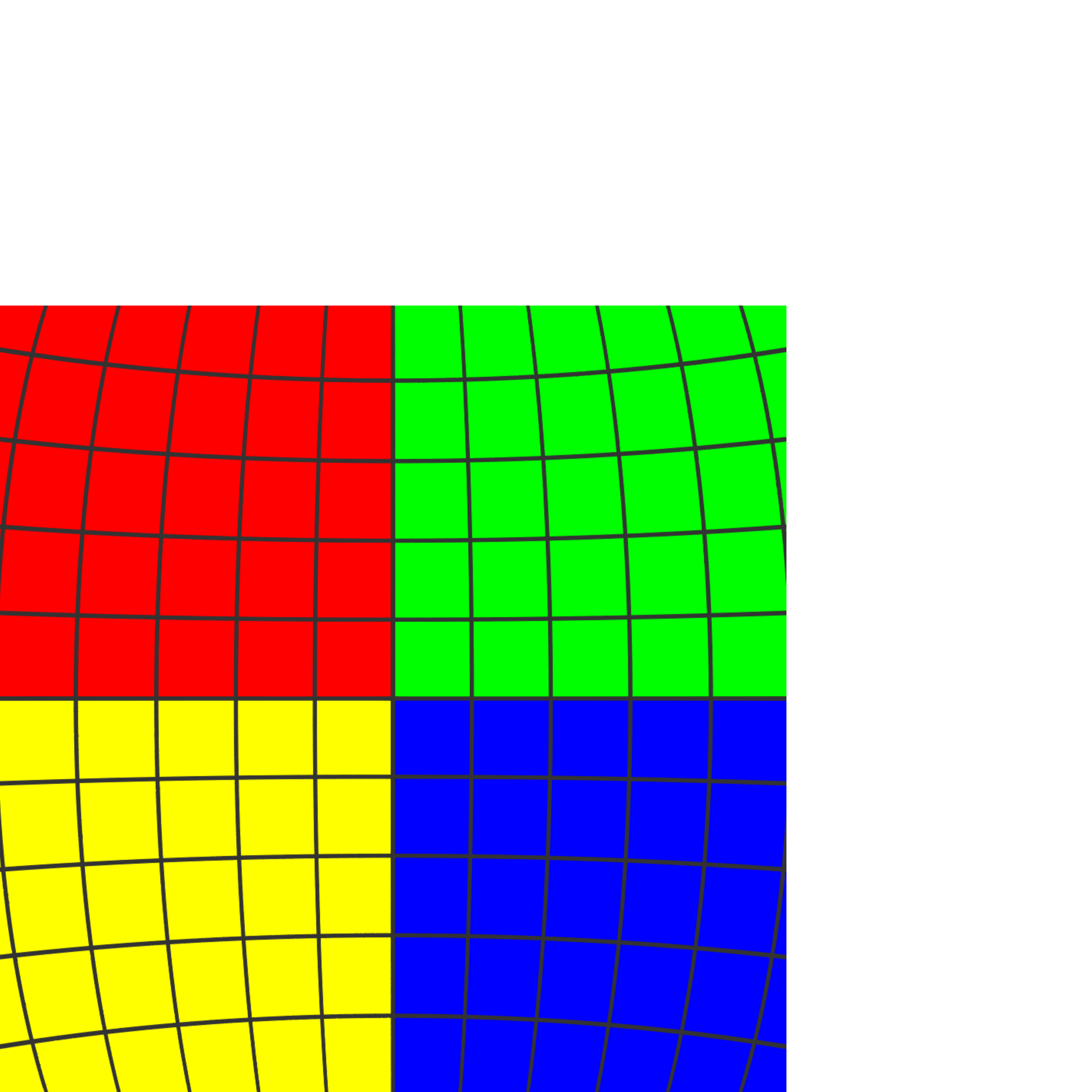}\
\includegraphics[height=.14\textheight, angle =0]{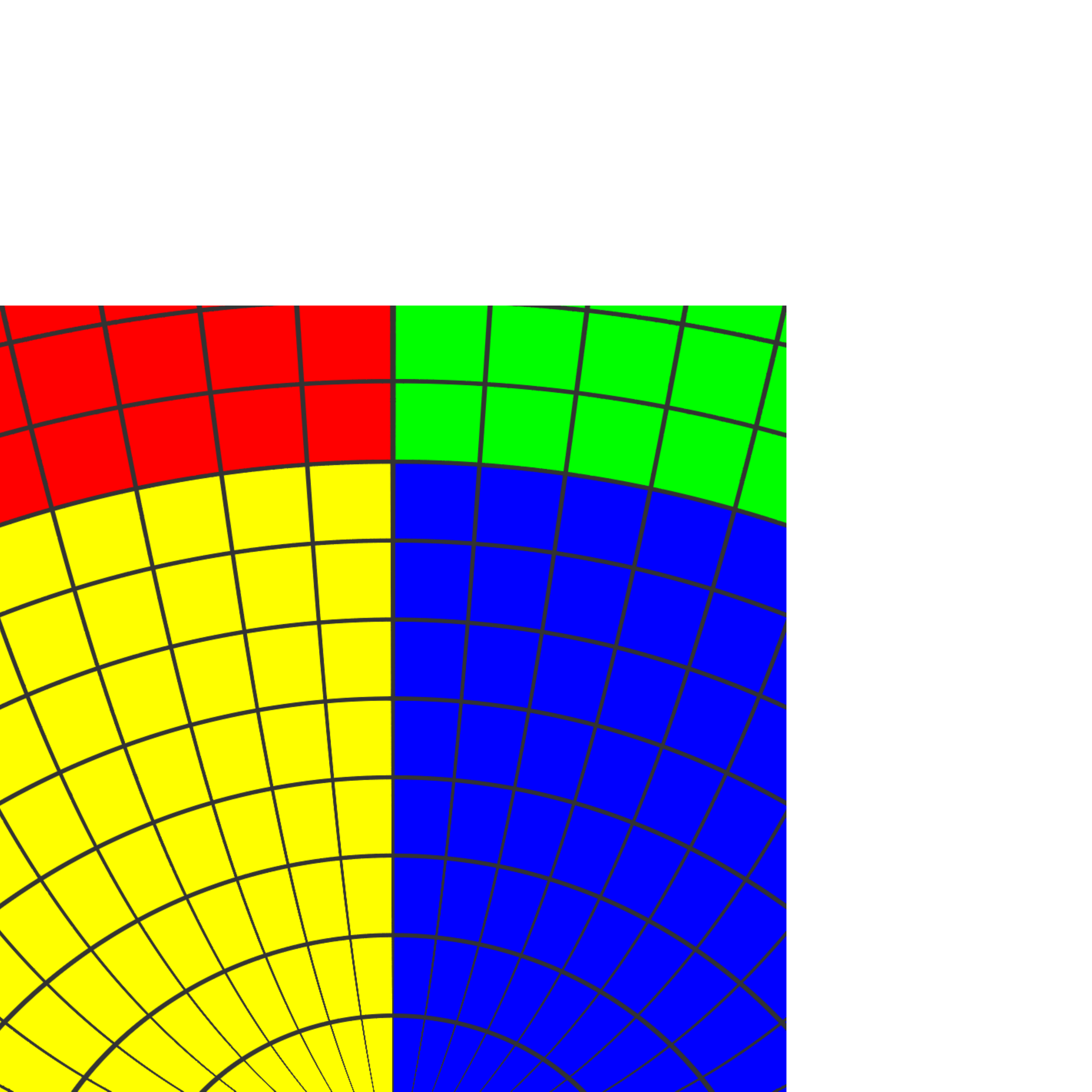}\
\includegraphics[height=.14\textheight, angle =0]{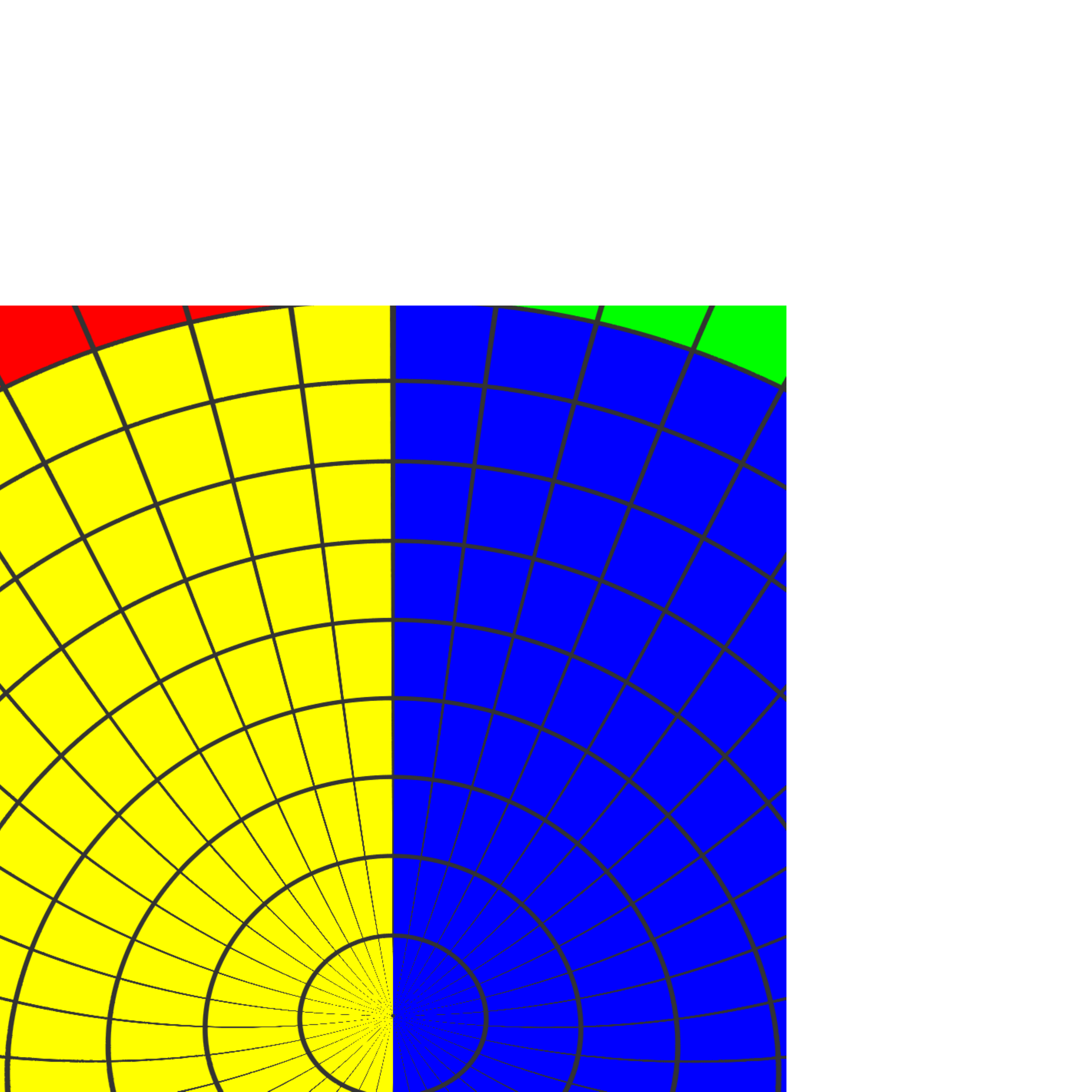}\
\includegraphics[height=.14\textheight, angle =0]{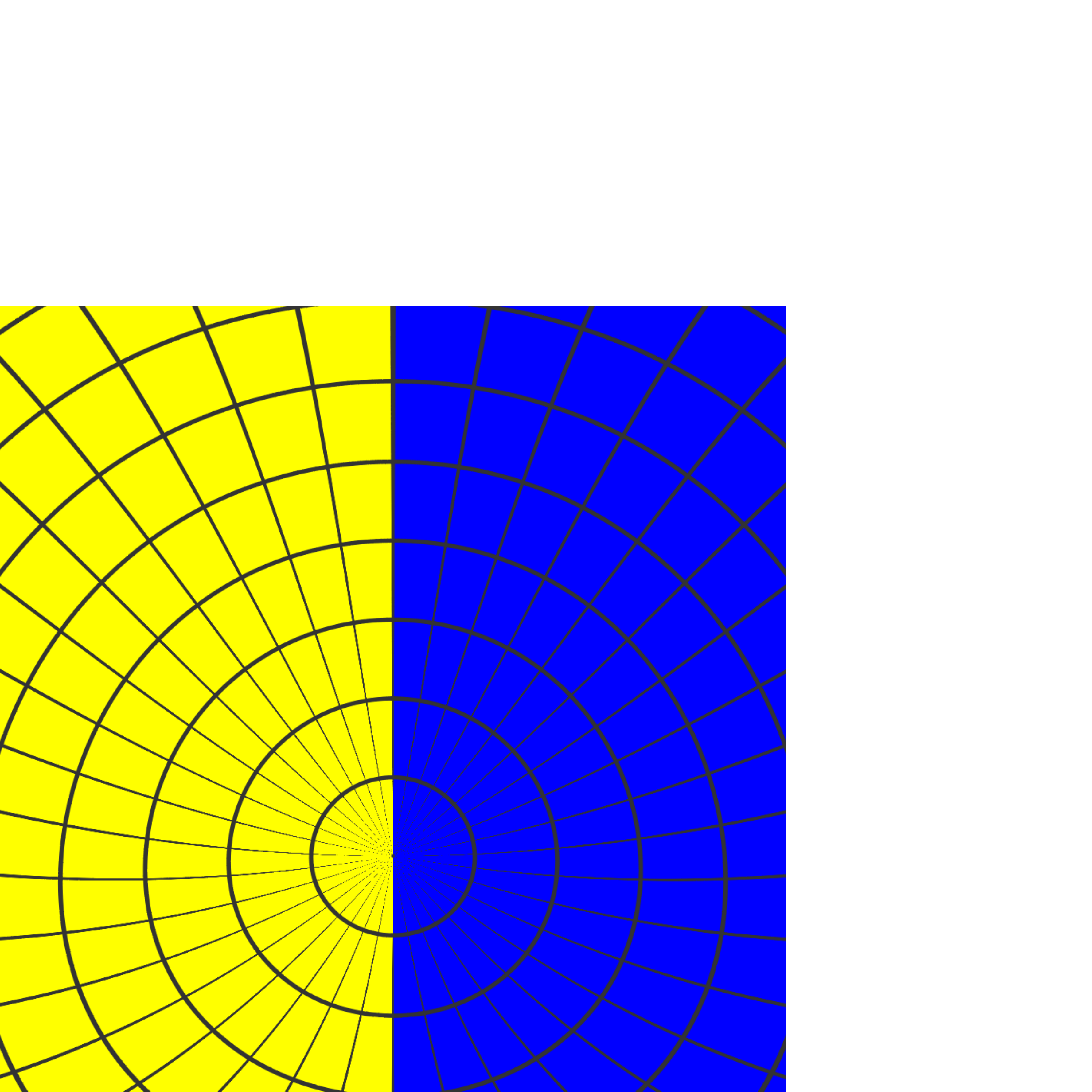}\
\includegraphics[height=.14\textheight, angle =0]{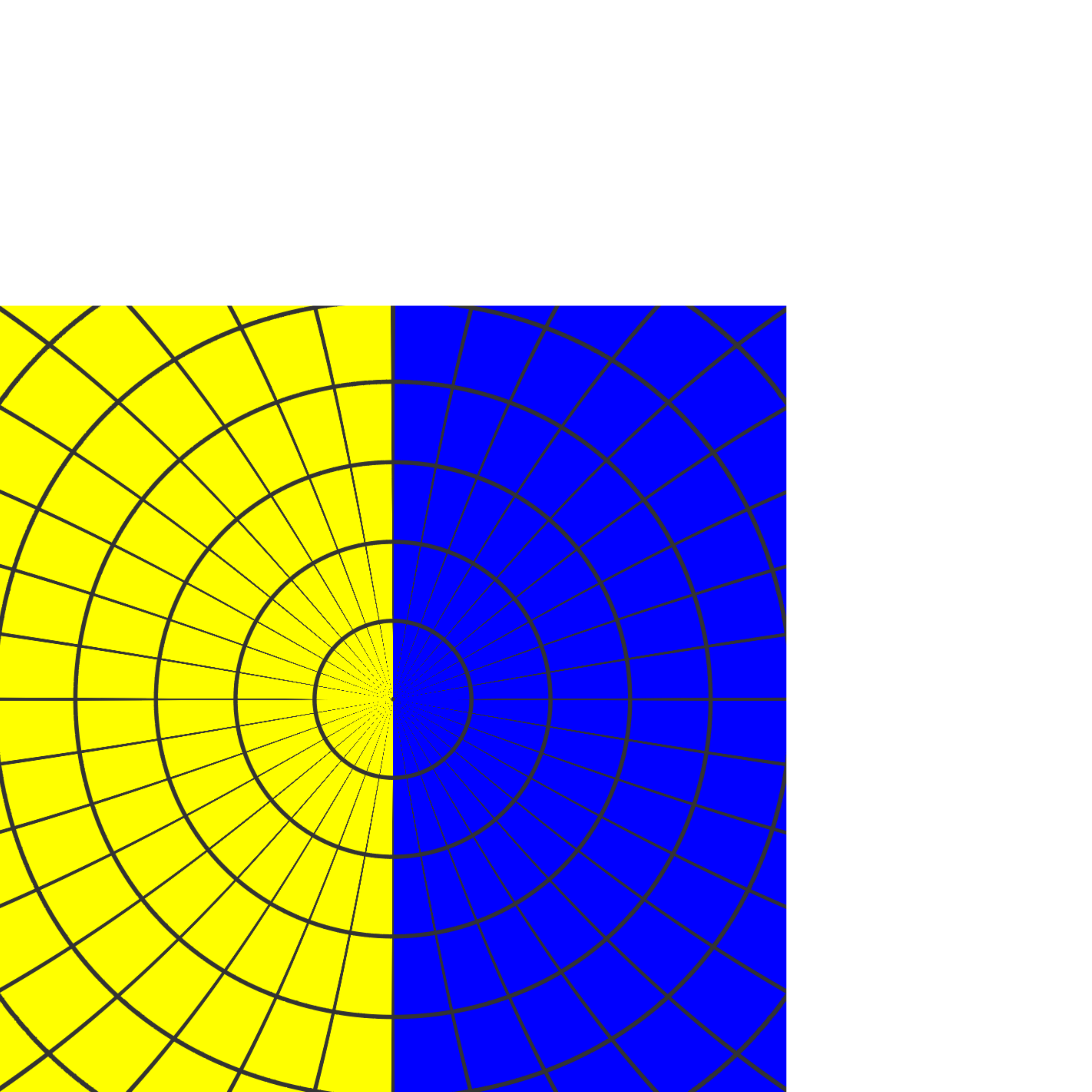}\\ \vspace{0.1cm}
\includegraphics[height=.14\textheight, angle =0]{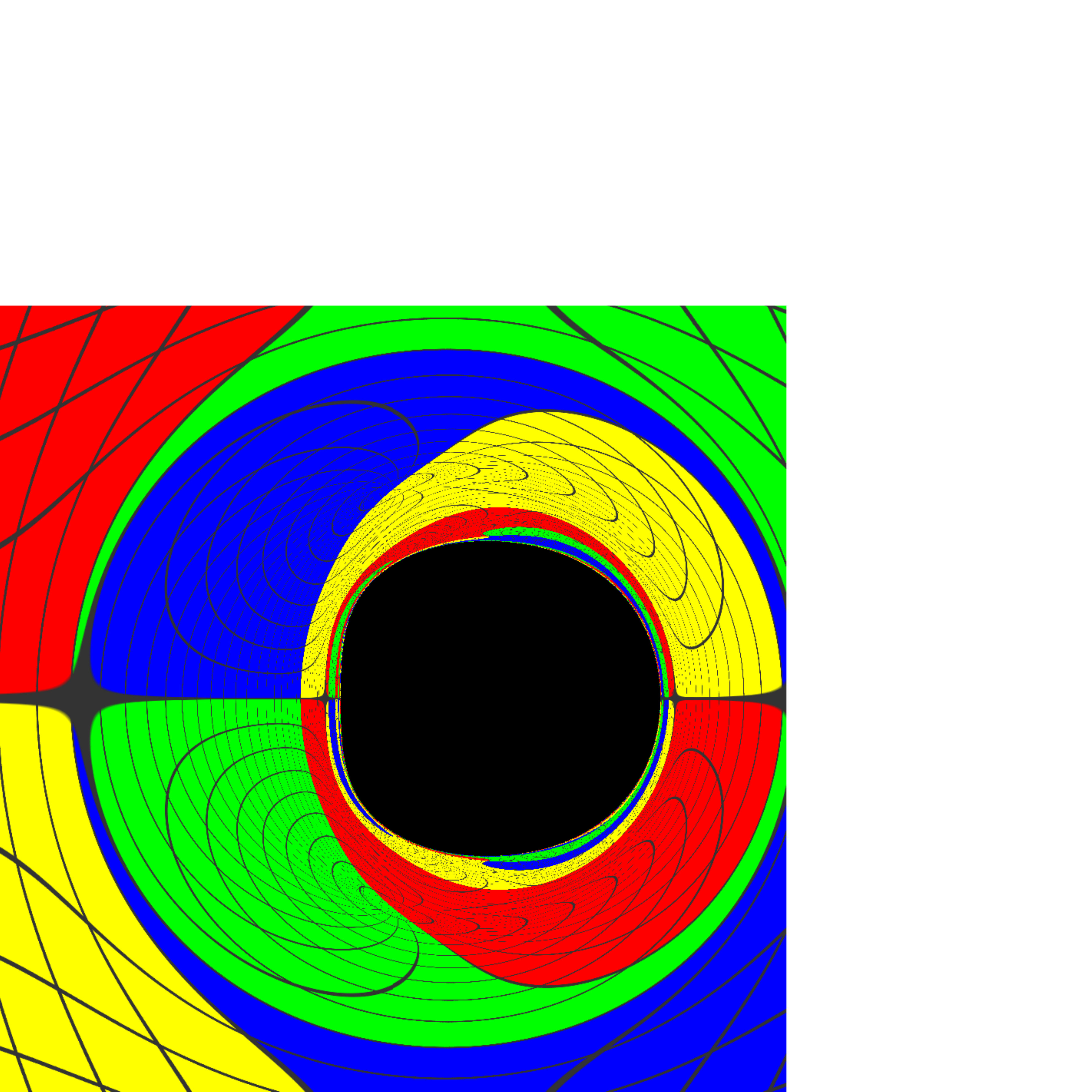}\
\includegraphics[height=.14\textheight, angle =0]{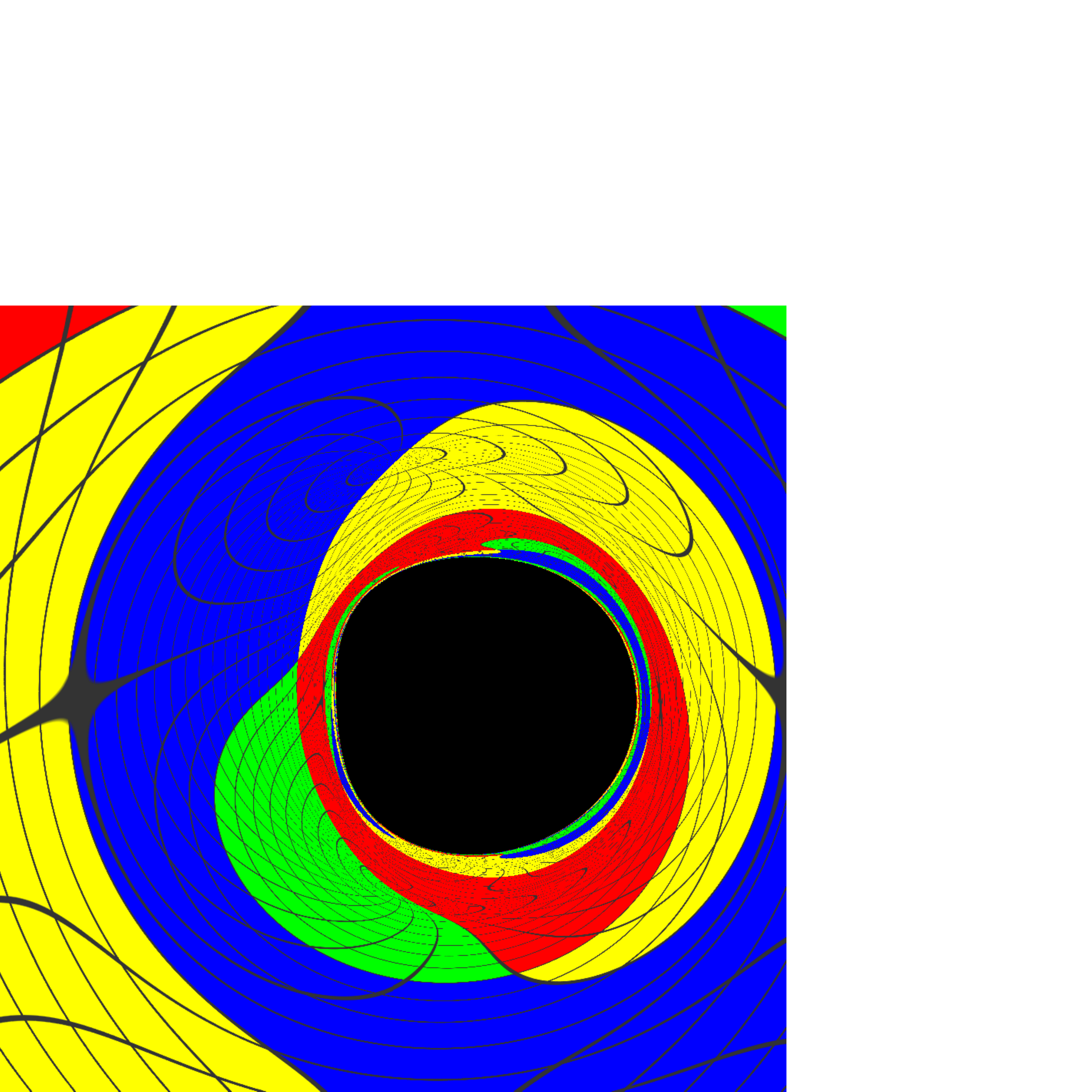}\
\includegraphics[height=.14\textheight, angle =0]{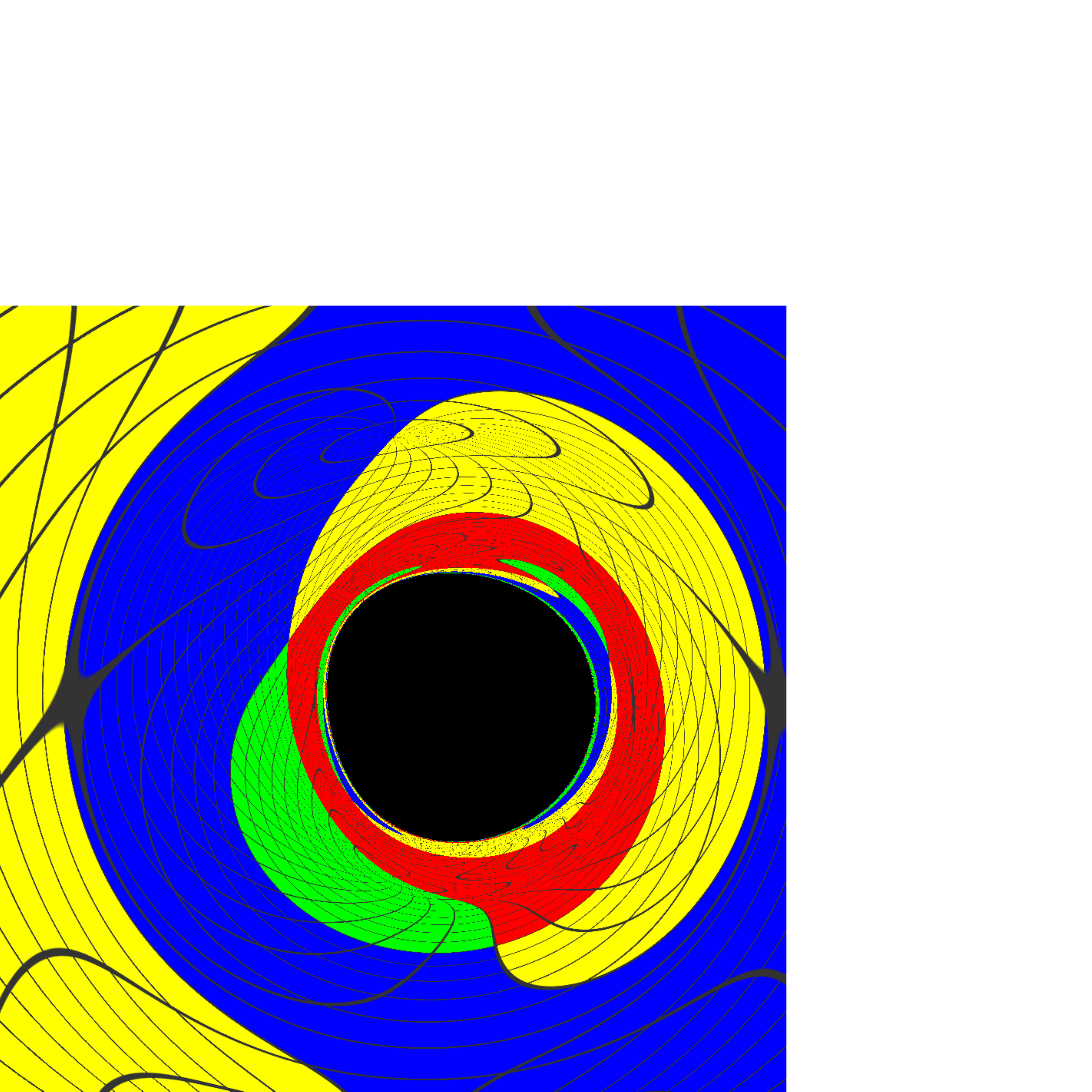}\
\includegraphics[height=.14\textheight, angle =0]{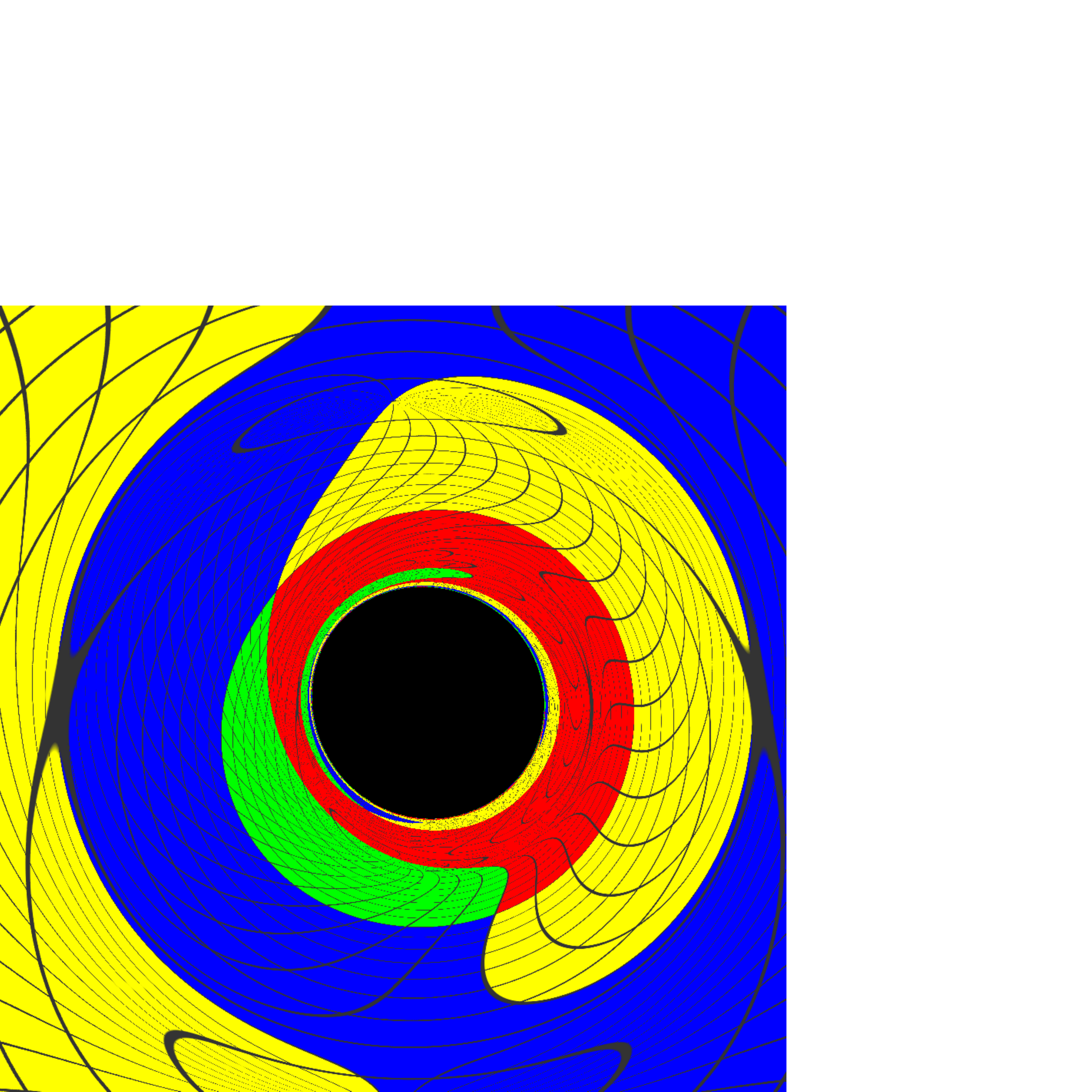}\
\includegraphics[height=.14\textheight, angle =0]{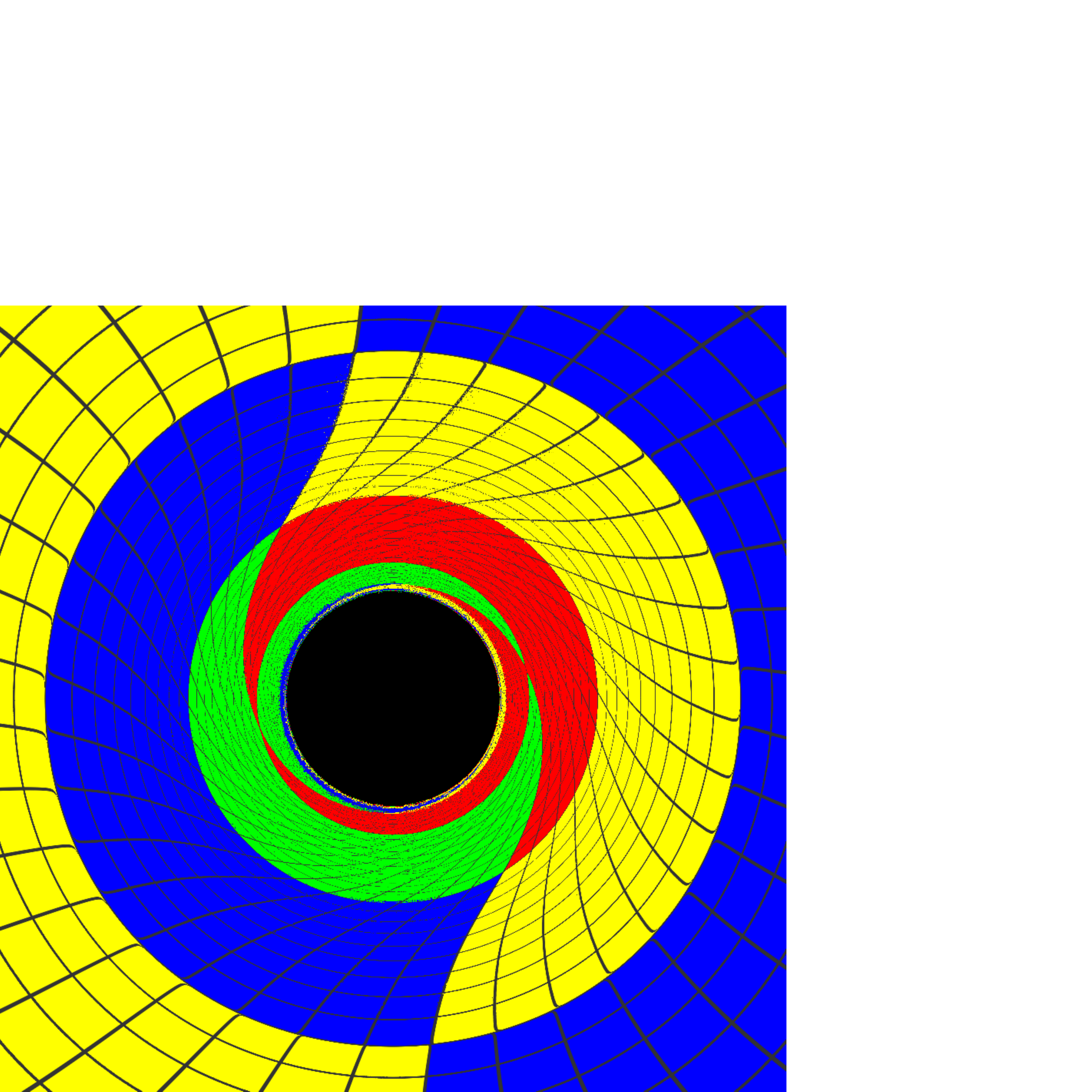} \\ \vspace{0.1cm}

\includegraphics[height=.23\textheight, angle =0]{view-J-1024x1024-90-V-NR-P.pdf}\
\includegraphics[height=.23\textheight, angle =0]{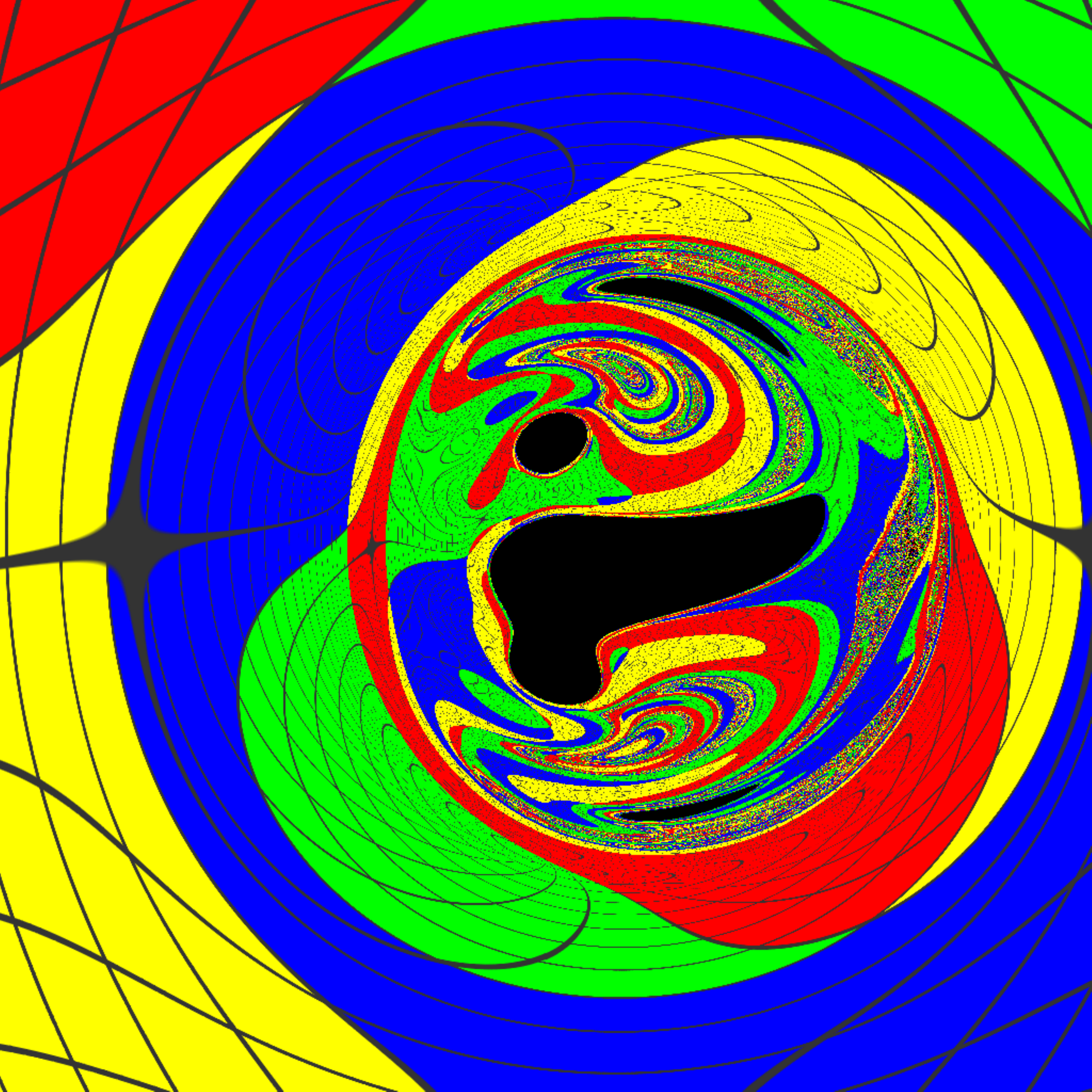}\
\includegraphics[height=.23\textheight, angle =0]{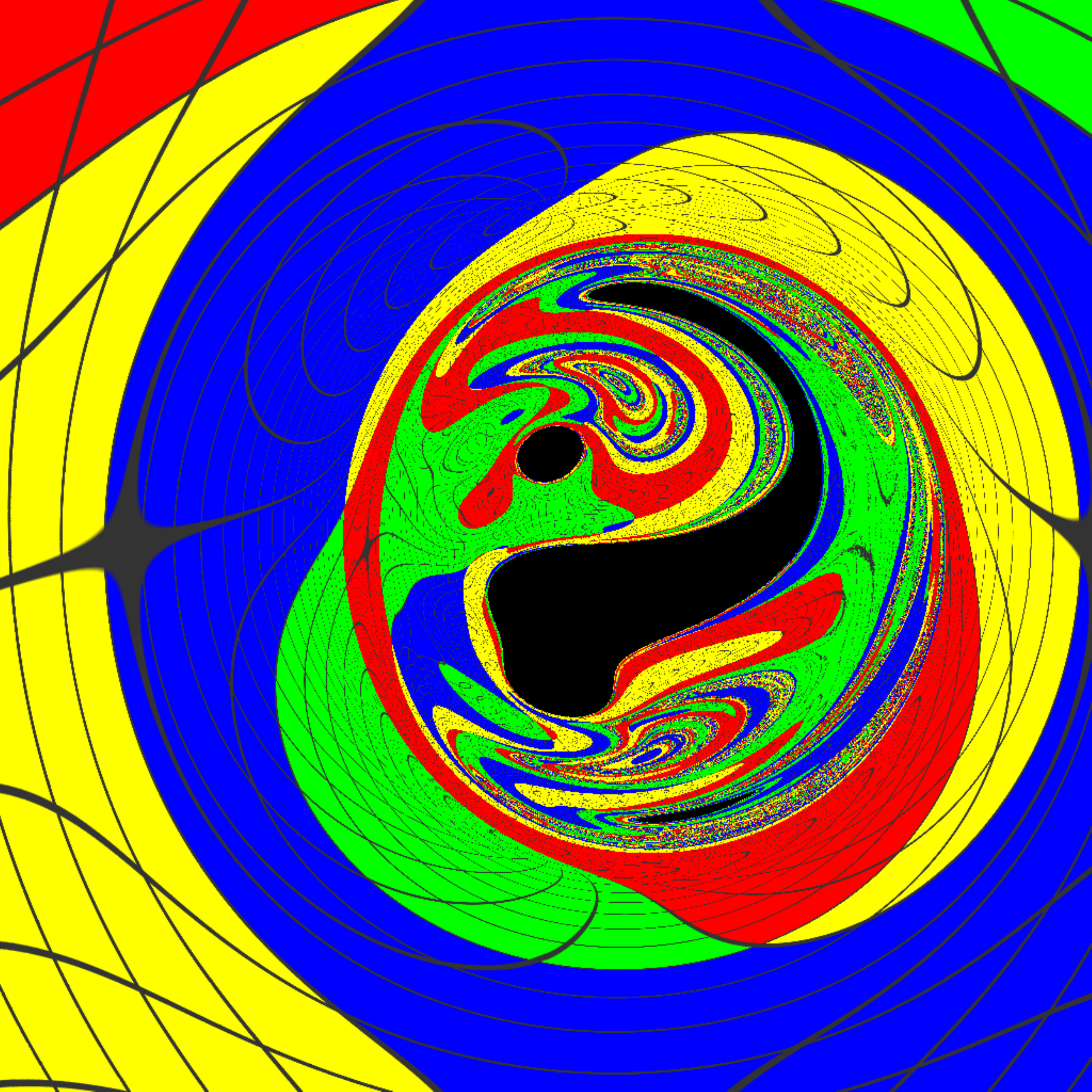}\
\includegraphics[height=.23\textheight, angle =0]{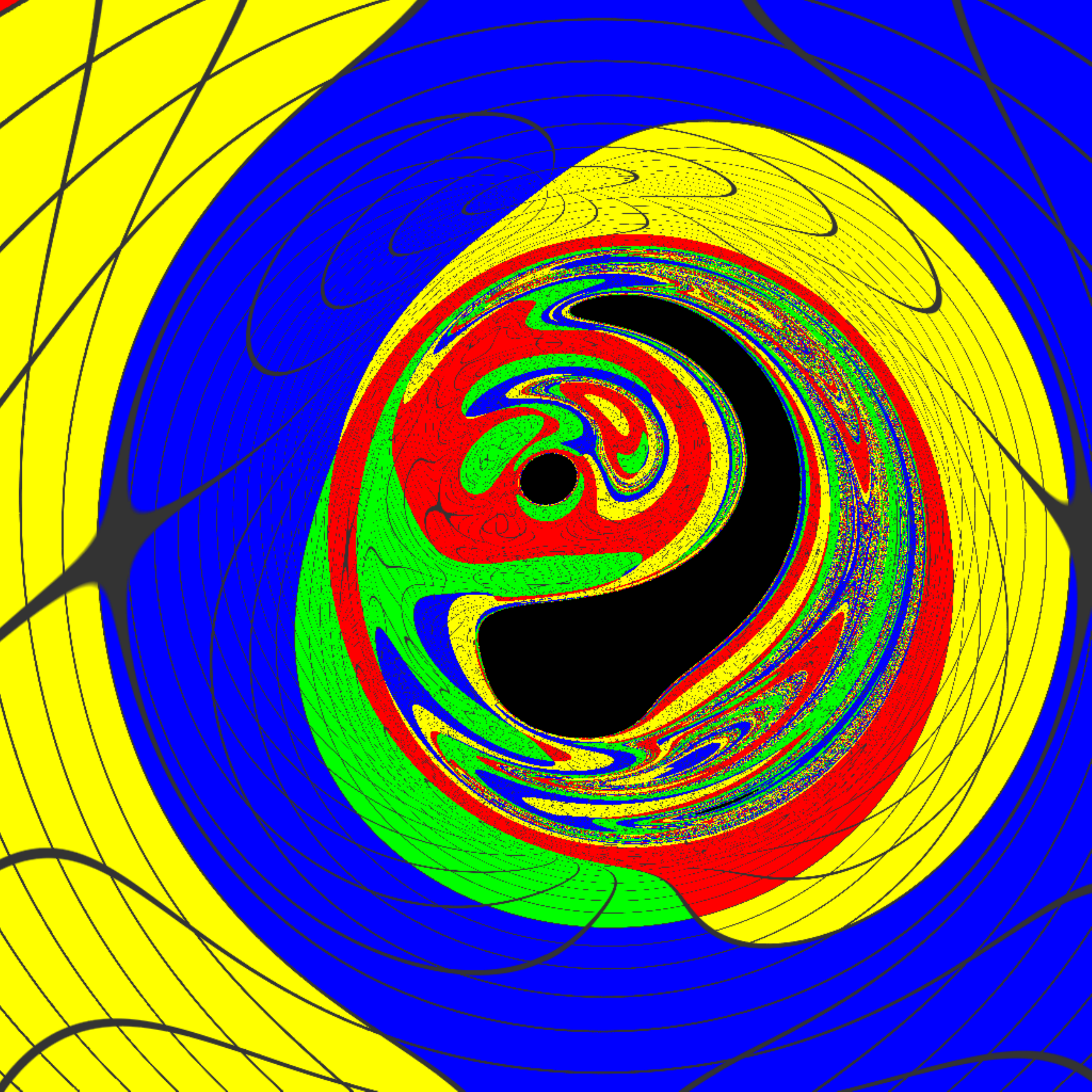}\
\includegraphics[height=.23\textheight, angle =0]{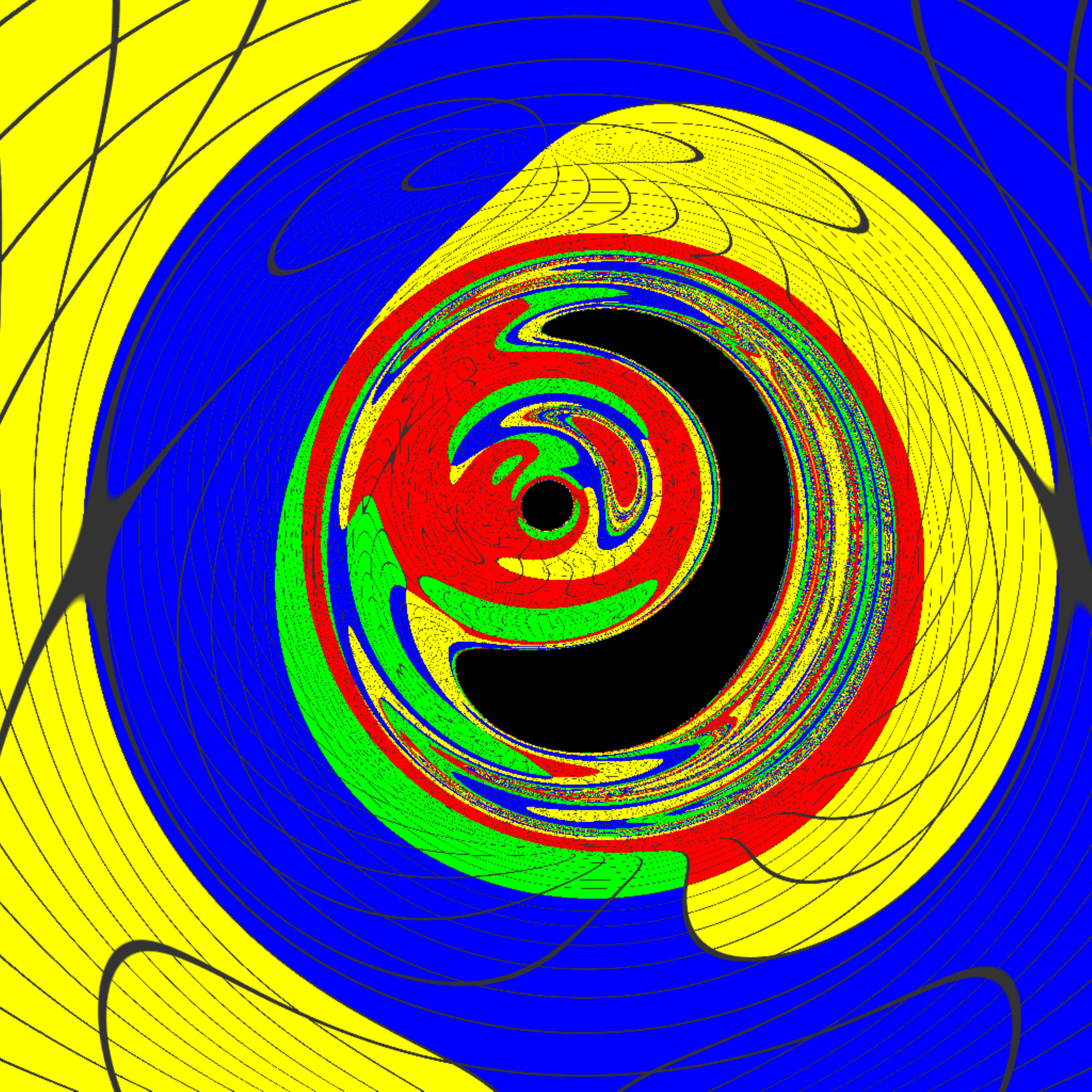}\
\includegraphics[height=.23\textheight, angle =0]{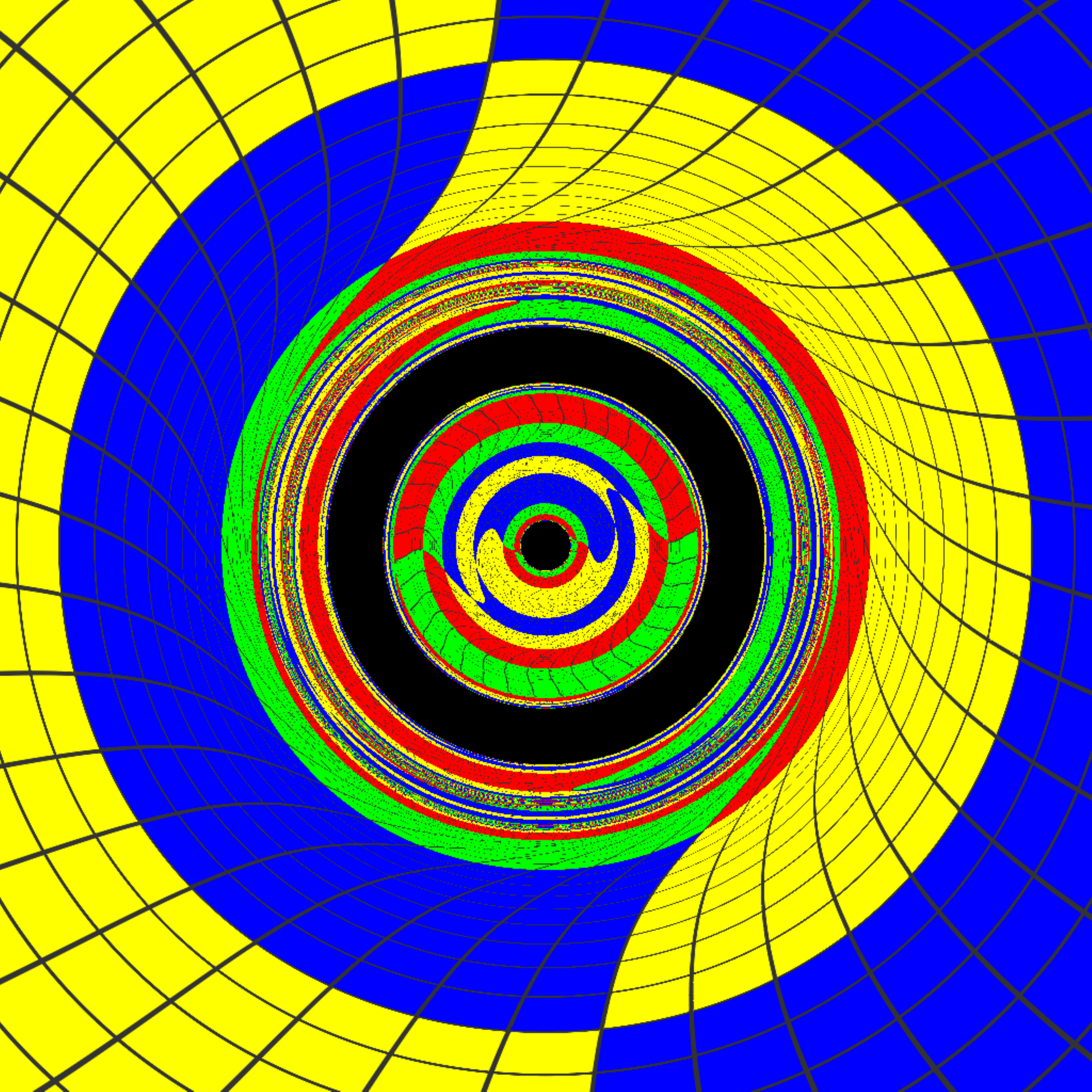}%
\end{center}
\caption{(From left to right) View from the camera of an empty space (top row) and configuration II in~\cite{Cunha:2015yba} (second row) with observer at $\theta=90,60,40,20$ and $0$. (Third and bottom rows) Shadows and lensing of configuration III in~\cite{Cunha:2015yba} with observer at $\theta=90,80,70,50,30$ and $0$. (1024$\times$1024 pixels).}
\label{emptyangles}
\end{figure}

As a test to \textsc{pyhole} we have performed the lensing of the same configurations used in~\cite{Cunha:2015yba} and obtained results in good agreement. In Fig.~\ref{emptyangles} we show the image in the camera when the metric is flat (top panels), along with configuration II  (second row) and III (third and fourth rows) in~\cite{Cunha:2015yba}. For all of these the observer is placed at different polar angles, from the equator to the pole. The images of configuration II show an expected transition from the ``squared" shadow shape observed along the equatorial plane to an axi-symmetric shadow when observed along the polar axis ($\theta=0$). A more spectacular gallery is provided by configuration III. For the latter, as we move away from the equatorial plane the main shadow splits into (at least) two disjoint pieces, and the largest of the two eventually merges with one of the (initial) eyebrows. As we approach the polar axis, the latter structure becomes annular, whereas the other piece of the main shadow that had separated from it becomes a central eye. At the pole, we obtain a \textit{Saturn-like shadow}, with the whole structure displaying axial symmetry, as it should be.

%
%

\subsection{Parallelization}
\label{pyper}

On modern computers typically many cores are available for parallel computation. Even simple consumer grade computers typically have at least 4 cores, while more powerful workstations easily reach 16 or 32 cores.
As the problem at hand is trivially parallelisable, \textsc{pyhole} has been designed to allow parallel execution via parallel processes or the MPI interface. The pixels of the image to be computed are split into chunks of roughly equal size, each of which is then propagated by a separate instance of the code. Eventually, the results are recombined in the main process to yield the full picture. To even the load on all available processors, the image is not simply subdivided into contiguous stripes, but rows are interleaved. Given $N$ available processors, the $n$-th processor computes lines $n,n+N,n+2N,n+3N,...$ of the image. This way each processor samples all regions, evening out the effect of more difficult (and hence time consuming) integrations, such as ones closer to the BH.

Furthermore, we have implemented a version of the propagator that is capable of running on modern graphical processing units (GPUs) using OpenCL\cite{opencl:2015}. These GPUs used on modern graphics cards and specialized high performance computing equipment are capable of performing parallel operations in thousands of threads at once. When using suitable algorithms, they can outperform classical CPUs by a significant margin. Furthermore, OpenCL includes the capability to run the exact same code used for GPU computing also for highly optimized parallel CPU computations. This provides a huge simplification over the manual parallelization of C or C++ based codes written for CPUs. Our OpenCL code is interfaced with \textsc{pyhole} via the PyOpenCL package\cite{kloeckner_pycuda_2012}.

Running the OpenCL implementation on an AMD W8100 FirePro graphics processor allows us to compute an entire $1024\times1024$ pixel image even of highly chaotic images such as the RBS configuration 12 in less than one hour, depending on the integrator accuracy settings.

\end{appendix}

\bibliographystyle{h-physrev4}
\bibliography{long}

\end{document}